\documentclass[twocolumn,trackchanges]{aastex631}

\usepackage{CJKutf8}
\usepackage{amsmath}
\usepackage{enumitem}
\usepackage{booktabs}

\newcommand{\HI}{\ion{H}{1}}

\newcommand{\CO}[2]{\mbox{$\mathrm{CO}\,(#1\text{--}#2)$}}

\graphicspath{{./}{./Figures/}{./figureset/}{./heightfigures/}{./weightfigures/}}

\begin{document}


\title{Modeling the Mass Distribution and Gravitational Potential of Nearby Disk Galaxies:\\ Implications for the ISM Dynamical Equilibrium}

\correspondingauthor{Vivek~Vijayakumar}
\email{vivekvijayakumar@arizona.edu}

\author[0009-0003-6076-8791]{Vivek~Vijayakumar}
\affiliation{Department of Astrophysical Sciences, Princeton University, 4 Ivy Lane, Princeton, NJ 08544, USA}

\author[0000-0003-0378-4667]{Jiayi~Sun \begin{CJK*}{UTF8}{gbsn}(孙嘉懿)\end{CJK*}}
\altaffiliation{NASA Hubble Fellow}
\affiliation{Department of Astrophysical Sciences, Princeton University, 4 Ivy Lane, Princeton, NJ 08544, USA}

\author[0000-0002-0509-9113]{Eve~C.~Ostriker}
\affiliation{Department of Astrophysical Sciences, Princeton University, 4 Ivy Lane, Princeton, NJ 08544, USA}

\author[0000-0003-4019-0673]{Enrico~M.~Di Teodoro}
\affiliation{Departement of Physics and Astronomy, University of Florence, via G. Sansone 1, 50019 Sesto Fiorentino, Italy}
\affiliation{INAF -- Arcetri Astrophysical Observatory, Largo Enrico Fermi 5, 50125 Firenze, Italy}

\author[0009-0007-7808-4653]{Konstantin~Haubner}
\affiliation{INAF -- Arcetri Astrophysical Observatory, Largo Enrico Fermi 5, 50125 Firenze, Italy}
\affiliation{Departement of Physics and Astronomy, University of Florence, via G. Sansone 1, 50019 Sesto Fiorentino, Italy}

\author[0000-0003-2896-3725]{Chang-Goo~Kim}
\affiliation{Department of Astrophysical Sciences, Princeton University, 4 Ivy Lane, Princeton, NJ 08544, USA}

\author[0000-0002-2545-1700]{Adam~K.~Leroy}
\affiliation{Department of Astronomy, The Ohio State University, 140 West 18th Avenue, Columbus, OH 43210, USA}
\affiliation{Center for Cosmology and Astroparticle Physics (CCAPP), 191 West Woodruff Avenue, Columbus, OH 43210, USA}

\author[0000-0002-0472-1011]{Miguel~Querejeta}
\affiliation{Observatorio Astron{\'o}mico Nacional (IGN), C/ Alfonso XII 3, E-28014 Madrid, Spain}

\begin{abstract}
We characterize stellar, gas, and dark matter mass distributions for 17 nearby massive disk galaxies from the PHANGS sample.
This allows us to compute the gravitational potential that vertically confines the interstellar gas and determines its equilibrium scale height and weight.
We first combine dynamical mass constraints from existing CO and H\textsc{i} rotation curves together with stellar and gas mass estimates from near-infrared, CO, and H\textsc{i} data.
These estimates incorporate current best practices in modeling stellar mass-to-light ratios and CO-to-H$_2$ conversion factor variations.
Then, we fit joint stellar--gas--dark matter mass models to the rotation curves, adopting the classic maximal disk assumption to account for remaining zero-point uncertainties on the stellar mass-to-light ratio.
After obtaining three-component radial mass profiles, we calculate the vertical equilibrium gas scale height and ISM weight in the combined gravitational potential.
We find the gas scale height $H_\mathrm{gas}$ increases from ${\lesssim}100$~pc in the inner disks to ${>}500$~pc at large radii, consistent with observations of our Galaxy and other edge-on galaxies.
The gas weight is dominated by stellar gravity at small radii, but the gas and dark matter gravity often become important beyond 3--6 times the stellar disk radial scale length.
Both our gas scale height and weight estimates are dependent on the treatment of stellar disk scale height $H_\star$, with $H_\mathrm{gas}$ varying by 30--40\% when $H_\star$ varies by a factor of 3.
The relationship between our refined ISM weight estimates and local star formation surface density generally agrees with previous observations and predictions from theory and simulations.
\end{abstract}


\section{Introduction} \label{sec:intro}

A galaxy's stellar, gas, and dark matter distribution encodes its evolutionary history in cosmological contexts \citep{Kormendy_Kennicutt_2004,Conselice_2014}. 
The gravitational potential of these components controls both stellar and gas dynamical processes \citep{vanderKruit_Freeman_2011}, contributes to regulation of the star formation rate averaged over large scales \citep[SFR; see][]{2022ApJ...936..137O,Schinnerer_Leroy_2024}, and affects the gas outflow rate \citep{Thompson_Heckman_2024}.
Thus, the gravitational potential plays a prominent role in determining galactic evolution.

Observationally, the galaxy-wide, multi-component mass distribution can be constrained with spatially resolved imaging and kinematics, which provides estimates of baryonic mass (from stars and gas) and the total dynamical mass.
This exercise has a long history starting in the early 1970s using both 21~cm and optical observations \citep[see review by][]{vanderKruit_Freeman_2011}, 
leading to the hypothesis that dark matter dominates the gravitational potential in outer regions \citep[e.g.,][]{Rubin,Bosma_1981,vanAlbada_etal_1985}.  
Since then, our understanding of mass distribution in galaxies has been greatly improved thanks to systematic surveys probing both the stellar and gaseous components, including WHISP  \citep{vanderHulst_etal_2001}, THINGS \citep{things,Walter2008}, and DiskMass surveys \citep{Bershady_etal_2010,diskmass,Martinsson_etal_2013a}.

A recurring theme in these studies is the challenge of determining an accurate stellar mass given an uncertain stellar mass-to-light ratio (M/L).
This uncertainty stems from the limitations of stellar population synthesis models \citep[SPS;][]{Bell_deJong_2001,Kauffmann_etal_2003}, and it complicates the decomposition of galaxy rotation curves.
Common remedies include assuming a maximum disk \citep{vanAlbada_etal_1985,maximaldisk} or invoking independent dynamical equilibrium considerations based on disk thickness and velocity dispersion \citep{vanderKruit_Freeman_1984,diskmass}.
With recent developments in stellar population models \citep[e.g.,][]{Sanchez_2020}, it is worth revisiting this topic in the context of galaxy mass modeling.

Another important aspect of disk galaxy mass modeling is the stellar and gas distribution perpendicular to the galaxy plane.
Direct constraints on the vertical scale heights of stellar and gas disks have been obtained primarily through photometric observations of edge-on galaxies \citep[e.g.,][]{vanderKruit_Searle_1981,Kregel2002,Yim_etal_2011,Yim_etal_2014,Salo_etal_2015}.
These studies suggested that the stellar disk scale height $H_\star$ strongly correlates with the disk's radial exponential scale length $\ell_\star$, but remains relatively constant with radius within galaxies \citep[also see][but see \citealt{Yim_etal_2011} for counter-evidence]{vanderKruit_Freeman_1984,Herrmann_etal_2008,diskmass}.
Intriguingly, the typical scale height to length ratio derived by these extragalactic photometric studies agrees with photometric measurements for the Milky Way \citep[e.g.,][]{Guerrette_etal_2024}, but these measurements disagree with dynamical modeling results for the stellar populations in the solar neighborhood \citep[e.g.,][]{2008ApJ...673..864J,Zhang2013,2022MNRAS.511.3863E}.
This tension may be due to different measurements preferentially reflecting different disk components (e.g., photometric results may be more sensitive to the thick disk) and will hopefully be resolved by future Galactic and extragalactic studies with refined modeling methods.

The galaxy mass distributions along both the radial and vertical directions have many important implications.
Among them, the dynamical equilibrium of the interstellar medium (ISM) along the vertical direction, together with its implied steady-state SFR, have been emphasized recently by both theoretical and observational studies \citep{OML2010,2015ApJ...815...67K,Sun2020,Sun_etal_2023}, especially in the context of the pressure-regulated, feedback-modulated star formation theory \citep[PRFM;][]{2022ApJ...936..137O}.
In this framework, the three-dimensional gas, stellar, and dark matter distribution determines the ISM's weight in the galaxy's gravitational potential.
This weight is balanced by the combined thermal, turbulent, and magnetic pressure (or stress) in the ISM, which requires a certain level of SFR and stellar feedback to maintain.
As such, accurate mass modeling is of critical importance when applying the PRFM theory to observed galaxies, especially for obtaining SFR predictions.
In particular, the gas scale height and weight depend on the vertical distribution of stars, and therefore are subject to potential uncertainty in the stellar scale height.

This work aims at modeling the stellar, gas, and dark matter distribution for a sample of nearby, massive, star-forming disk galaxies.
These galaxies all have high-quality, multiwavelength coverage, in particular CO line imaging from PHANGS--ALMA \citep{Leroy2021}, \HI\ data from THINGS \citep{Walter2008} and PHANGS--VLA \citep[see][]{Chiang_etal_2024}, and ancillary near-infrared imaging by \textit{Spitzer} and \textit{WISE} \citep[as aggregated in][also see \citealt{Querejeta_etal_2021}]{Sun2022}.
These datasets together provide the necessary information for reconstructing the total radial mass distribution as well as the mass distribution of gaseous components.
The same galaxies also have empirical stellar M/L ratio estimates from \citet{Leroy2019}, which are anchored to modern SPS-based results \citep{Salim_etal_2016}, enabling construction of stellar mass profiles.  
The dark matter contribution can then be constrained by combining these measurements. 
Last but not least, having three-dimensional mass models for these galaxies can critically support many active science endeavors, especially those examining the ISM dynamical equilibrium \citep{Sun2020,Eibensteiner_etal_2024} or modeling feedback-driven gas outflows (D.~Pathak et al., in preparation).

We structure this paper as follows. \S\ref{sec:data} describes several key datasets and high-level data products used in our analysis. \S\ref{sec:method} details our method for modeling the galaxy mass distributions, gravitational potential, and the ISM vertical equilibrium weight. \S\ref{sec:results} presents the modeling results and discusses their implications in the PRFM framework. \S\ref{sec:summary} summarizes all findings and provides an outlook for future work.

\section{Data} \label{sec:data}

We use existing multiwavelength data for 17 massive star-forming disk galaxies (see \autoref{table:galprops}) selected from the PHANGS--ALMA parent sample \citep{Leroy2021}.
These galaxies satisfy the follow criteria:
\begin{itemize}[itemsep=0.2em,topsep=0.5em,leftmargin=1em]
\item High-quality CO, \HI, and near-IR imaging data have been obtained from various sources (see references in \S\ref{sec:intro}). These data have been compiled and converted into radial profiles of molecular gas, atomic gas, and stellar mass surface densities by \citet[see \S\ref{sec:data:nir}--\ref{sec:data:co} for more details]{Sun2022}.
\item CO and \HI\ rotation curves are also available from either the literature \citep{rotation,DiTeodoro_Peek_2021} or our own kinematic modeling (see \S\ref{sec:data:rotcurve}).
\item For galaxies hosting stellar bars \citep[as cataloged in][]{Querejeta_etal_2021}, the semi-major axis of the bar should not exceed $2.2\,\ell_\star$, where $\ell_\star$ is the radial exponential scale length of the galaxy disk. This criterion is to mitigate the substantial imprint of stellar bars on gas kinematics by requiring reasonable radial coverage \emph{outside} their footprint for robust dynamical mass modeling (see further discussion in \S\ref{sec:result:rcmodel}).
\end{itemize}

These criteria ensure sufficient data coverage for constraining the distribution of stars, gas, and dark matter across each galaxy.
We provide high-level information about the key observational measurements involved in this process in \S\ref{sec:data:nir}--\ref{sec:data:rotcurve} and refer the reader to the original references for full descriptions.
Visual presentations of these measurements for all galaxies can be found in \autoref{apdx:gallery}.

\begin{deluxetable}{cccccc}
\small
\centering
\tablecaption{Galaxy Properties\label{table:galprops}}
\tablehead{
Galaxy & $d$ & $\log_{10}\!M_\star$ & $i$ & $\phi$ & \HI\ data \\
& [Mpc] & [$M_\odot$] & [deg] & [deg] & 
}
\colnumbers
\startdata
NGC~1087& 15.9 & 9.9 & 42.9 & 359.1 & PHANGS-VLA \\
NGC~1300& 19.0 & 10.6 & 31.8 & 278.0 & PHANGS-VLA \\
NGC~1385& 17.2 & 10.0 & 44.0 & 181.3 & PHANGS-VLA \\
NGC~1512& 18.8 & 10.7 & 42.5 & 261.9 & PHANGS-VLA \\
NGC~2283& 13.7 & 9.9 & 43.7 & 355.9 & PHANGS-VLA \\
NGC~2835& 12.2 & 10.0 & 41.3 & 1.0 & PHANGS-VLA \\
NGC~2903& 10.0 & 10.6 & 66.8 & 203.7 & THINGS \\
NGC~2997& 14.1 & 10.7 & 33.0 & 108.1 & PHANGS-VLA \\
NGC~3137& 16.4 & 9.9 & 70.3 & 359.7 & PHANGS-VLA \\
NGC~3507& 23.6 & 10.4 & 21.7 & 55.8 & PHANGS-VLA \\
NGC~3511& 13.9 & 10.0 & 75.1 & 256.8 & PHANGS-VLA \\
NGC~3521& 13.2& 11.0 & 68.8 & 343.0 & THINGS \\
NGC~3621& 7.1 & 10.1 & 65.8 & 343.8 & THINGS \\
NGC~4303& 17.0 & 10.5 & 23.5 & 312.4 & PHANGS-VLA \\
NGC~4571& 14.9 & 10.1 & 32.7 & 217.5 & PHANGS-VLA \\
NGC~4781& 11.3 & 9.6 & 59.0 & 290.0 & PHANGS-VLA \\
NGC~5042& 16.8 & 9.9 & 49.4 & 190.6 & PHANGS-VLA \\
\enddata
\tablecomments{Column descriptions: (2) galaxy distance \citep{Anand_etal_2021};
(3) global stellar mass \citep{Leroy2021}; 
(4-5) inclination and position angles \citep[derived from CO kinematics by][also consistent with \citealt{Sun2022}]{rotation};
(6) Source of \HI\ data -- 
THINGS \citep{Walter2008},
PHANGS--VLA \citep[see][]{Chiang_etal_2024}.}
\vspace{-2\baselineskip}
\end{deluxetable}

\begin{figure*}[ht!]
\gridline{
\fig{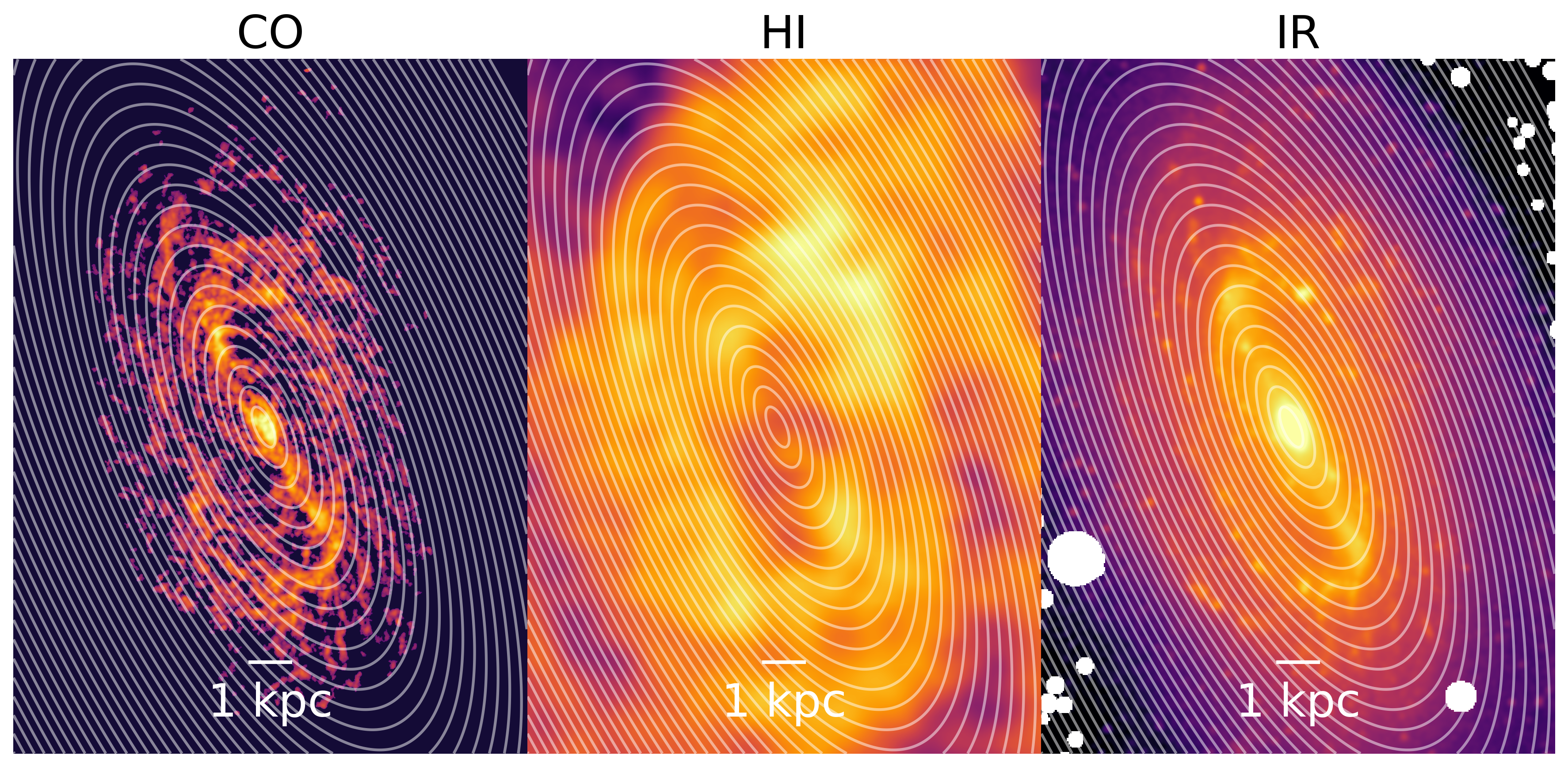}{\textwidth}{}
}
\vspace{-2.5\baselineskip}
\caption{Multiwavelength observations of NGC~2903 used in our analysis. \textit{Left:} PHANGS–ALMA \CO21\ line intensity map tracing the molecular gas distribution (\S\ref{sec:data:co}). The white ellipses mark the radial bins used for averaging \citep{Sun2022}. \textit{Middle:} THINGS \HI\ 21~cm line intensity map tracing atomic gas distribution (\S\ref{sec:data:hi}). \textit{Right:} S$^4$G 3.6~\micron\ image tracing stellar mass distribution (\S\ref{sec:data:nir}).
The CO and \HI\ data were also used to derive rotation curves and velocity dispersions (\S\ref{sec:data:rotcurve}).
\label{fig:observations}}
\end{figure*}

\subsection{Near-IR Data Tracing Stellar Distribution} \label{sec:data:nir}

Our analysis leverages near-IR imaging data taken by the IRAC instrument on the \emph{Spitzer Space Telescope}, some processed and made public as part of the S$^4$G survey \citep{Sheth2010} and others from \citet{Querejeta_etal_2021}. In particular, the measurements we use are based on a processed version of the IRAC 3.6~$\micron$ images with the background subtracted and foreground stars masked \citep[see \autoref{fig:observations}]{Sheth2010,Querejeta_etal_2021}. From there, \citet{Sun2022} created radial profiles of the 3.6~$\micron$ surface brightness for each galaxy (with a fixed radial bin width of 500~pc) and converted them into stellar mass surface densities via
\begin{equation}
\frac{\Sigma_\star}{M_\odot \textrm{ pc}^{-2}} = 350 \left(\frac{\Upsilon_\mathrm{3.6\mu m}}{0.5\,M_\odot\,L_\mathrm{\odot,3.6\mu m}^{-1}}\right) \left(\frac{I_\mathrm{3.6\mu m}}{\textrm{MJy sr}^{-1}}\right) \cos i~,
\label{eq:Sigma_star}
\end{equation}
where $I_\mathrm{3.6\mu m}$ is the 3.6~$\micron$ surface brightness within a given radial bin, $\Upsilon_\mathrm{3.6\mu m}$ is the stellar M/L ratio at 3.6~$\micron$ (see below), and $i$ is the inclination angle of the galactic disk (see \autoref{table:galprops}). The derived stellar mass surface density, $\Sigma_\star$, is what we directly use in our analyses.

The treatment of the stellar M/L ratio is long known to be of critical importance for dynamical mass modeling \citep{vanderKruit_Freeman_2011}.
In this work, we start with the empirically calibrated, spatially varying $\Upsilon_\mathrm{3.6\mu m}$ values derived by \citet{Leroy2019,Leroy2021} for all our targets.
These values are anchored against modern SPS models with \textit{GALEX}, SDSS, and \textit{WISE} data \citep{Salim_etal_2016,Salim_etal_2018}.
In essence, the $\Upsilon_\mathrm{3.6\mu m}$ values are derived for each radial bin based on the ratio of SFR surface density to near-IR surface brightness, which traces the local specific SFR and thus stellar population age.
In this way, the varying $\Upsilon_\mathrm{3.6\mu m}$ values account for systematically varying stellar age as a function of galactocentric radius within each galaxy.

While our approach allows for a radially varying M/L ratio, it could still introduce a ${\sim}0.1$~dex (or $25\%$) error on the \textit{global} stellar mass \citep[determined relative to SPS results, see appendix in][]{Leroy2019}, which is still substantial for determining the stellar contribution to the total dynamical mass.
We thus allow for a multiplicative factor on top of the existing $\Upsilon_\mathrm{3.6\mu m}$ estimates to account for any residual ``zero point offset.''
This multiplicative factor is a fixed number for each galaxy, and we describe its derivation in \S\ref{sec:method:scaling}.
In this way, our M/L ratio treatment attempts to combine the advantage of SPS-based and maximum disk approaches, simultaneously accounting for radial stellar age variations within galaxies and ensuring reasonable decomposition of total dynamical mass into stellar and non-stellar (gas and dark matter) contributions.

\subsection{\texorpdfstring{\HI}{HI} Data Tracing Atomic Gas Distribution} \label{sec:data:hi}

We make use of \HI\ 21~cm line observations taken by the Very Large Array (VLA), as part of the THINGS \citep{Walter2008} and PHANGS--VLA surveys \citep[see][]{Chiang_etal_2024}. 
From these data, \citet{Sun2022} derived radial profiles of the \HI\ line integrated intensity and subsequently the atomic gas surface density via
\begin{equation}
\frac{\Sigma_\mathrm{atom}}{M_\odot \textrm{ pc}^{-2}} = 0.020 \left(\frac{I_\mathrm{HI}}{\textrm{K km s}^{-1}}\right) \cos{i}~,
\label{eq:Sigma_atom}
\end{equation}
where $I_\mathrm{HI}$ is the average \HI\ line integrated intensity in each radial bin. 
We use the derived atomic gas surface density ($\Sigma_\mathrm{atom}$, including the helium contribution) in our analyses.

\subsection{CO Data Tracing Molecular Gas Distribution} \label{sec:data:co}

We also make use of \CO21\ line data from the Atacama Large Millimeter/sub-millimeter Array (ALMA), obtained as part of the PHANGS-ALMA survey \citep{Leroy2021}. 
These data allowed \citet{Sun2022} to derive radial profiles of the \CO21\ line integrated intensity, $I_\mathrm{CO(2-1)}$.
We then convert these $I_\mathrm{CO(2-1)}$ values into molecular gas surface densities via
\begin{equation}
\Sigma_\mathrm{mol} = \alpha_\mathrm{CO(2-1)}\,I_\mathrm{CO(2-1)} \cos{i}~,
\label{eq:Sigma_mol}
\end{equation}
where $\alpha_\mathrm{CO(2-1)}$ is a varying CO(2--1)-to-H$_2$ conversion factor, derived following current best practices suggested by \citet{Schinnerer_Leroy_2024}.
These are based on empirically calibrated scaling relations that attempt to account for CO-dark gas, CO emissivity, and CO excitation variations within and among galaxies. The molecular gas surface density $\Sigma_\mathrm{mol}$ derived from \autoref{eq:Sigma_mol} is the measurement that enters the following analyses.

\subsection{CO and \texorpdfstring{\HI}{HI} Kinematic Measurements} \label{sec:data:rotcurve}

We use CO and \HI\ rotation curves derived from the aforementioned ALMA and VLA datasets. 
The CO rotation curves were measured by \citet{rotation}, assuming fixed position and inclination angles across each galaxy disk.
These position and inclination angles are quoted in \autoref{table:galprops} and were also used in \citet{Sun2022} for radial binning and deriving profiles of $\Sigma_\star$, $\Sigma_\mathrm{atom}$, and $\Sigma_\mathrm{mol}$ (\S\ref{sec:data:nir}--\ref{sec:data:co}).
Overall, the CO rotation curves and associated orientation parameters are fairly consistent with literature measurements for a subset of our targets \citep[e.g.,][based on older HERACLES CO data]{Frank2016}.

The \HI\ rotation curves come from two sources.
For galaxies in the THINGS sample, we use the rotation curves derived by \citet{DiTeodoro_Peek_2021} with 3D-Barolo \citep{DiTeodoro_Fraternali_2015}.
We have verified that these rotation curves are quantitatively consistent with those derived and presented by \citet{things} and \citet{Frank2016} based on the same THINGS data.
For the other galaxies, we perform similar kinematic modeling with 3D-Barolo (version~1.7) on the PHANGS--VLA \HI\ data cubes.
Our modeling followed a standard two-step approach, with the first step devoted to determining the galaxy geometry (i.e.\ its center, inclination and position angles, systemic velocity) and the second step aimed to derive the kinematical parameters (the rotation velocity and the intrinsic gas velocity dispersion).
The inclination and position angles are allowed to vary as a function of galactocentric radius, which is appropriate for extended \HI\ disks which often warp at large radii.
For further technical details on the modeling procedure, we refer to \citet{DiTeodoro+2023}. 

Beside rotation curves, we also use measurements of CO and \HI\ velocity dispersion for the purpose of modeling the vertical thickness of the gas disks (\S\ref{sec:method:height}).
The CO velocity dispersion is measured at fixed 150~pc scales \citep{Sun2018,Sun2020a} and then averaged over each radial bin in a CO intensity-weighted fashion \citep{Sun2022}.
A moderate inclination correction is also applied to the CO velocity dispersion following \citet{Sun2022}.
The \HI\ velocity dispersion per radial bin is provided by 3D-BAROLO as an output parameter.

\section{Methodology} \label{sec:method}

Based on the observational data described in \S\ref{sec:data}, we model the mass distributions of stars, gas, and dark matter, and from these we derive their combined gravitational potential, the weight of the ISM in this potential, and the thickness of the ISM disk (assuming vertical dynamical equilibrium), for each of our target galaxies.
To achieve these goals, we first model the radial mass profiles of the stars and gas in each galaxy with the azimuthally-averaged surface density profiles from \citet{Sun2022}.
We then determine the dark matter radial distribution by calculating the stellar and gas gravitational potentials and subtracting their contributions from the rotation curves.
To obtain a smooth dark matter potential, we assume the dark matter halos follow a spherically symmetric, Navarro-Frenk-White profile \citep[NFW;][]{NFW}. 
After obtaining the radial mass profiles for stars, gas, and dark matter, we compute the equilibrium weight of the gas within the combined gravitational potential as well as the equilibrium vertical scale height of the gas disk.
We explore different assumptions regarding the vertical scale height of the stellar disk, and the impacts of this choice on the modeling results. 

\subsection{Stellar and Gas Disk Potentials}
\label{sec:method:potential}

To prepare for gravitational potential calculations for the stellar and gas components, we first create smooth models of their radial mass profiles from the observed, discrete, bin-averaged radial profiles (\S\ref{sec:data:nir}--\ref{sec:data:co}).

\begin{figure}[t!]
\gridline{
\fig{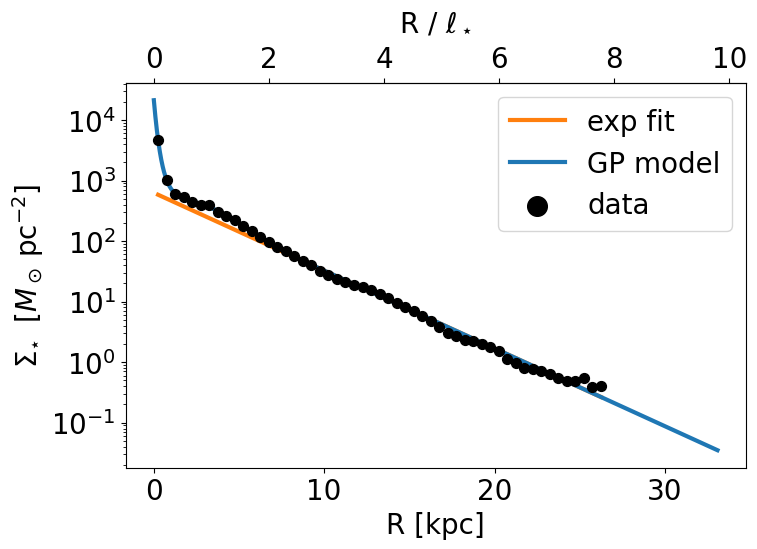}{0.47\textwidth}{}
}
\vspace{-2\baselineskip}
\caption{The radial profile of stellar mass surface density for NGC~2903. The data points represent the observed average surface density in radial bins from \citet{Sun2022}. The blue curve shows a smooth model created with Gaussian process regression and used in gravitational potential calculations (\S\ref{sec:method:potential}). The orange line shows an exponential fit for the outer disk, with a radial scale length of $\ell_\star=3.38$~kpc. This scale length is used for inferring the stellar disk vertical scale height (\S\ref{sec:method:height}) and for extrapolating the stellar mass surface density (\S\ref{sec:method:potential}).}
\label{fig:starprofiles}
\end{figure}

The observed stellar and gas mass surface density profiles show obvious fluctuations between adjacent radial bins.
This is not unexpected given measurement uncertainties and the true inhomogeneity in the distributions, but it can cause issues in the subsequent calculations of gravitational potential and circular velocity (as they involve numerical integrals and derivatives).
To mitigate these issues, we perform a Gaussian process regression to damp out small-scale fluctuations in the surface density profiles (see \autoref{fig:starprofiles}).
This effective ``smoothing'' is achieved by adopting a radial basis function kernel with a relatively large length scale to ensure covariance over that scale.
We test different length scales on our data and visually inspect the outcomes, finding that the optimal length scales for the stellar and gas profiles are 3~kpc and 1.5~kpc, respectively.

For a subset of galaxies, the radial coverage of their stellar surface density profiles are more limited than their \HI\ rotation curves.
This is primarily due to an outer radius threshold imposed by \citet[out to 1.5 times their $\rm 25\,mag/arcsec^2$ isophotal radii]{Sun2022} when creating the surface density profiles.
Considering that these galaxies are well described by exponential profiles near the outskirts, we fit exponential models to their outer parts and extrapolate them to match the corresponding \HI\ rotation curve coverage.
This allows us to fully utilize the \HI\ kinematic information at large radii, which is especially important for constraining the dark matter halo profiles (also see \autoref{sec:method:rcfit}).

From the interpolated and smoothed surface density profiles for the stars and gas, we compute their gravitational potentials and corresponding contributions to the galaxy rotation curve. For the stellar disk, we follow \citet[Eq.~8.72]{Bovy} to derive the potential near the galaxy midplane as a function of radius, assuming an exponential vertical distribution with a constant vertical scale height $H_\star$,
\begin{align}
\Phi(R, z{=}0) &= -\frac{2\pi G}{H_\star}\!\int_{0}^{\infty}\!dk \frac{H_\star}{1+kH_\star} \ J_0(kR) \ S_0(k)\\
\text{where } &S_0(k) = \int_{0}^{\infty}\!dR'\ J_0(kR') \ R'  \ \Sigma(R') ~.
\label{eq:thickpotential}
\end{align}
We further discuss choices of the stellar disk vertical scale height $H_\star$ and their rationale in \ref{sec:method:height}.

For the gas disk, we assume a geometrically thin disk and follow \citet[Eq.~2.265]{2008gady.book.....B} to derive the potential,
\begin{equation}
\Phi(R, z{=}0) = -4G\!\int_{0}^{\infty}\!dR'\frac{R'\Sigma(R')}{R+R'}K\!\left(\frac{2\sqrt{RR'}}{R+R'}\right)~,
\label{eq:thinpotential}
\end{equation}
where $K$ represents the complete elliptic integral of the first kind.

We derive the potentials for each component, $\Phi_\star(R)$ and $\Phi_\mathrm{gas}(R)$, by numerically integrating \autoref{eq:thickpotential} and \autoref{eq:thinpotential} using the corresponding surface density functions, $\Sigma_\star(R)$ and $\Sigma_\mathrm{gas}(R)$, respectively. From the potentials we can derive the circular velocity contributed by each mass component via
\begin{align}
    V^2_\star(R) &= R \frac{d\Phi_\star}{dR}~, \label{eq:vcirc_star}\\
    V^2_\mathrm{gas}(R) &= R \frac{d\Phi_\mathrm{gas}}{dR}~. \label{eq:vcirc_gas}
\end{align}

\subsection{Rescaling of the \texorpdfstring{\HI}{HI} Rotation Curves and the Stellar Potential}
\label{sec:method:scaling}

When comparing the CO and \HI\ rotation curves (\S\ref{sec:data:rotcurve}) and the derived $V_\star(R)$ profiles (from \autoref{eq:vcirc_star}), we notice two issues.
The first issue is that the CO and \HI\ rotation curves do not always agree within the radius range where both are available.
This may be caused by the different spatial and velocity resolutions of the CO and \HI\ data and the sometimes inconsistent inclination angles derived from them.
To address this, we scale the entire \HI\ rotation curve by a multiplicative factor $f_\mathrm{HI}$:
\begin{equation}
V_\mathrm{\mathrm{HI},scaled}(R) = f_\mathrm{HI} V_\mathrm{HI}(R)~,
\label{eq:HIscaling}
\end{equation}
\noindent such that it matches the CO rotation velocity in the outer part of the CO radial coverage to ensure continuity.
Here we use the CO-based results as our ``anchoring points'' for internal consistency, as the surface density and velocity dispersion measurements from \citet{Sun2022} were all derived assuming the CO-based inclination angles.
The resulting correction is small for most galaxies, typically $|1-f_\mathrm{HI}|\,{\lesssim}\,10\%$ (see \S\ref{sec:result:rcmodel}).
After that, we merge the CO rotation curve with the scaled \HI\ curve at larger radii to form a combined rotation curve, $V_\mathrm{c}(R)$, to be used in further analyses\footnote{Since the \HI\ rotation curves have larger radial bin width than the CO curves, we also assign proportionally larger weights to each of the \HI\ data points when using the hybrid rotation curves in further analyses.}. 

The second (perhaps more important) issue is that the $V_\star(R)$ profile derived directly from \autoref{eq:vcirc_star} appears inconsistent with the measured $V_\mathrm{c}(R)$ in many cases.
That is, we expect stellar gravity be dominant in the inner parts of massive disk galaxies, so that $V_\star(R)$ should be smaller than but very close to $V_\mathrm{c}(R)$.
However, we find $V_\star(R)$ to be much lower than $V_\mathrm{c}(R)$ in some cases (such that stellar contribution become very subdominant) and to exceed $V_\mathrm{c}(R)$ in a few other cases (which is unphysical).
This likely reflects uncertainties from various sources, including the stellar M/L ratio (see \S\ref{sec:data:nir}), galaxy distance (which affects the overall normalization of the stellar contribution), and galaxy inclination angle (which affects the normalization of the rotation curve).
To address this issue, we re-scale the stellar contribution to the circular velocity with a multiplicative factor, $f_\star$.

\begin{equation}
V_\mathrm{\star,scaled}(R) = f_\star V_\star(R)
\label{eq:stellarscaling}
\end{equation}

The value of $f_\star$ is determined for each galaxy from the classic maximum disk assumption \citep{maximaldisk}, such that
\begin{equation}
\sqrt{V^2_\mathrm{\star,scaled} + V^2_\mathrm{gas}} = 0.85 \times V_\mathrm{c}\quad \mathrm{at} \quad R=2.2\ell_\star~.
\label{eq:maximumdisk}
\end{equation}
\noindent Here $\ell_\star$ is the radial exponential scale length of the stellar disk, which we measure for each galaxy by fitting an exponential function to the $\Sigma_\star$ profile within the disk-dominated radius range, where the range is determined by outlier rejection (see \autoref{fig:starprofiles}, and \autoref{table:scalingpars} for the measured $\ell_\star$ values).
\autoref{eq:maximumdisk} uses $2.2 \times \ell_\star$ as the ``anchoring radius'' because it is where $V_\star(R)$ would reach its maximum for a pure exponential disk.
The choice to normalize the total baryonic contribution to 0.85 times the dynamical mass implied by the circular velocity is empirically determined in previous studies \citep[e.g.,][]{maximaldisk}.

\subsection{Rotation Curve Fitting}
\label{sec:method:rcfit}

After merging the CO and \HI\ rotation curves and scaling $V_\star(R)$ according to the maximum disk assumption, we fit a combined stellar--gas--dark matter model to the rotation curve:
\begin{equation}
V_\mathrm{c}^2(R) = V_\mathrm{\star,scaled}^2(R) + V_\mathrm{gas}^2(R) + V_\mathrm{DM}^2(R)~.
\label{eq:rcmodel}
\end{equation}
\noindent In this model, the stellar and gas components do not have free parameters as they are given by gravitational potential calculations (\S\ref{sec:method:potential}), with the stellar component rescaled according to the maximum disk assumption (\S\ref{sec:method:scaling}).
The dark matter component is parametrized according to the cumulative version of the NFW profile \citep{NFW},
\begin{align}
V_\mathrm{DM}^2(r)
&= \frac{GM_\mathrm{DM}({<}r)}{r} \nonumber\\
&= \frac{4\pi G\rho_0 R_s^3}{r}\left[\ln\!\left(1+\frac{r}{R_s}\right) - \frac{r/R_s}{1+r/R_s}\right]~.
\label{eq:NFW_vsquared}
\end{align}
Here $R_s$ is the dark matter halo scale radius and $\rho_0$ is the normalization factor.
With the best-fit $\rho_0$ and $R_s$, the dark matter density profile can be expressed as
\begin{equation}
\rho_\mathrm{DM}(r) = \frac{\rho_0}{\frac{r}{R_s}\left(1+\frac{r}{R_s}\right)^2}~.
\label{eq:NFW_rho}
\end{equation}

In some of our galaxies, we do not have sufficient coverage at larger radii (where the dark matter dominates the potential) to separately constrain $R_s$ and $\rho_0$. To remedy this, we compare the fitting results between a two-parameter model with varying $R_s$ and a one-parameter model with a fixed fiducial $R_s = 9H_\star$ \citep[following empirical scaling relations reported in][]{Somerville2025}. We use the Bayesian information criterion (BIC) to determine which model best fits the rotation curve of each galaxy.

\subsection{Disk Scale Height}
\label{sec:method:height}

To convert the 2D, in-plane stellar and gas surface density profiles into full 3D mass distributions, we also need the vertical scale heights of the stellar and gas disks.
For the gas disk, we compute its scale height $H_\mathrm{gas}$ from the vertical dynamical equilibrium condition, which relates $H_\mathrm{gas}$ to the gas effective velocity dispersion $\sigma_\mathrm{eff}$ and the galaxy gravitational potential.
In practice, this requires solving a cubic equation \citep[see][equation~15]{Hassan_etal_2024}:
\begin{align}
H_\mathrm{gas}^3 &\left(  \frac{2\zeta\Omega_\mathrm{DM}^2  }{\pi G \Sigma_\mathrm{gas}} \right) +
H_\mathrm{gas}^2 \left(1 + \frac{2\Sigma_\star}{\Sigma_\mathrm{gas}} + \frac{2\zeta\Omega_\mathrm{DM}^2  H_\star}{\pi G \Sigma_\mathrm{gas}} \right)\nonumber\\  
&+ H_\mathrm{gas} \left(H_\star - \frac{\sigma_{\rm eff}^2 }{\pi G\Sigma_\mathrm{gas}} \right) - H_\star \frac{\sigma_{\rm eff}^2 }{\pi G\Sigma_\mathrm{gas}}=0.
\label{eq:Hg_cubic}
\end{align}
\noindent Here, $\zeta \approx 1/3$ is a geometric factor that is insensitive to the exact vertical distribution of the gas density \citep{Ostriker+Shetty};
$\Omega_\mathrm{DM}=V_\mathrm{DM}/R$ quantifies the dark matter gravity;
$H_\star$ is the vertical scale height\footnote{We adopt the \emph{exponential} disk scale height convention, such that the volume density near the mid-plane is $\rho_\star=\Sigma_\star/(2H_\star)$.} of the stellar disk (see below);
and $\sigma_\mathrm{eff}$ represents the effective gas velocity dispersion.
\autoref{eq:Hg_cubic} is exact if the gas disk and stellar disk both have exponential profiles along the vertical direction, and is correct within $\sim 25\%$ more generally \citep{Hassan_etal_2024}.

For the effective gas velocity dispersion $\sigma_\mathrm{eff}$, we use the mass-weighted average of the molecular and atomic gas velocity dispersion\footnote{This does not take into account the magnetic contribution to $\sigma_\mathrm{eff}$, which could increase its value by $\sim 30\%$ \citep{2022ApJ...936..137O,2024ApJ...972...67K}.} (see \S\ref{sec:data:rotcurve}) following \citet{Sun2020}.
Similar to our treatments for the stellar and gas surface density profiles in \autoref{sec:method:potential}, we perform a Gaussian process regression also on the $\sigma_\mathrm{eff}$ profiles to suppress small-scale fluctuations.

For the stellar disk scale height $H_\star$, we consider five different assumptions both for calculating the stellar disk potential in \ref{sec:method:potential} and for deriving the gas scale height here:
\begin{enumerate}[leftmargin=1.2em,topsep=0.5em,itemsep=0em]
\item The conventional estimate of $H_\star = 0.27\,\ell_\star$, where $\ell_\star$ is the radial scale length of the stellar surface density profile \citep{Kregel2002,Sun2020}
\item A rescaled version of the above, $H_\star = 0.09\,\ell_\star$, based on the ratio of the Milky Way's stellar disk scale length \citep[$l_*=2.2$~kpc;][]{Bovy_Rix_2013} and scale height \citep[$H_\star=0.2$~kpc from dynamical models of the stellar populations in the solar neighborhood;][]{Zhang2013}.
\item An assumption that $H_\star=H_\mathrm{gas}$. The purpose of considering this assumption is to test how much the results differ from those in other cases when a specific value of $H_\star$ is adopted. This is particularly relevant to situations where $H_\star$ is uncertain (e.g. in cosmological simulations where the vertical structure of the disk is not resolved). If one assumes a constant $H_\star/H_g$ ratio, \autoref{eq:Hg_cubic} reduces to a quadratic form with a single unknown variable \citep{Hassan_etal_2024}, from which $H_\star$ or $H_g$ can be directly solved. Here we assume $H_\star=H_g$, which would be appropriate for early stages of galactic evolution prior to secular thickening of the stellar disk, and is also not too far from observations of the Milky Way solar neighborhood.
\item A constant value of $H_\star = 800$~pc. This value is motivated by the $H_\star = 0.27\,\ell_\star$ relation mentioned above and the observation that the galaxy scale radii in our sample are clustered around $\sim$3~kpc. This simple assumption is easy to implement and may be appropriate for local massive galaxy disks in dynamical equilibrium.
\item An alternative constant value, $H_\star = 300$~pc. The motivation is similar to the assumption above, but instead based on the $H_\star = 0.09\,\ell_\star$ scaling.
\end{enumerate}

\subsection{Gas Equilibrium Weight}
\label{sec:method:weight}

With the aforementioned treatments of the stellar and gas disk vertical scale heights, we now have the full 3D information for the stellar, gas, and dark matter distribution.
This allows us to compute the total weight of the gas (per unit area) in the combined galaxy potential, which should be balanced by the average gas pressure near the disk midplane under the assumption of vertical dynamical equilibrium.
This total weight is equal to the sum of the weight contributed by the gravity of the gas disk, stellar disk, and dark matter halo, as given by \citet[][see equations~5, 8, and 9 therein]{Hassan_etal_2024},
\begin{align}
\mathcal{W}_\mathrm{gas} &= \frac{\pi}{2}G\Sigma_\mathrm{gas}^2~,
\label{eq:Wgas}\\
\mathcal{W}_\star &= \pi G\Sigma_\mathrm{gas} \Sigma_\star \frac{H_\mathrm{gas}}{H_\mathrm{gas} + H_\star}~,
\label{eq:Wstar}\\
\mathcal{W}_\mathrm{DM} &= 
\zeta \Sigma_\mathrm{gas} \Omega_\mathrm{DM}^2 H_\mathrm{gas},
\label{eq:Wdm}
\end{align}
where $H_\mathrm{gas}$ is the solution of \autoref{eq:Hg_cubic}.
The total weight of the gas per unit area is the sum of these terms,
\begin{equation}
\mathcal{W} = \mathcal{W}_\mathrm{gas} + \mathcal{W}_\star + \mathcal{W}_\mathrm{DM}~.
\label{eq:Wtot}
\end{equation}

In the limit of $H_\mathrm{gas} \ll H_\star$ and negligible dark matter contribution, we obtain a result very similar to the expression 
\begin{align}
\mathcal{W} &=\frac{\pi G \Sigma_\mathrm{gas}^2}{2} + \Sigma_\mathrm{gas} (2 G \rho_\star)^{1/2}\sigma_\mathrm{eff}
\label{eq:Wtot_lim}
\end{align}
that is commonly adopted in the nearby galaxy observational literature \citep[e.g.][]{2004ApJ...612L..29B,2017ApJ...835..201H,Sun2020,Sun_etal_2023,2021MNRAS.503.3643B},
with the only difference being a factor $\pi$ instead of $2$ in the square root.  
For galaxy models with similar properties to normal spirals, \autoref{eq:Wtot_lim} has also been validated in numerical hydrodynamic and magnetohydrodynamic simulations \citep[e.g.][]{2011ApJ...743...25K,2013ApJ...776....1K,2015ApJ...815...67K,2022ApJ...936..137O,2024ApJ...972...67K}.

\section{Results} \label{sec:results}

\subsection{Rotation Curve Decomposition}
\label{sec:result:rcmodel}

\begin{figure*}[p!]
\gridline{
\fig{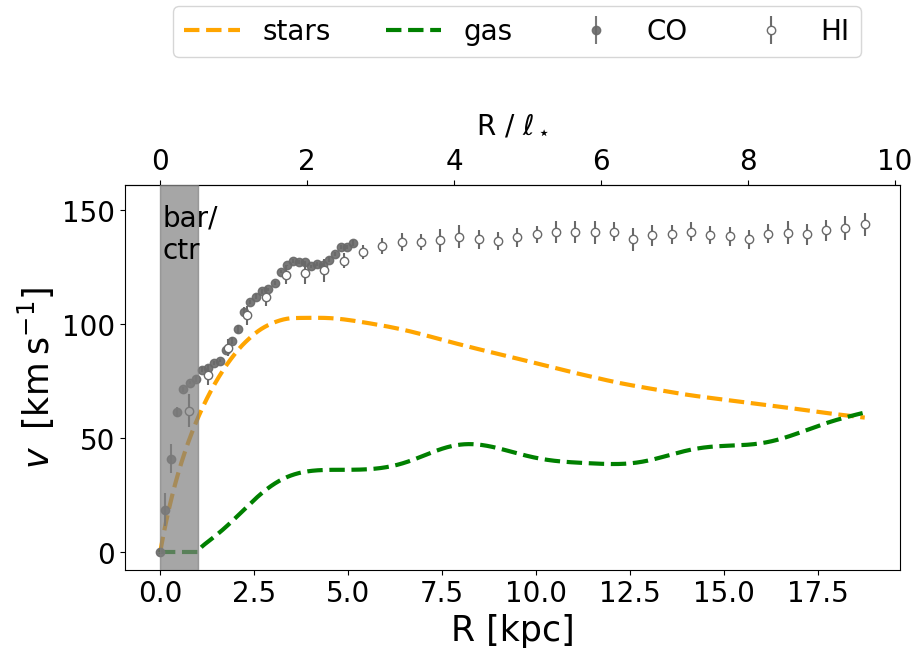}{0.49\textwidth}{(a1) NGC~3621 (original data)}
\fig{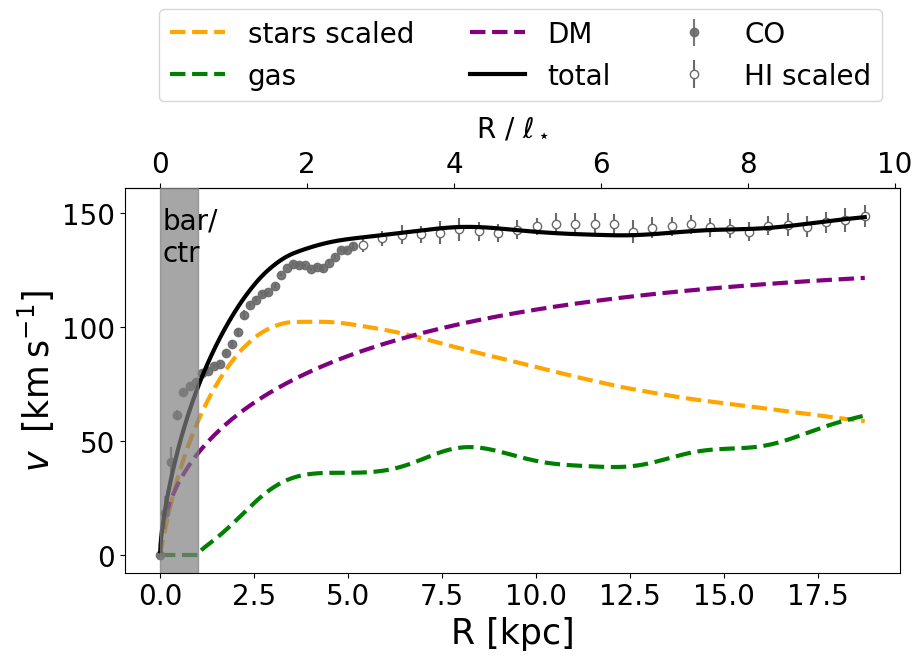}{0.49\textwidth}{(a2) NGC~3621 (scaled data and joint rotation curve fit)}
}
\vspace{-1\baselineskip}
\gridline{
\fig{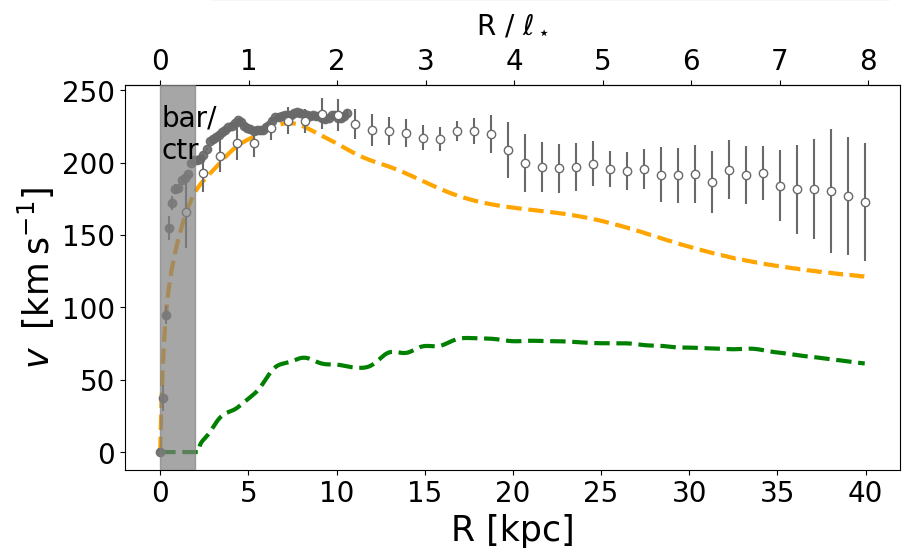}{0.49\textwidth}{(b1) NGC~3521 (original data)}
\fig{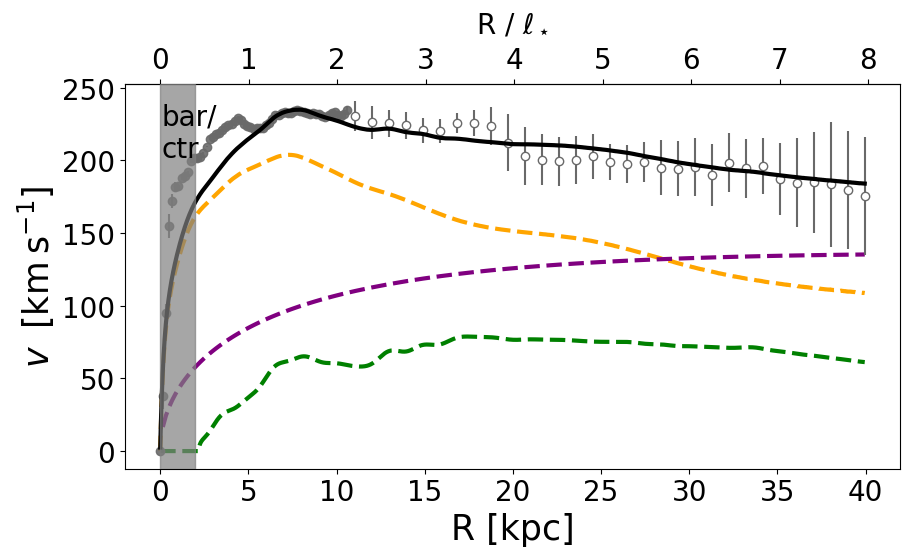}{0.49\textwidth}{(b2) NGC~3521 (scaled data and joint rotation curve fit)}
}
\vspace{-1\baselineskip}
\gridline{
\fig{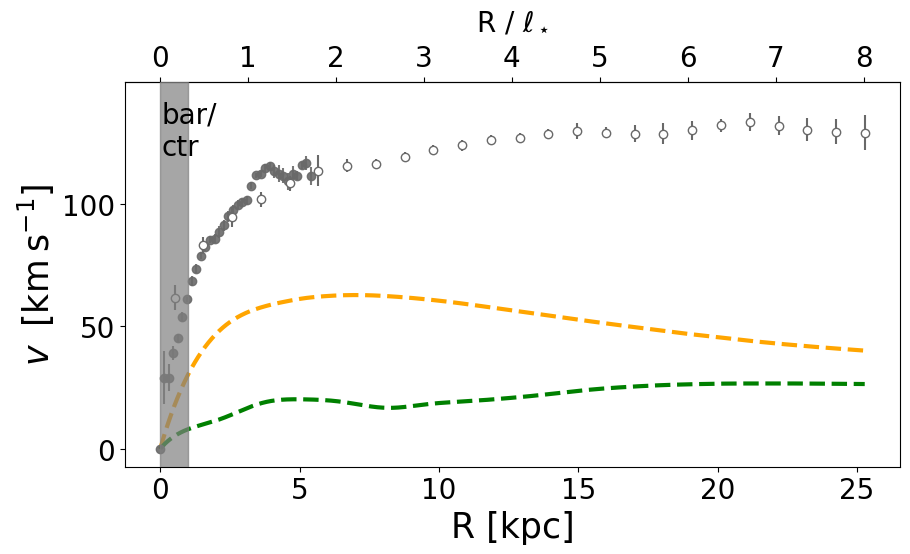}{0.49\textwidth}{(c1) NGC~3137 (original data)}
\fig{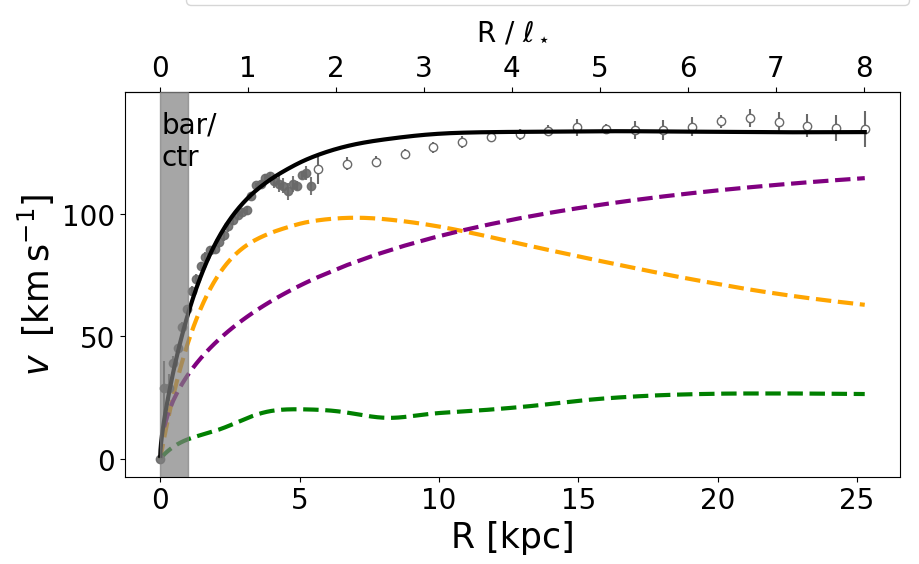}{0.49\textwidth}{(c2) NGC~3137 (scaled data and joint rotation curve fit)}
}
\vspace{-0.5\baselineskip}
\caption{Rotation curve fit results for three example galaxies. The left panels show the original CO and \HI\ rotation curves and the estimated contributions from stellar and gas potential without any rescaling. The right panels show the merged rotation curves and the joint stellar--gas--dark matter fitting results, with the stellar component rescaled according to the maximal disk assumption (see \S\ref{sec:method:scaling}). Note that the radius range covered by a stellar bar or central stellar structure (gray shaded area) are masked and ignored during fitting. The three rows show three examples for which the scaling factor applied to the stellar component is close to unity (NGC~3621), significantly below unity (NGC~3521), or significantly above unity (NGC~3137). For all cases results shown here, we adopt $H_\star/\ell_\star = 0.27$. Note that the rotation curve fit results for all 17 galaxies in our sample are available in \autoref{apdx:gallery}.}
\label{fig:vsquared}
\end{figure*}

A key part of this work is decomposing the observed galaxy rotation curves into stellar, gas, and dark matter contributions for mass modeling.
We highlight the main results and caveats in \autoref{fig:vsquared}.
Specifically, the left panels show the original CO and \HI\ rotation curves derived from kinematic modeling as well as the stellar and gas circular velocity contributions computed without any rescaling (\S\ref{sec:method:scaling}).
The three galaxies shown in \autoref{fig:vsquared} reflect the diversity of situations we see across the whole sample -- that is, for some galaxies the amplitudes of the rotation curves and the stellar component are consistent with each other as they are; but for the remaining majority, the CO and \HI\ rotation curves show small relative offsets in the overlapping radius range, and the stellar component either exceeds the rotation curves or appears much smaller than expected for these massive disk galaxies.
These issues motivate us to rescale the \HI\ rotation curve relative to CO and the stellar contribution relative to the combined rotation curve (as described in \S\ref{sec:method:scaling}), such that we can achieve reasonable rotation curve decomposition results for mass modeling.

\begin{deluxetable*}{lcccccc}
\small
\centering
\tablecaption{Derived Parameters\label{table:scalingpars}
}
\tablehead{
\colhead{Galaxy} &
\colhead{$\ell_\star$} &
\colhead{$f_\mathrm{HI}$} &
\colhead{$\log_{10}\rho_0$} &
\colhead{$R_s$} &
\multicolumn{2}{c}{$f_\star$}\\
\cmidrule(lr){6-7}
\colhead{} & \colhead{[kpc]} & \colhead{} & \colhead{[$\rm M_\odot/pc^3$]} & \colhead{[kpc]} &
\colhead{$H_\star = 0.27\ell_\star$} &
\colhead{$H_\star = 0.09\ell_\star$}
}
\colnumbers
\startdata
NGC~1087 & 1.54 & 1.10 & -2.2 & 14 & 1.71 & 1.55\\
NGC~1300 & 3.73 & 1.19 & -2.6 & 34 & 1.94 & 1.70\\
NGC~1385$^\dagger$ & 1.97 & 1.08 & -2.7 & 18 & 1.10 & 1.00\\
NGC~1512 & 6.18 & 1.12 & -3.1 & 56 & 1.07 & 0.99\\
NGC~2283 & 2.19 & 1.05 & -2.7 & 20 & 1.83 & 1.67\\
NGC~2835 & 2.53 & 1.08 & -2.8 & 23 & 1.97 & 1.79\\
NGC~2903 & 3.38 & 1.01 & -2.6 & 30 & 1.09 & 0.98\\
NGC~2997 & 4.25 & 1.08 & -2.5 & 38 & 1.68 & 1.48\\
NGC~3137 & 3.15 & 1.02 & -2.8 & 28 & 2.46 & 2.21\\
NGC~3507 & 2.17 & 1.19 & -2.3 & 20 & 0.67 & 0.60\\
NGC~3511 & 2.49 & 1.03 & -2.5 & 22 & 1.25 & 1.14\\
NGC~3521 & 5.02 & 1.01 & -2.5 & 23 & 0.80 & 0.74\\
NGC~3621 & 1.95 & 1.02 & -2.3 & 18 & 0.99 & 0.90\\
NGC~4303 & 2.77 & 1.09 & -2.4 & 25 & 0.95 & 0.86\\
NGC~4571$^\dagger$ & 2.14 & 0.94 & -2.6 & 19 & 1.61 & 1.46 \\
NGC~4781 & 0.97 & 1.06 & -1.9 & 8.7 & 1.15 & 1.05\\
NGC~5042 & 2.13 & 1.10 & -2.4 & 19 & 1.51 & 1.37\\
\enddata
\tablecomments{Column descriptions: (2) stellar disk radial scale length measured from the $\Sigma_\star$ radial profile; (3) multiplicative factor applied to $V_\mathrm{HI}$ to match $V_\mathrm{CO}$ (\autoref{eq:HIscaling});
(4) \& (5) the dark matter NFW profile normalization factor $\rho_0$ and scale radius $R_s$; (6) \& (7) scaling factor applied to $\Sigma_\star$ to satisfy the maximal disk criterion (\autoref{eq:stellarscaling}), assuming either $H_\star = 0.27\ell_\star$ or $H_\star = 0.09\ell_\star$.}
\tablenotetext{\dagger}{For these two galaxies the observational data have very limited radial coverage (only out to $\approx3.5\,\ell_\star$). Their dark matter NFW profile fit results should be taken with caution.}
\vspace{-2\baselineskip}
\end{deluxetable*}
\vspace{-2\baselineskip}

\begin{figure}[ht!]
\gridline{
\fig{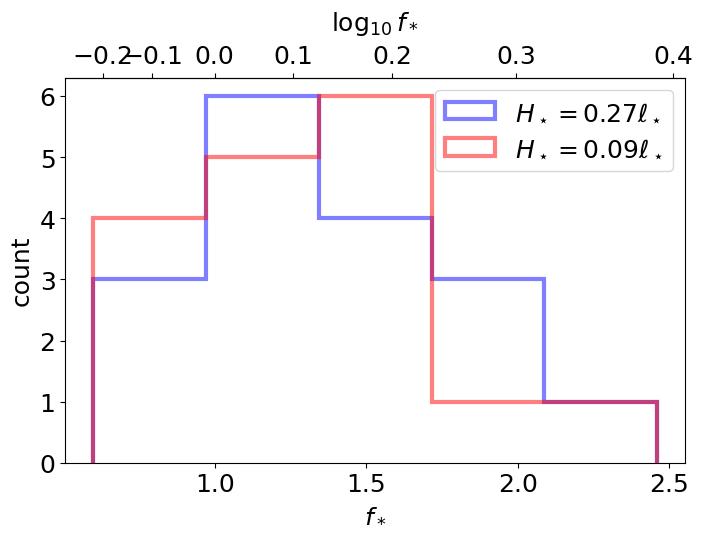}{0.47\textwidth}{}
}
\vspace{-2\baselineskip}
\caption{Distribution of the scaling factor applied to the stellar mass and potential to satisfy the maximal disk criterion for each galaxy ($f_\star$; see \S\ref{sec:method:scaling}). $f_\star$ mildly depends on the assumed stellar disk scale height $H_\star$ (shown by different colors), but in either case the majority of galaxies have $f_\star$ within $\pm0.15$~dex.}
\label{fig:scaling}
\end{figure}

We report in \autoref{table:scalingpars} the scaling factors applied to the \HI\ rotation curves using \autoref{eq:HIscaling} ($f_\mathrm{HI}$) and the stellar components using \autoref{eq:stellarscaling} ($f_\star$).
The $f_\mathrm{HI}$ values for most galaxies are in the range of $0.94{-}1.19$, reflecting the small relative offset (${\lesssim}19\%$) between the CO and \HI\ rotation curves.
The fact that most $f_\mathrm{HI}$ values are above 1 indicates that the estimated \HI\ rotation velocity tends to be smaller than CO rotation velocity at the same radius.
This is in part due to systematic differences in the inclination angles derived from \HI\ and CO kinematic modeling, such that the estimated inclination is often higher for \HI, resulting in a lower circular rotation velocity than CO.
Another possibility is that the \HI\ circular velocity is indeed smaller than CO on average as the \HI\ gas disk is often thicker, so most of the \HI\ gas is located farther away from the galaxy mid-plane.
This effect has been reported by previous studied when comparing the centroid of \HI\ and CO line profiles in nearby galaxies \citep[e.g.,][in M33]{Koch_etal_2018}, but the amplitude of this effect in literature studies tends to be smaller than the offsets that we find here.

The scaling factor applied to the stellar mass and potential, $f_\star$, tends to be more substantial than $f_\mathrm{HI}$.
As mentioned in \S\ref{sec:method:scaling}, this likely reflects a much larger uncertainty on the stellar M/L ratio, although uncertainties on the galaxy distance and inclination angle can also play some role.
\autoref{fig:scaling} shows the distribution of $f_\star$ across our sample.
Assuming $H_\star = 0.27\ell_\star$, 13 out of 17 galaxies have $f_\star$ values between $0.8{-}1.5$ (or $-0.10$ to $0.15$~dex), comparable to the expected level of uncertainty for our stellar M/L ratio treatment \citep[see \S\ref{sec:data:nir};][]{Leroy2019}.
The remaining four galaxies have $f_\star$ values of $1.6{-}2.1$, i.e., the whole stellar mass profile and potential needs to be scaled up by the square of that factor to satisfy the maximum disk assumption.
Adopting $H_\star = 0.09\ell_\star$ yields mildly different $f_\star$ values, but the qualitative behaviors and the fraction of ``outlier'' galaxies remain similar.

We note that our analyses cannot conclusively confirm or rule out the maximal disk scenario.
It is possible that galaxies with large $f_\star$ values actually have sub-maximal disks, though one of them (NGC~1300) is among the most massive galaxies in our sample ($10^{10.6}\,M_\odot$) and is expected to be stellar gravity dominated at least within the inner disk.
On the other hand, if we assume a maximal disk and attribute the $f_\star$ scaling entirely to uncertain M/L ratio estimates, then they would imply a modified M/L ratio with a median value and 1$\sigma$ range of $0.47\pm0.14\;M_\odot\,L_\mathrm{\odot,3.6\mu m}^{-1}$ across our sample, which also seem quite plausible.
In any case, the maximal disk (or submaximal disk) scenario simply dictates some modified M/L ratio in this work, and we have verified that adopting any reasonable alternative M/L ratio (e.g., between 0.2 and 0.6) would not change our main results on a qualitative level.

After rescaling the \HI\ rotation curve and stellar potential, we perform the joint stellar--gas--dark matter fit on the rotation curve and illustrate the results in the right panels in \autoref{fig:vsquared}.
Note that the fit ignores all measurements in the radius range covered by a stellar bar or a central stellar structure (shaded area), where we expect, and indeed often see, strong non-circular motions affecting the rotation curve measurements and non-axisymmetric structures distorting the stellar potential.

We find the combination of the rescaled stellar component, the gas component, and the NFW dark matter component can fit the joint CO-\HI\ rotation curve very well, especially outside the bar and/or center region.
The rescaled stellar component always dominates at small radii, which is partly by construction with the maximum disk assumption, but also does align with our expectation for massive disk galaxies at $z\sim0$.
The dark matter component often surpasses the stellar component at the largest radius probed by the \HI\ data, except for very few galaxies where the \HI\ rotation curves are too limited in their radius coverage (i.e., only out to $4{-}5\;\ell_\star$) to reach the dark matter-dominant region.
For all galaxies, the gas gravity has a negligible contribution to the circular velocity at all radii, as one would expect given their small contribution to the overall mass budget in these systems.

\subsection{Disk Scale Heights}
\label{sec:result:height}

\begin{figure*}[ht!]
\gridline{
\fig{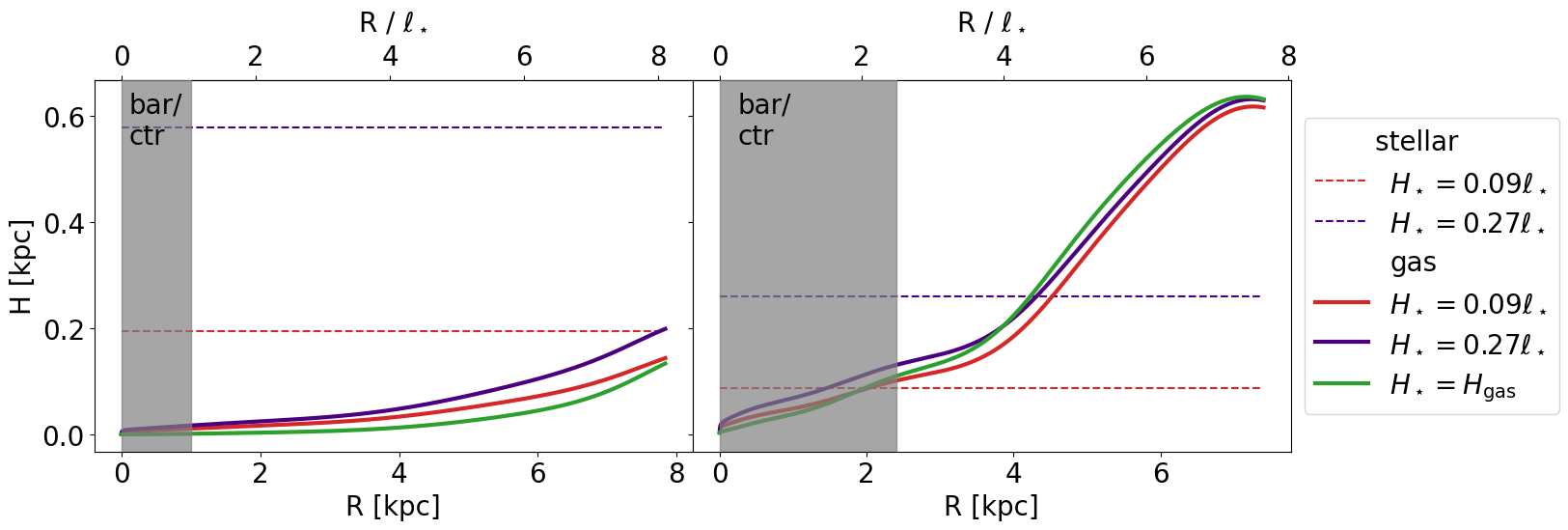}{\textwidth}{}
}
\vspace{-2\baselineskip}
\caption{Gas scale heights (solid lines) and the corresponding stellar scale height assumptions (dashed lines) as functions of galactocentric radius for two galaxies. Different line colors correspond to different assumptions (see legend), and when assuming $H_\star=H_\mathrm{gas}$ the solid and dashed lines are identical by construction.
\textit{Left:} NGC~4571, for which the gas scale height never exceeds the stellar scale height over the radius range probed by the observations.
\textit{Right:} NGC~4781, for which the gas scale height exceeds the stellar scale height at $R\approx4$~kpc when assuming $H_\star=0.27\ell_\star$, or at $R\approx2$~kpc when assuming $H_\star=0.09\ell_\star$.}
\label{fig:scaleheights}
\end{figure*}

\begin{figure*}[ht!]
\gridline{
\fig{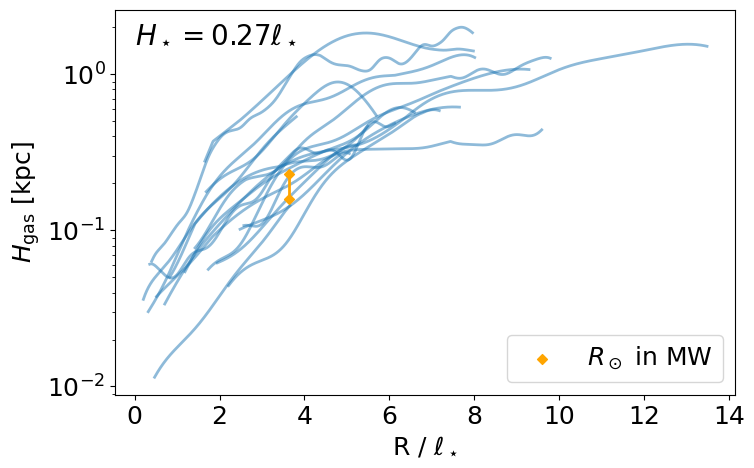}{0.49\textwidth}{}
\fig{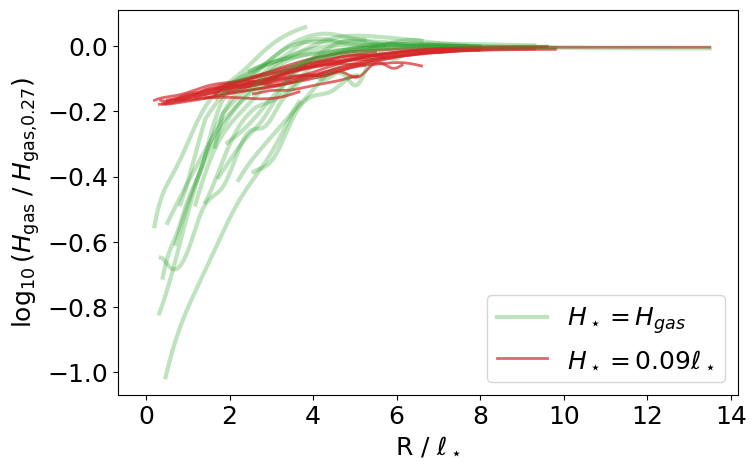}{0.49\textwidth}{}
}
\vspace{-2\baselineskip}
\caption{\textit{Left:} Gas vertical scale height as a function of normalized galactocentric radius for all 17 galaxies (blue curves), under the assumption $H_\star = 0.27\ell_\star$. For comparison, the orange diamonds mark the measured scale heights of the neutral ISM in the Milky Way near the solar neighborhood \citep[$230$~pc and $160$~pc for the warm and cold phases, respectively; see][]{McClure-Griffiths_etal_2023}.
\textit{Right:} The fractional changes in the derived gas scale heights when adopting two alternative stellar scale height assumptions ($H_\star = 0.09\ell_\star$ and $H_\star = H_{\mathrm{gas}}$) as opposed to $H_\star = 0.27\ell_\star$. The discrepancies between $H_\star = 0.09\ell_\star$ versus $H_\star = 0.27\ell_\star$ diminish with radius, from $-0.15$~dex at $R\lesssim \ell_\star$ to $-0.05$~dex at $R\gtrsim5\ell_\star$. In contrast, the $H_\star = H_\mathrm{gas}$ assumption can lead to discrepancies as large as $-0.5$ to $-1.0$~dex in the inner disk in comparison to $H_\star = 0.27\ell_\star$.}
\label{fig:allgasscaleheights}
\end{figure*}

Based on the stellar, gas, and dark matter mass profiles obtained via rotation curve decomposition, we derive the equilibrium gas disk scale height ($H_\mathrm{gas}$) from \autoref{eq:Hg_cubic} at all galactocentric radii and show the results in \autoref{fig:scaleheights} and \autoref{fig:allgasscaleheights}.
As discussed in \S\ref{sec:method:height}, the $H_\mathrm{gas}$ calculation is dependent on and sensitive to the assumption for the stellar disk scale height $H_\star$.
We thus compare the results under three different assumptions, namely $H_\star=0.27\ell_\star$ \citep[common choice in extragalactic literature;][]{Kregel2002,Sun2020}, $H_\star=0.09\ell_\star$ \citep[matching solar neighborhood dynamical modeling results;][]{Zhang2013}, and $H_\star=H_\mathrm{gas}$ (to test how accurate an $H_\mathrm{gas}$ could be obtained if $H_\star$ is unknown).

\autoref{fig:scaleheights} shows the derived gas scale height $H_\mathrm{gas}$ as a function of galactocentric radius in two galaxies: NGC~4571 and NGC~4781.
A general trend is that $H_\mathrm{gas}$ increases strongly with radius, from ${\lesssim}100$~pc in the inner disk to ${>}500$~pc in the outer disk.
Considering that $H_\mathrm{gas}$ mostly reflects the scale height of the mass-dominating ISM phase at each radius, this trend is consistent with measurements for the Milky Way and other edge-on galaxies \citep{Yim_etal_2011,Yim_etal_2014,Heyer_Dame_2015,McClure-Griffiths_etal_2023}.
On top of this overall trend are small-scale fluctuations that can be traced back to residual bin-to-bin variations in stellar surface density and gas velocity dispersion.

The derived $H_\mathrm{gas}$ values directly depend on the assumption for $H_\star$.
Adopting a larger fixed $H_\star$ value (i.e., $H_\star=0.27\ell_\star$ rather than $0.09\ell_\star$) would lead to larger $H_\mathrm{gas}$ values at any given radius, as one would expect from the reduced ``confining force'' by stellar gravity near the disk mid-plane.
This also affects the radius at which the gas scale height exceeds the stellar scale height -- in this example, for NGC 4781, $H_\mathrm{gas}$ exceeds $H_\star$ at $r\approx4$~kpc when assuming $H_\star=0.27\ell_\star$ (blue dashed line), but it happens at $r\approx2$~kpc when assuming $H_\star=0.09\ell_\star$ (red dashed line).
The assumption $H_\star=H_\mathrm{gas}$ leads to a solution consistent with the solution obtained for smaller $H_\star$ at small $R$, and larger $H_\star$ at large $R$.

\autoref{fig:allgasscaleheights} summarizes our gas scale height measurements across the entire sample.
The left panel shows all the $H_\mathrm{gas}$ radial profiles assuming $H_\star = 0.27\ell_\star$, with the $x$-axis normalized by the stellar disk radial scale length of each galaxy.
Aside from the general trend of increasing $H_\mathrm{gas}$ with radius, the $H_\mathrm{gas}$ profiles show some variety as each galaxy has its unique radial profiles of $\Sigma_\star$, $\Sigma_\mathrm{gas}$, and $\sigma_\mathrm{eff}$.
Most galaxies have $H_\mathrm{gas}$ in the range of 100--500~pc over several disk scale lengths.
For a comparison, in the Milky Way solar neighborhood, the scale heights of the mass-dominating cold and warm neutral media are $160$~pc and $230$~pc, respectively \citep[after conversion between scale height conventions;][]{McClure-Griffiths_etal_2023}.
With the Sun located at $R=8.3\;\mathrm{kpc}$ \citep[or $\approx3.8\ell_\star$ for $\ell_\star=2.2$~kpc;][]{Bovy_Rix_2013} in the Milky Way, the local gas disk scale heights appear normal among nearby star-forming disk galaxies studied in this work.
Our results also seem to agree with previous extragalactic studies \citep[e.g.,][]{Yim_etal_2011,Yim_etal_2014,Bacchini_etal_2019,ManceraPina_etal_2022} in both the gas scale height range and the quantitative trends.

The right panel in \autoref{fig:allgasscaleheights} shows the fractional changes in the gas scale heights derived from the alternative stellar scale height assumptions ($H_\star=0.09\ell_\star$ and $H_\star = H_\mathrm{gas}$) relative to $H_\star=0.27\ell_\star$.
Overall, assuming $H_\star=0.09\ell_\star$ leads to ${\sim}0.15$~dex (or $30{-}40\%$) smaller gas scale heights than $H_\star=0.27\ell_\star$ in the inner disk ($R \lesssim \ell_\star$), with the discrepancy reducing to ${\lesssim}0.05$~dex (or ${\lesssim}\,10\%$) towards the outer disk ($R \gtrsim 5\ell_\star$).
This radial trend is expected because (a) stellar gravity is most important at small radii (also see \S\ref{sec:result:weight} below), and (b) the gas disk is much thinner than the stellar disk there (\autoref{fig:scaleheights}).
Under these conditions, the stellar volume density near the disk mid-plane ($\rho_\star$) is most important for setting the gas equilibrium scale height.
Since $\rho_\star$ is directly set by $H_\star$ via $\rho_\star = \Sigma_\star/2H_\star$, it is not surprising the $H_\star$ assumption matters the most for setting $H_\mathrm{gas}$ in the inner disk.
Then as the gas scale height approaches and even exceeds the stellar scale height towards larger radii, the $H_\star$ assumption matters less, hence the trend seen in the right panel of \autoref{fig:allgasscaleheights}.
The same arguments hold for the $H_\star = H_\mathrm{gas}$ case as well, except that we see much larger discrepancies with $H_\star=0.27\ell_\star$ there, up to $0.5{-}1.0$~dex in the inner disk.

\subsection{Gas Equilibrium Weight}
\label{sec:result:weight}

\begin{figure}[htp]
\gridline{
\fig{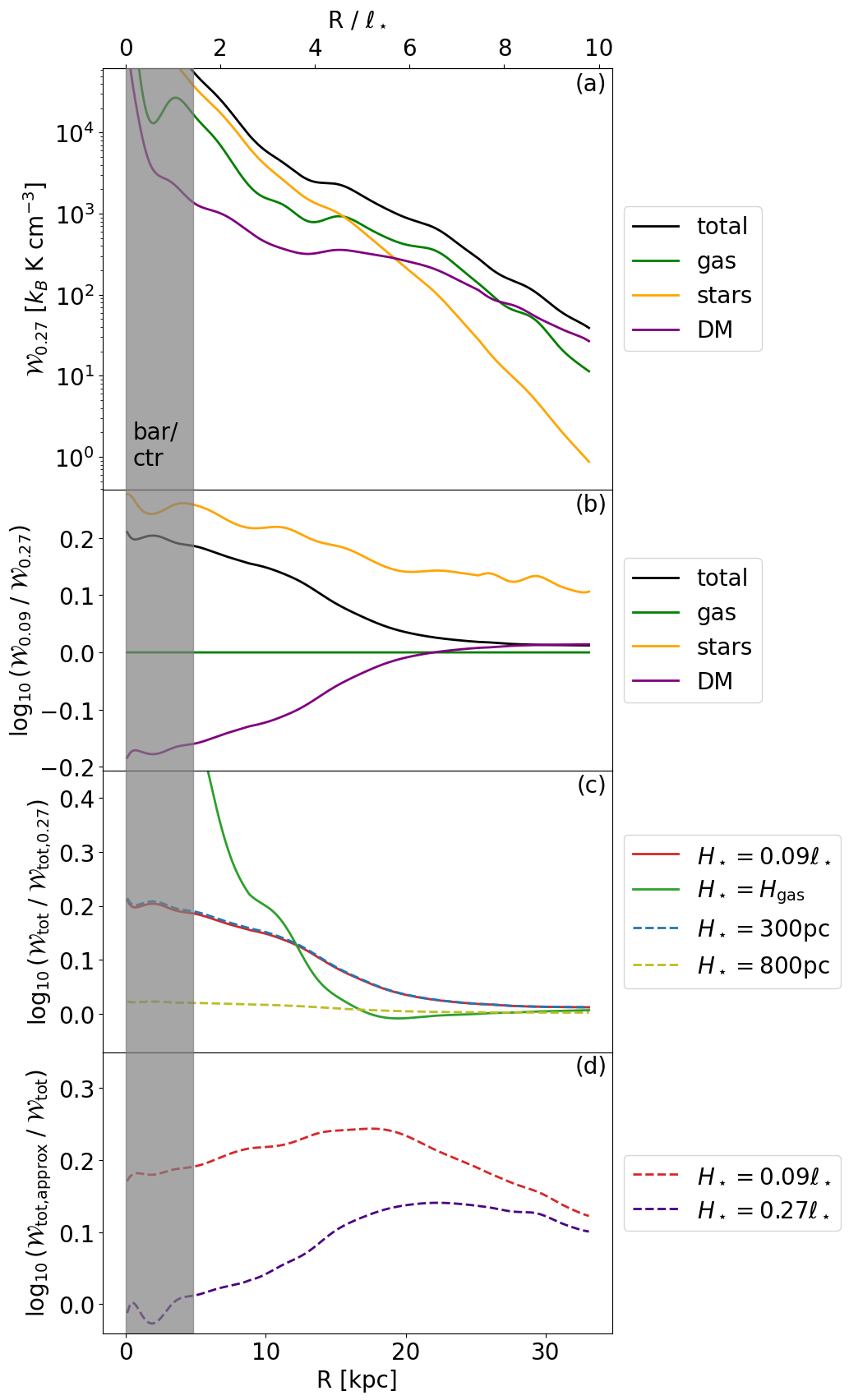}{0.49\textwidth}{}
}
\vspace{-2\baselineskip}
\caption{
The gas equilibrium weight calculations for NGC~2903, comparing different components and assumptions (results for all other galaxies are available in \autoref{apdx:weight}).
\textit{(a)} Radial profiles of total gas weight and its individual components associated with stellar, gas, and dark matter gravity, assuming $H_\star = 0.27\ell_\star$. The stellar component dominates the total weight in the inner region, until $r\gtrsim15$~kpc where the gas and dark matter gravity become more important.
\textit{(b)} Changes in gas weight components when assuming $H_\star = 0.09\ell_\star$ versus $H_\star = 0.27\ell_\star$. The stellar component is strongly affected by ${\sim}0.1{-}0.2$~dex as a tighter vertical distribution of stars leads to a steeper stellar potential. The dark matter component is indirectly affected due to constraints imposed by the rotation curve. 
\textit{(c)} Changes in \emph{total} gas weight when adopting four alternative assumptions for stellar scale height as opposed to $H_\star = 0.27\ell_\star$. Assuming $H_\star = 300\text{pc}$ ($H_\star = 800\text{pc}$) leads to nearly identical results as $H_\star = 0.09\ell_\star$ ($H_\star = 0.27\ell_\star$) for this galaxy.
\textit{(d)} Difference in total gas weight when adopting a common approximation (\autoref{eq:Wtot_lim}) rather than the more accurate solution (\autoref{eq:Wgas}--\ref{eq:Wtot}). For $H_\star = 0.09\ell_\star$, we see a difference of $\sim$0.2~dex in the inner disk and $\sim$0.1~dex towards the outskirts; the difference is smaller when adopting $H_\star = 0.27\ell_\star$.
}
\label{fig:weight}
\end{figure}

Combining the mass profiles and gas equilibrium scale heights derived in \S\ref{sec:result:rcmodel}--\ref{sec:result:height}, we compute the weight of the gas (along the vertical direction) in the combined galaxy potential via \autoref{eq:Wgas}--\ref{eq:Wtot} and showcase the results for one galaxy (NGC~2903) in \autoref{fig:weight}.
The total gas weight and all of its components generally decreases with galactocentric radius, as both the gas surface density drops and the vertical gravitational field decreases towards the outer galaxy.
Among all the weight components (as defined by Equations~\ref{eq:Wgas}--\ref{eq:Wdm}), the component associated with stellar gravity often dominates over a wide radius range across the inner galaxy, and in a few cases over the entire radius range probed by the CO and \HI\ data (out to $4{-}5\;\ell_\star$ in those cases).
However, the gas and dark matter components can become more important at large radii: for many galaxies, the gas component becomes dominant beyond $3{-}6\;\ell_\star$; for some, we even see the dark matter component taking over beyond $5{-}8\;\ell_\star$.
The latter means that ignoring the dark matter gravity can sometimes be problematic, especially when examining the spatially extended \HI\ gas.

We note that the relative contribution of stellar, gas, and dark matter gravity to the gas weight is a related but distinct issue from their relative contribution to the galaxy rotation curve.
The latter mostly depends on the cumulative mass of each component within a given three-dimensional radius, whereas the former is more closely related to the local mass surface density near the galaxy mid-plane within the scale height of the gas disk.
Considering the different geometries, radial profiles, and scale heights of each component, we can understand why gas gravity can be much more prominent in the local weight budget (especially at large radius) even though its contribution to the rotation curve is always negligible.
Similarly, the dark matter gravity already becomes quite important for the rotation curve at intermediate radius, but its effect on the local gas weight can be less pronounced until further out in the galaxy disk.

Similar to the gas scale height results discussed in \S\ref{sec:result:height}, the gas weight also depends on the assumption for the \emph{stellar} disk scale height.
As shown by \autoref{fig:weight}(b-c), assuming a smaller stellar scale height often results in a larger total weight, because the stellar mass surface density within the gas disk scale height is larger in that case. 
\autoref{fig:weight}(b) shows that the stellar and dark matter contribution to the weight are affected in opposing ways (due to additional constraints imposed by the rotation curve), though the tighter stellar distribution is primarily responsible for the increased total weight.
These effects can in turn alter the relative importance of stellar gravity versus gas and dark matter gravity in the total weight budget, which can sometimes change the physical picture at a qualitative level (i.e., whether the gas component dominates over any radius range at all).

We further compare the total gas weight calculated based on five different stellar scale height assumptions (see \autoref{sec:method:height}). That is, in addition to $H_\star = 0.27\ell_\star$, $H_\star = 0.09\ell_\star$, and $H_\star = H_\mathrm{gas}$ (shown in \autoref{fig:allgasscaleheights} right panel), we also consider $H_\star = 800$~pc and $300$~pc as easily implementable alternatives.
For many galaxies including NGC~2903 in our sample, adopting $H_\star = 0.27\ell_\star$ ($0.09\ell_\star$) leads to similar results as $H_\star = 800$~pc ($300$~pc). This is partly because the $\ell_\star$ values are close to $\sim3$~kpc for most galaxies in our sample, and partly due to the somewhat mild sensitivity of gas weight to stellar scale height assumptions. Nonetheless, galaxies with scale radii significantly different from $\sim 3$~kpc see deviations of up to 0.1~dex in the inner disk.

We also test the impact of gas weight calculation with different formulae in \autoref{fig:weight}(d), comparing the more accurate and broadly applicable Equations~\ref{eq:Wgas}--\ref{eq:Wtot} with the commonly adopted approximation of \autoref{eq:Wtot_lim}.
While in most cases the two options give almost identical results, in a few cases they can deviate on a non-trivial level.
This is mostly because the gas scale height can approach or even exceed the stellar scale height towards the outer disk in some galaxies (see \autoref{fig:scaleheights}), which would violate the assumption of $H_\mathrm{gas} \ll H_\star$ behind \autoref{eq:Wtot_lim}.
We thus advocate for the use of Equations~\ref{eq:Wgas}--\ref{eq:Wtot} in combination with \autoref{eq:Hg_cubic} for gas weight calculations, especially in the \HI-dominated outer disks.

\subsection{Gas Weight and Star Formation}
\label{sec:result:sfr}

\begin{figure*}[bt]
\gridline{
\fig{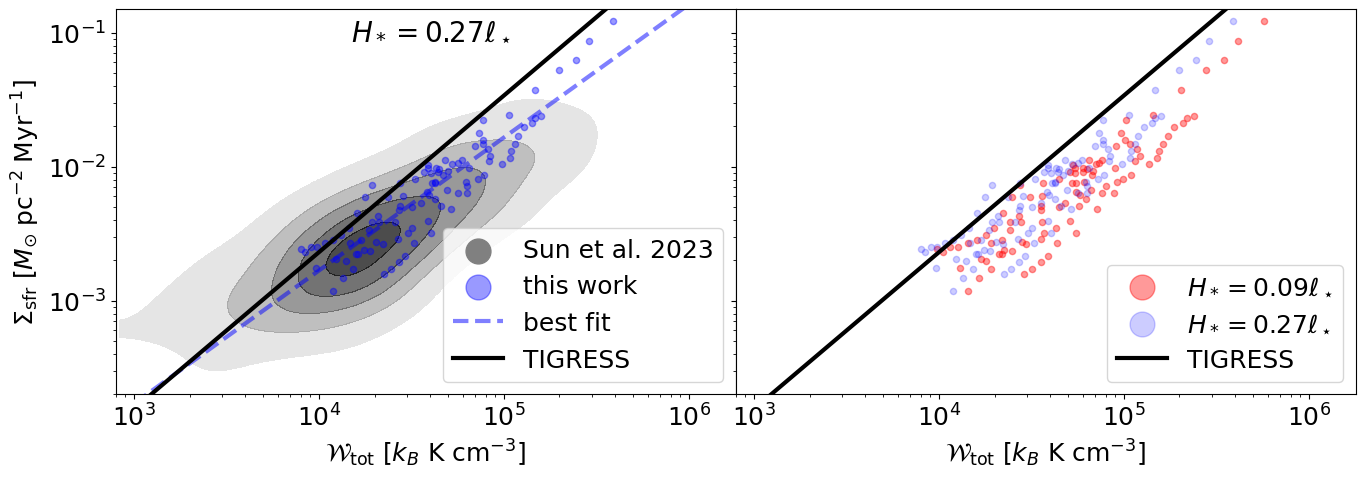}{\textwidth}{}
}
\vspace{-2\baselineskip}
\caption{Local SFR surface density as a function of gas weight under different assumptions. For our sample, the $\Sigma_\mathrm{SFR}$ measurements come from \citet{Sun_etal_2023}, whereas the $\mathcal{W}_\mathrm{tot}$ values are calculated in this work and represent improvements over previous calculations by \citet{Sun2020,Sun_etal_2023}. We plot the best fit power law relation for these measurements as a blue dashed line. We also plot \autoref{eq:SFRWeight} and the distribution of previous observational measurements \citep[gray contours showing data density;][]{Sun_etal_2023}. We see good agreement between this work and literature results when assuming $H_\star = 0.27\ell_\star$ \citep[matching the assumption in][]{Sun_etal_2023}. Our results show more deviation from the simulation prediction when assuming $H_\star = 0.09\ell_\star$.}
\label{fig:sfr}
\end{figure*}

\begin{figure}[ht!]
\gridline{
\fig{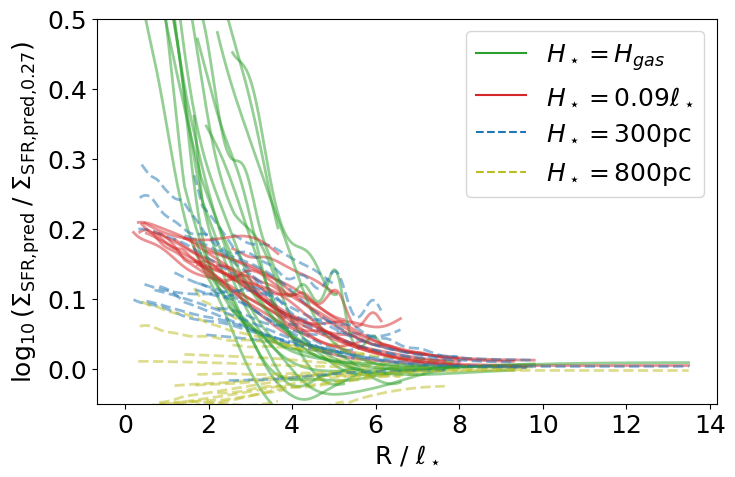}{0.47\textwidth}{}
}
\vspace{-2\baselineskip}
\caption{Predicted SFR surface density for four different assumptions of stellar scale height as functions of normalized galactocentric radius, normalized with the predicted SFR surface density for $H_\star = 0.27\ell_\star$. The predicted SFR surface density is derived by converting our calculated gas weight using the power-law relation in \autoref{eq:SFRWeight}, which is why the trends shown here mirror those in \autoref{fig:weight}(c). We can see differences of $\sim$0.1-0.2 dex between $H_\star = 0.09\ell_\star$ and $H_\star = 0.27\ell_\star$ in the inner disk ($\sim$$2\ell_\star$), and ${>}$0.2 dex with $H_\star = H_\mathrm{gas}$. We also plot inferred SFR surface density for two assumed scale heights of $H_\star = 300\mathrm{pc}$ and $H_\star = 800\mathrm{pc}$, which provide results similar to $H_\star = 0.09\ell_\star$ and $H_\star = 0.27\ell_\star$ respectively.}
\label{fig:sfrinferred}
\end{figure}

We additionally reassess the relationship between the gas weight (which in equilibrium balances total pressure, ${\cal W} = P_\mathrm{tot}$) and the local SFR surface density, which is predicted by PRFM theory and simulations to follow a close-to-linear relationship \citep{OML2010,Ostriker+Shetty,2011ApJ...743...25K,2013ApJ...776....1K,2015ApJ...815...67K,2022ApJ...936..137O}.
In particular, \citet{2022ApJ...936..137O} predicts the following power-law relation based on the TIGRESS simulations:
\begin{equation}
\frac{\Sigma_\text{SFR}}{M_\odot \mathrm{\; pc^{-2} \;Myr^{-1}}} =  10^{-7.32}\left(\frac{\cal W}{k_B\;\mathrm{K \; cm^{-3}}}\right)^{1.17}~.
\label{eq:SFRWeight}
\end{equation}
\noindent This ``Ostriker--Kim'' relationship between $\mathcal{W}$ and $\Sigma_\mathrm{SFR}$ has previously been examined for the full PHANGS--ALMA galaxy sample by \citet{Sun2020,Sun_etal_2023}, using the approximate expression in \autoref{eq:Wtot_lim} for the weight.
While we use existing SFR surface density measurements for the these galaxies by \citet{Sun_etal_2023}, our detailed gravitational potential modeling and gas weight calculations provide an opportunity to revisit this topic with more rigor, in particular including the dark matter contribution to the weight and using a formulation that does not assume $H_\mathrm{gas} \ll H_\star$.

The left panel in \autoref{fig:sfr} shows $\cal W$ versus $\Sigma_\mathrm{SFR}$ across our sample, in comparison with previous observational measurements by \citet{Sun_etal_2023} and \autoref{eq:SFRWeight}.
Our results agree well with previous observational and numerical studies when assuming $H_\star = 0.27\ell_\star$, i.e., matching the assumption used by \citet{Sun_etal_2023} and other observational works.
It is worth noting that applying the scaling factor on the stellar mass ($f_\star$) resulted in a tighter $\mathcal{W}$--$\Sigma_\mathrm{SFR}$ relation in comparison to the same relation calculated with the raw, unscaled measurements (rms scatter decreases from 0.18 to 0.16~dex).
This reduction in scatter possibly means that our rescaling of the stellar mass also helps reduce a source of systematic uncertainty (namely the stellar M/L ratio) in the gas weight calculation.

The right panel shows the comparison for the two choices of $H_\star$.
Because weight is higher for lower $H_\star$, and this primarily affects the higher-pressure, inner regions of galaxies, the $\Sigma_\mathrm{SFR}$ versus weight relation has a slightly shallower slope for smaller $H_\star$. 
At high pressure, both choices of $H_\star$ lead to a $\Sigma_\mathrm{SFR}$ versus weight relation that is shallower than that obtained from simulations. 
This discrepancy may be partly attributable to known observational biases \citep[e.g., on the CO-to-H$_2$ conversion factor; see \S\ref{sec:data:co} and][]{Sun_etal_2023,Schinnerer_Leroy_2024} that would lead to overestimated gas mass (and therefore pressure) near galaxy centers.
However, it is also possible that effects other than feedback significantly contribute to driving pressure, which local-patch simulations such as TIGRESS do not capture.
We expect current and future galaxy-scale simulations with detailed ISM physics and feedback recipes to further our understanding of ISM dynamical equilibrium and star formation self-regulation in disk galaxies.

Lastly, we directly compare the predicted SFR surface densities from \autoref{eq:SFRWeight} for all of our stellar scale height assumptions in \autoref{fig:sfrinferred}.
The comparisons are all made against the fiducial $H_\star = 0.27\ell_\star$ assumption and are plotted as a function of the normalized galactocentric radius.
We can see differences of ${\sim}0.1{-}0.2$~dex between $H_\star = 0.09\ell_\star$ and $H_\star = 0.27\ell_\star$ in the inner disk ($\sim 2\ell_\star$), indicating that the choice of stellar scale height can change the predicted SFR by up to $\sim 60\%$ in the inner disk. 
With $H_\star = H_\mathrm{gas}$ we see larger differences of ${>}0.2$~dex in the inner disk.
These trends mirror those in \autoref{fig:allgasscaleheights} right panel due to an anticorrelation between gas scale height and weight.
They are also essentially the same trends as shown in \autoref{fig:weight}(c), but here recast in SFR surface density units.
We also plot inferred SFR surface density for two assumed scale heights of $H_\star = 300$~pc and $H_\star = 800$~pc, which provide results similar to $H_\star = 0.09\ell_\star$ and $H_\star = 0.27\ell_\star$ respectively.
This shows that for predicting SFR surface density for galaxies in our sample, their galaxy scale radii are similar enough that assuming these simple values would suffice for most galaxies in the sample, though there remains some galaxy-to-galaxy scatter of up to 0.1~dex.

\section{Summary} \label{sec:summary}

We combine high-quality multiwavelength data for a sample of 17 nearby massive disk galaxies to model their stellar, gas, and dark matter mass distribution as functions of galactocentric radius.
We use \textit{Spitzer} IRAC images, VLA \HI\ data, and ALMA CO data (\autoref{fig:observations}) to trace the stellar, atomic gas, and molecular gas mass, leveraging recent efforts in understanding and accounting for variations in stellar mass-to-light ratio and CO-to-H$_2$ conversion factor \citep{Salim_etal_2016,Leroy2019,Schinnerer_Leroy_2024}.
These measurements in combination with \HI\ and CO rotation curves allow us to decompose the total dynamical mass and constrain the dark matter mass profile.

From the multi-component radial mass profile, we further derive the gas disk vertical scale height and the gas equilibrium weight in the combined galaxy potential.
These calculations are dependent on the treatment of the stellar disk scale height, and we test several plausible options motivated by various Galactic and extragalactic studies \citep{Kregel2002,Zhang2013,Sun2020}.
We also assess the validity of a commonly used approximation for the weight calculation over a range of conditions present among our targets.
Finally, our refined gas weight measurements enable a reassessment of its relationship with local SFR surface density, in contexts of the PRFM theory \citep{2022ApJ...936..137O} and previous observational results \citep{Sun2020,Sun_etal_2023}.

Our key findings are as follows:
\begin{enumerate}[topsep=0.3\baselineskip,itemsep=0\baselineskip,leftmargin=2em]

\item We construct a joint stellar--gas--dark matter mass model for each galaxy that can successfully account for its measured rotation curve (\autoref{fig:vsquared}). This is achieved only after combining the CO and \HI\ rotation curves and rescaling the stellar mass component according to the maximal disk assumption \citep{maximaldisk}. The scaling factor applied to the stellar component is between $0.8{-}1.5$ for most (13 out of 17) galaxies (\autoref{fig:scaling}), consistent with the expected uncertainty of our stellar M/L ratio estimates relative to SPS-based results.

\item Our multi-component model suggests that the dark matter contribution to the circular velocity often exceeds the stellar contribution at $R\approx3{-}5\,\ell_\star$, except for a few cases where the radial coverage of the data does not reach the dark matter dominated outer disk. The gas contribution to the circular velocity remains subdominant at all radii.

\item We find that the gas disk vertical scale height is typically $H_\mathrm{gas}\lesssim100$~pc in inner disks and increases towards the outer disk to ${>}500$~pc (\autoref{fig:scaleheights}--\ref{fig:allgasscaleheights}), consistent with other Galactic and extragalactic studies. This increasing trend with radius is present regardless of the assumed stellar disk scale height $H_\star$. The stellar disk scale height assumption matters more in the inner disk, with a $3\times$ difference in $H_\star$ translating into a $30{-}40\%$ difference in $H_\mathrm{gas}$. Its impact becomes smaller (${<}10\%$) towards the outer disk, as stellar gravity becomes less important and the gas scale height approaches or even exceeds the stellar scale height.

\item For the gas equilibrium weight in the combined galaxy potential, the stellar gravity contribution is also dominant in the innermost disk, but the gas self-gravity contribution can often surpass it at $3{-}6\,\ell_\star$; for fewer galaxies, the dark matter gravity contribution becomes dominant beyond $5{-}8\,\ell_\star$ (\autoref{fig:weight}). The relative importance of these contributions in the gas weight budget differs from that in the circular velocity decomposition due to the different 3D geometry of these mass components.

\item The gas weight calculation also mildly depends on the stellar scale height assumption, with smaller stellar scale heights leading to larger gas weights (\autoref{fig:weight}). More importantly, the commonly adopted approximation for the weight calculation (\autoref{eq:Wtot_lim}) can fail when the gas and stellar disk scale height become comparable, which does happen in a few galaxies towards the outer disks.

\item When examining the gas weight versus SFR surface density relation ($\mathcal{W}_\mathrm{tot}$--$\Sigma_\mathrm{SFR}$, \autoref{fig:sfr}), our refined gas weight calculation still results in overall agreement with simulation predictions \citep{2022ApJ...936..137O} and previous observational measurements \citep{Sun2020,Sun_etal_2023}. The slope of this relation appears mildly shallower than the simulation prediction for both $H_\star$ assumptions, possibly due to observational biases at the high pressure end and/or physics not captured by subgalactic, local-patch simulations.

\end{enumerate}

This work represents one of several ongoing efforts in modeling the mass distribution of nearby disk galaxies for the purpose of assessing ISM dynamics and self-regulation.
We anticipate follow-up studies to improve upon this work in various ways, including joint CO-\HI\ kinematic modeling, dedicated analyses of the influence of stellar bars and central structures on gas kinematics, as well as obtaining independent constraints on stellar disk scale heights based on spectroscopic stellar velocity dispersion measurements.

\vspace{\baselineskip}
{

We thank Neta~Bahcall for helpful discussions.

JS acknowledges support by the National Aeronautics and Space Administration (NASA) through the NASA Hubble Fellowship grant HST-HF2-51544 awarded by the Space Telescope Science Institute (STScI), operated by the Association of Universities for Research in Astronomy, Inc., under contract NAS~5-26555. 
The work of ECO was partly supported by grant No.~510940 from the Simons Foundation.
EDT is supported by the European Research Council (ERC) under grant agreement no.\ 101040751.

This paper makes use of the following ALMA data, which have been processed as part of the PHANGS--ALMA \CO21 survey: \\
\noindent ADS/JAO.ALMA\#2015.1.00925.S, \linebreak 
ADS/JAO.ALMA\#2015.1.00956.S, \linebreak 
ADS/JAO.ALMA\#2017.1.00392.S, \linebreak 
ADS/JAO.ALMA\#2017.1.00886.L, \linebreak 
ADS/JAO.ALMA\#2018.1.01651.S. \linebreak 
ALMA is a partnership of ESO (representing its member states), NSF (USA), and NINS (Japan), together with NRC (Canada), NSC and ASIAA (Taiwan), and KASI (Republic of Korea), in cooperation with the Republic of Chile. The Joint ALMA Observatory is operated by ESO, AUI/NRAO, and NAOJ. The National Radio Astronomy Observatory (NRAO) is a facility of NSF operated under cooperative agreement by Associated Universities, Inc (AUI).

This work is based in part on observations made with NSF's Karl~G.~Jansky Very Large Array (VLA; project code: 
18B-184). 
VLA is also operated by NRAO.

This work is based in part on observations made with the \textit{Spitzer Space Telescope}, which is operated by the Jet Propulsion Laboratory, California Institute of Technology under a contract with NASA.

We acknowledge the usage of the SAO/NASA Astrophysics Data System.

\facilities{ALMA, Spitzer, VLA}

\software{
\texttt{NumPy} \citep{NumPy_2020},
\texttt{SciPy} \citep{SciPy_2020},
\texttt{Astropy} \citep{Astropy2013,Astropy2018,AstroPy_2022},
\texttt{Matplotlib} \citep{Matplotlib_2007},
\texttt{Scikit-learn} \citep{Sklearn_2011},
\texttt{hankel} \citep{Murray2019},
\texttt{Pandas} \citep{Pandas_2020},
\texttt{3D-Barolo} \citep{DiTeodoro_Fraternali_2015},
\texttt{MegaTable} \citep{MegaTable_4.0}.
}

}

\appendix
\twocolumngrid

\section{Rotation Curve Modeling Results}
\label{apdx:gallery}

\restartappendixnumbering

\begin{figure}[htb]
\centering
\gridline{
\fig{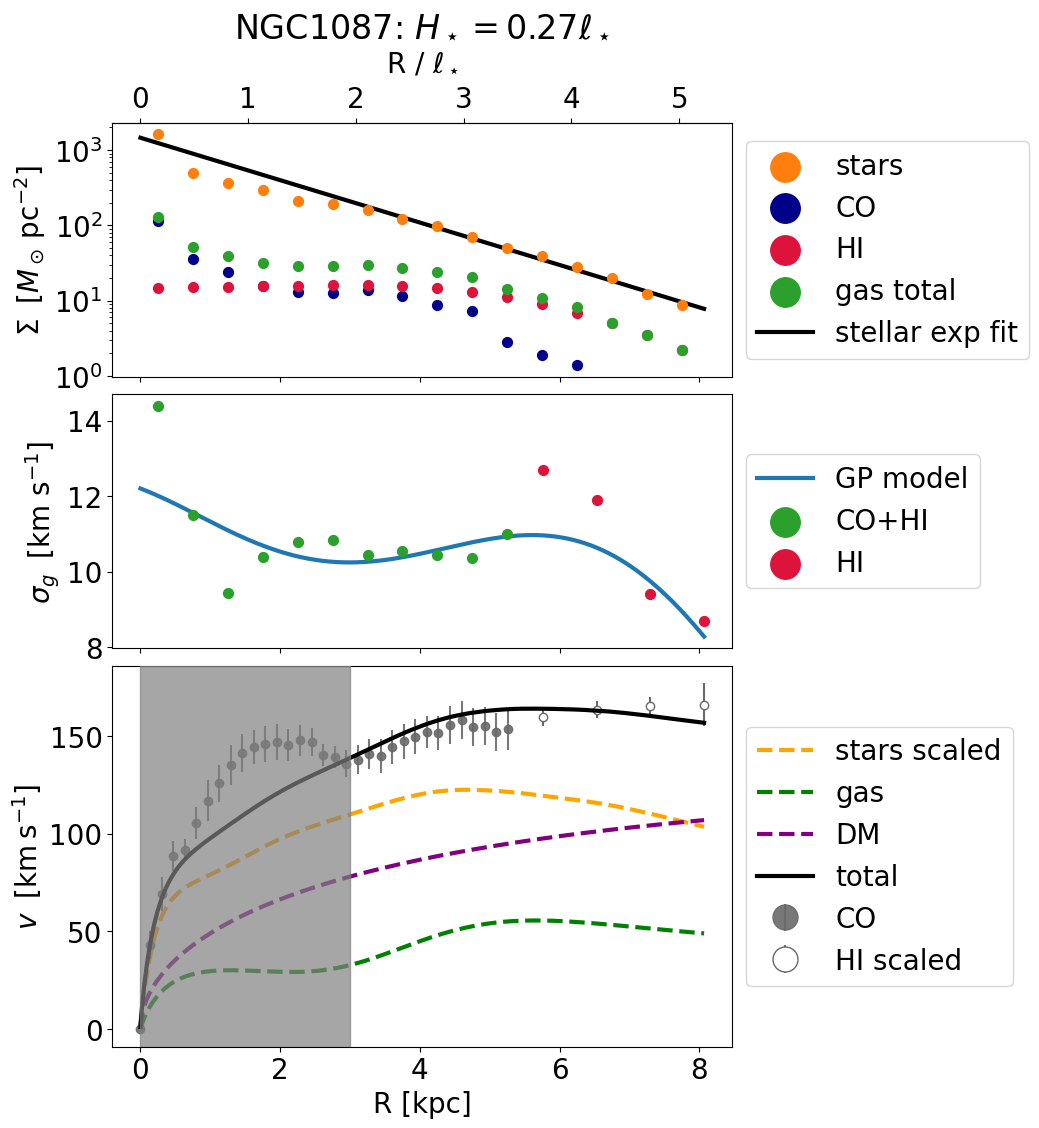}{0.49\textwidth}{}
}
\vspace{-2.0\baselineskip}
\caption{Gas and stellar surface density profiles (top), gas velocity dispersion profiles (middle), and rotation curve modeling results (bottom) for NGC~1087.
The results for all 17 targets are available online as a figure set.
}
\vspace{0.5\baselineskip}
\label{fig:gallery}
\end{figure}

We show radial profiles of gas and stellar surface densities, gas velocity dispersions, as well as rotation curve modeling results for all 17 galaxies in Figure Set~\ref{fig:gallery}.

\figsetstart
\figsetnum{A1}
\figsettitle{Rotation curve modeling results for all 17 galaxies.}

\figsetgrpstart
\figsetgrpnum{A1.1}
\figsetgrptitle{NGC~1087}
\figsetplot{figureset/NGC1087_figureset.png}
\figsetgrpnote{Input parameters and rotation curve modeling results for NGC~1087. From the top, the first panel plots the surface density profiles for the stars, molecular gas (CO), atomic gas (\HI), and total gas. The second panel plots the gas velocity dispersion from combined CO and \HI measurements, and the corresponding Gaussian process fit model. The third panel is identical to those shown in Figure 3.}
\figsetgrpend

\figsetgrpstart
\figsetgrpnum{A1.2}
\figsetgrptitle{NGC~1300}
\figsetplot{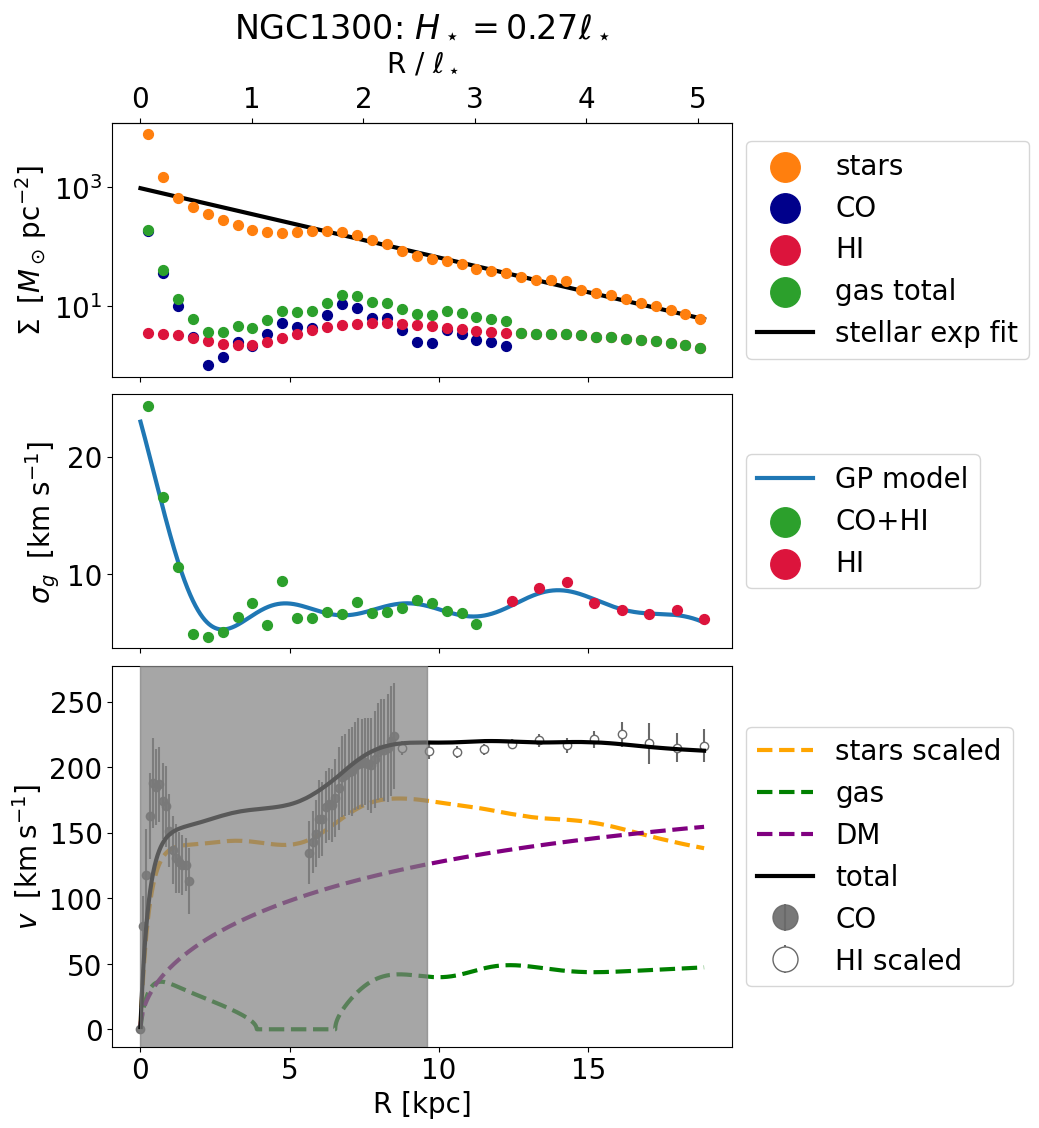}
\figsetgrpnote{Input parameters and rotation curve modeling results. From the top, the first panel plots the surface density profiles for the stars, molecular gas (CO), atomic gas (\HI), and total gas. The second panel plots the gas velocity dispersion from combined CO and \HI measurements, and the corresponding Gaussian process fit model. The third panel is identical to those shown in Figure 3.}
\figsetgrpend

\figsetgrpstart
\figsetgrpnum{12.3}
\figsetgrptitle{NGC~1385}
\figsetplot{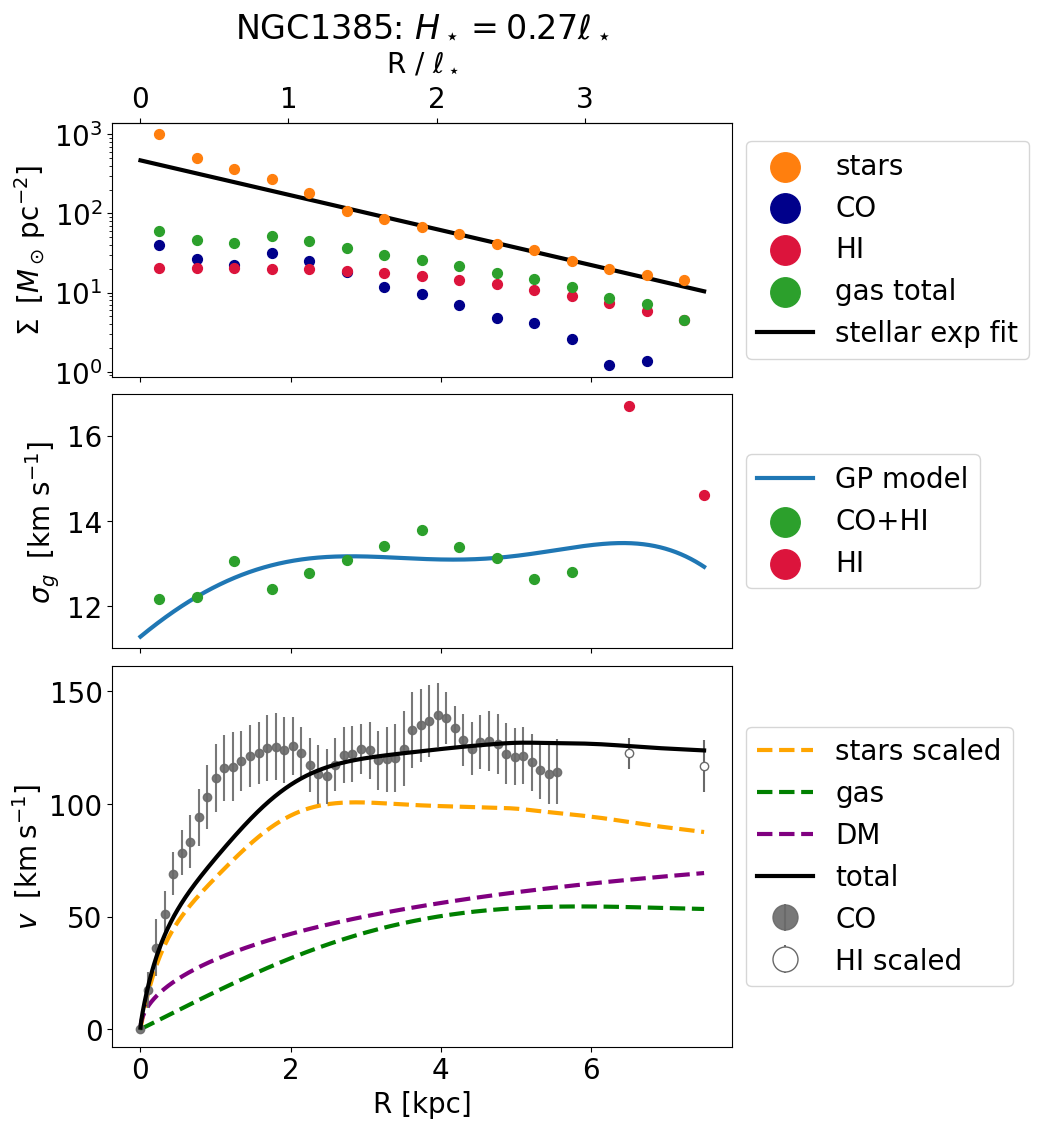}
\figsetgrpnote{Input parameters and rotation curve modeling results. From the top, the first panel plots the surface density profiles for the stars, molecular gas (CO), atomic gas (\HI), and total gas. The second panel plots the gas velocity dispersion from combined CO and \HI measurements, and the corresponding Gaussian process fit model. The third panel is identical to those shown in Figure 3.}
\figsetgrpend

\figsetgrpstart
\figsetgrpnum{12.4}
\figsetgrptitle{NGC~1512}
\figsetplot{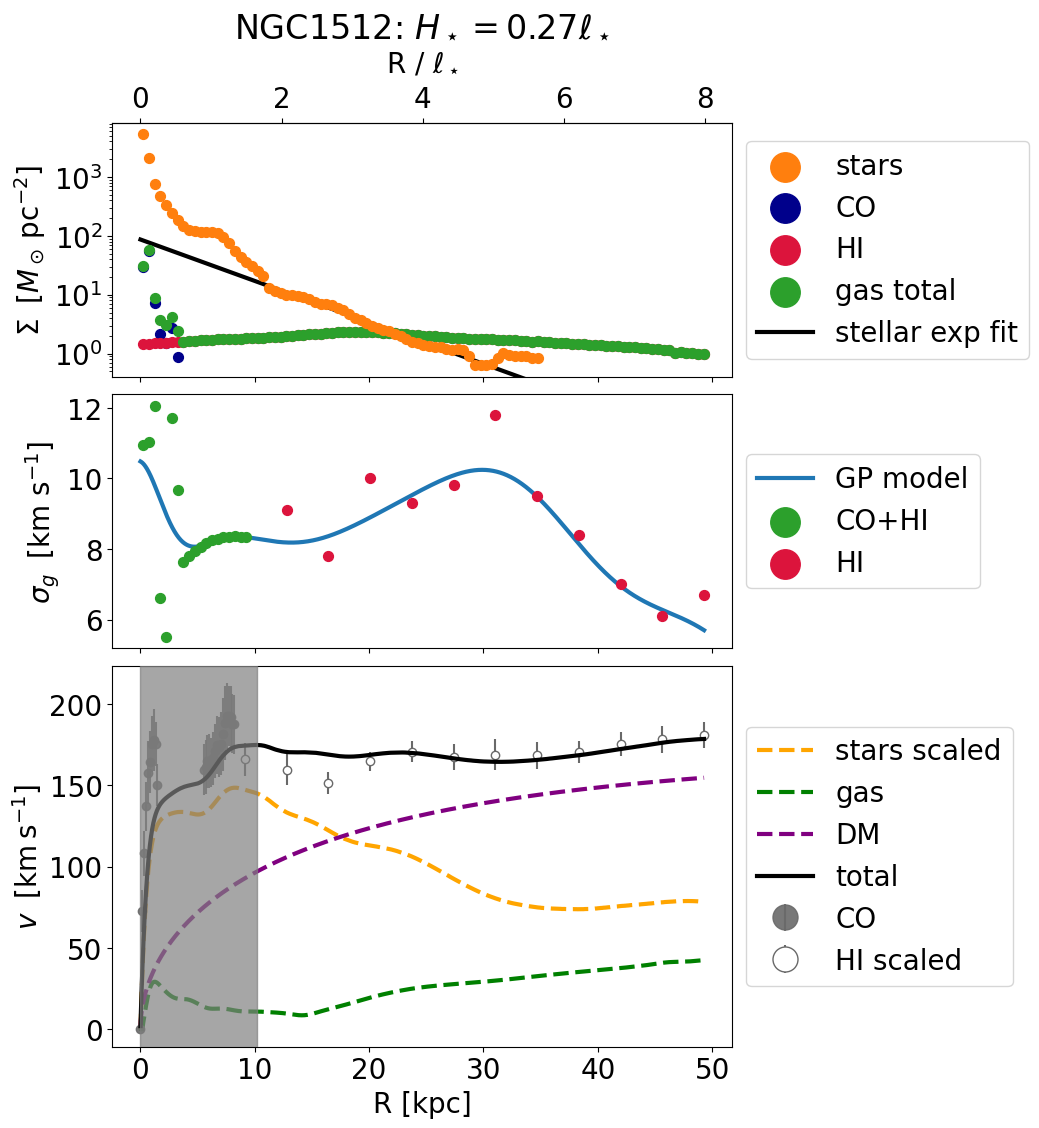}
\figsetgrpnote{Input parameters and rotation curve modeling results. From the top, the first panel plots the surface density profiles for the stars, molecular gas (CO), atomic gas (\HI), and total gas. The second panel plots the gas velocity dispersion from combined CO and \HI measurements, and the corresponding Gaussian process fit model. The third panel is identical to those shown in Figure 3.}
\figsetgrpend

\figsetgrpstart
\figsetgrpnum{12.5}
\figsetgrptitle{NGC~2283}
\figsetplot{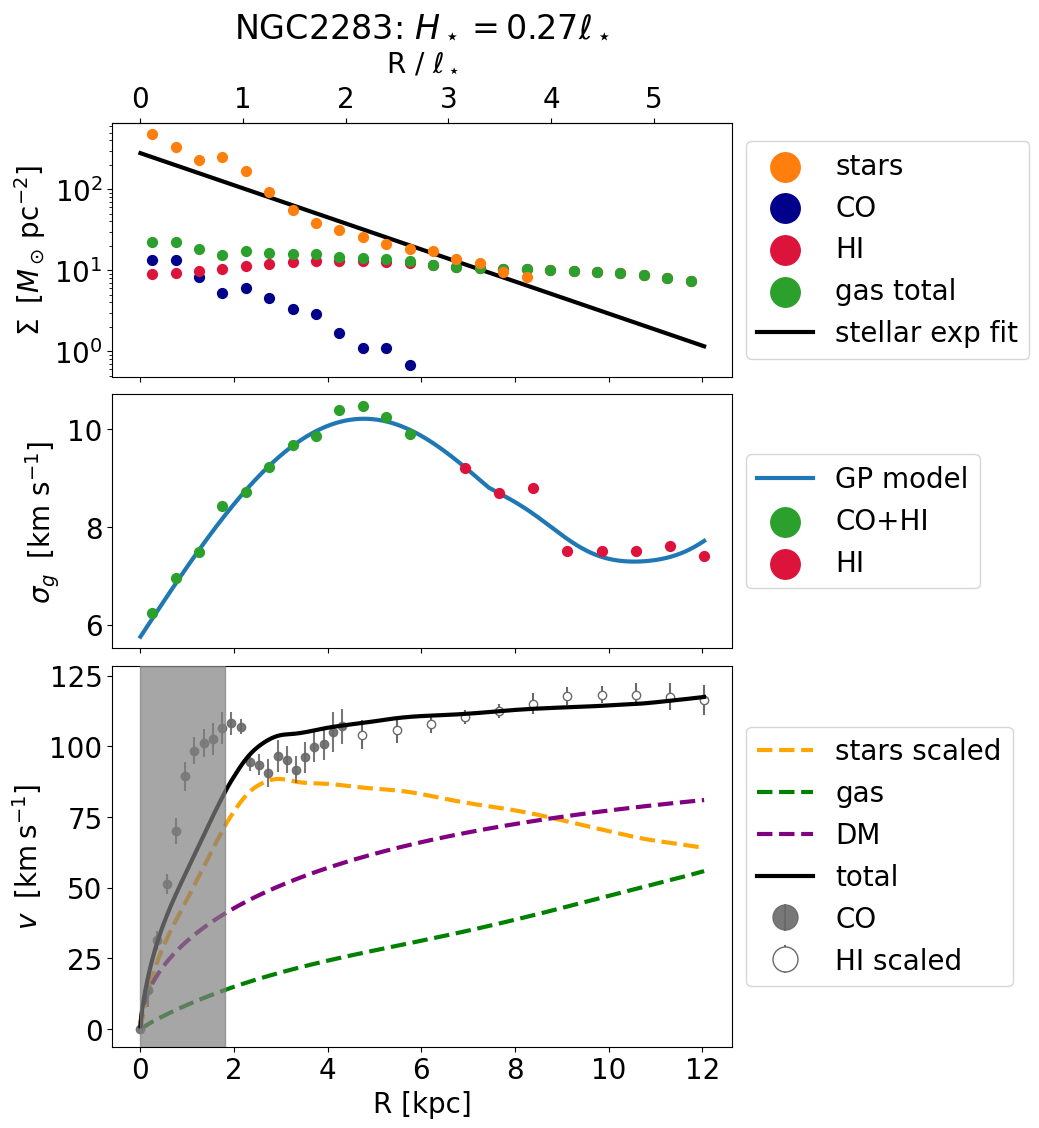}
\figsetgrpnote{Input parameters and rotation curve modeling results. From the top, the first panel plots the surface density profiles for the stars, molecular gas (CO), atomic gas (\HI), and total gas. The second panel plots the gas velocity dispersion from combined CO and \HI measurements, and the corresponding Gaussian process fit model. The third panel is identical to those shown in Figure 3.}
\figsetgrpend

\figsetgrpstart
\figsetgrpnum{12.6}
\figsetgrptitle{NGC~2835}
\figsetplot{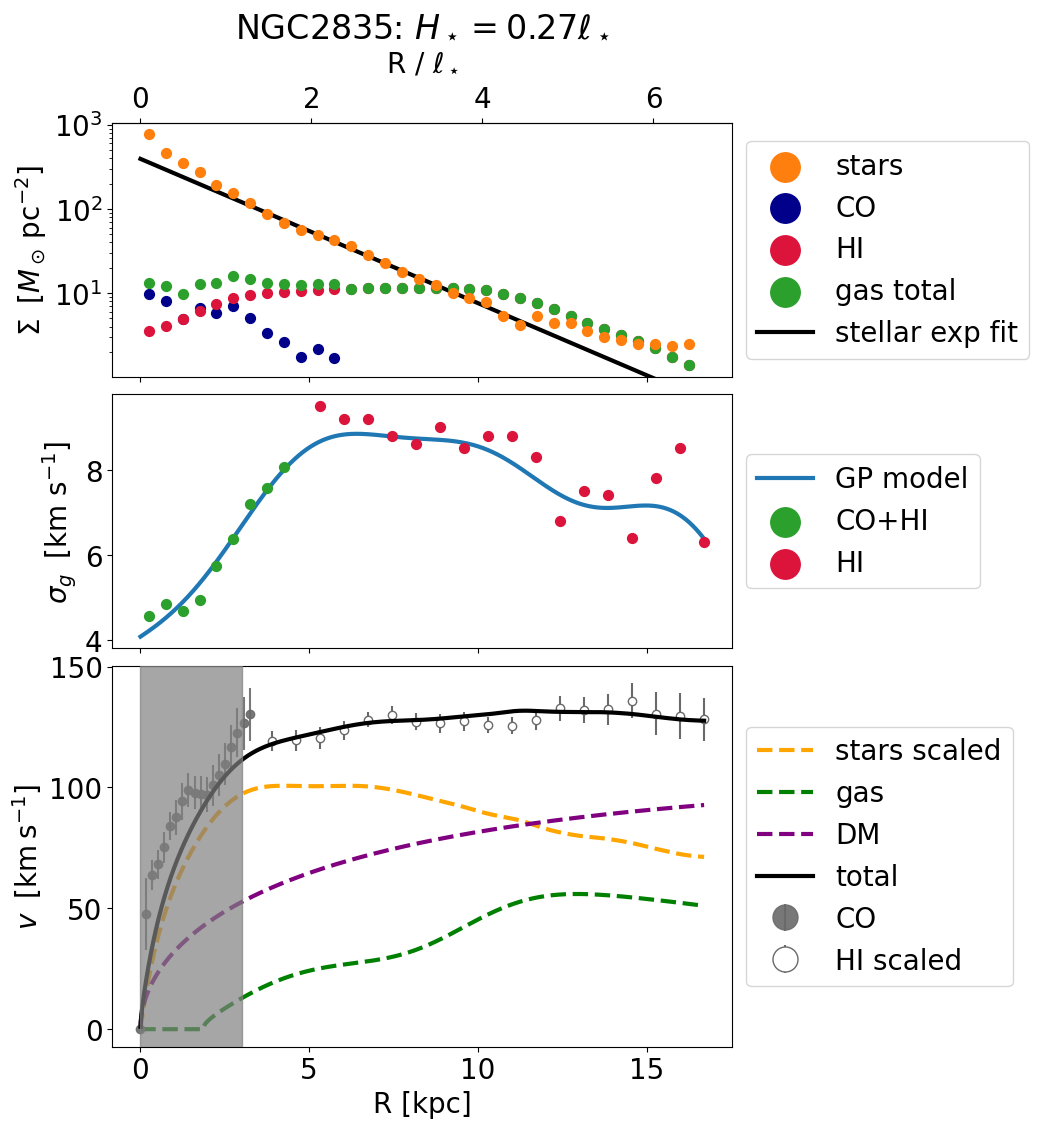}
\figsetgrpnote{Input parameters and rotation curve modeling results. From the top, the first panel plots the surface density profiles for the stars, molecular gas (CO), atomic gas (\HI), and total gas. The second panel plots the gas velocity dispersion from combined CO and \HI measurements, and the corresponding Gaussian process fit model. The third panel is identical to those shown in Figure 3.}
\figsetgrpend

\figsetgrpstart
\figsetgrpnum{12.7}
\figsetgrptitle{NGC~2903}
\figsetplot{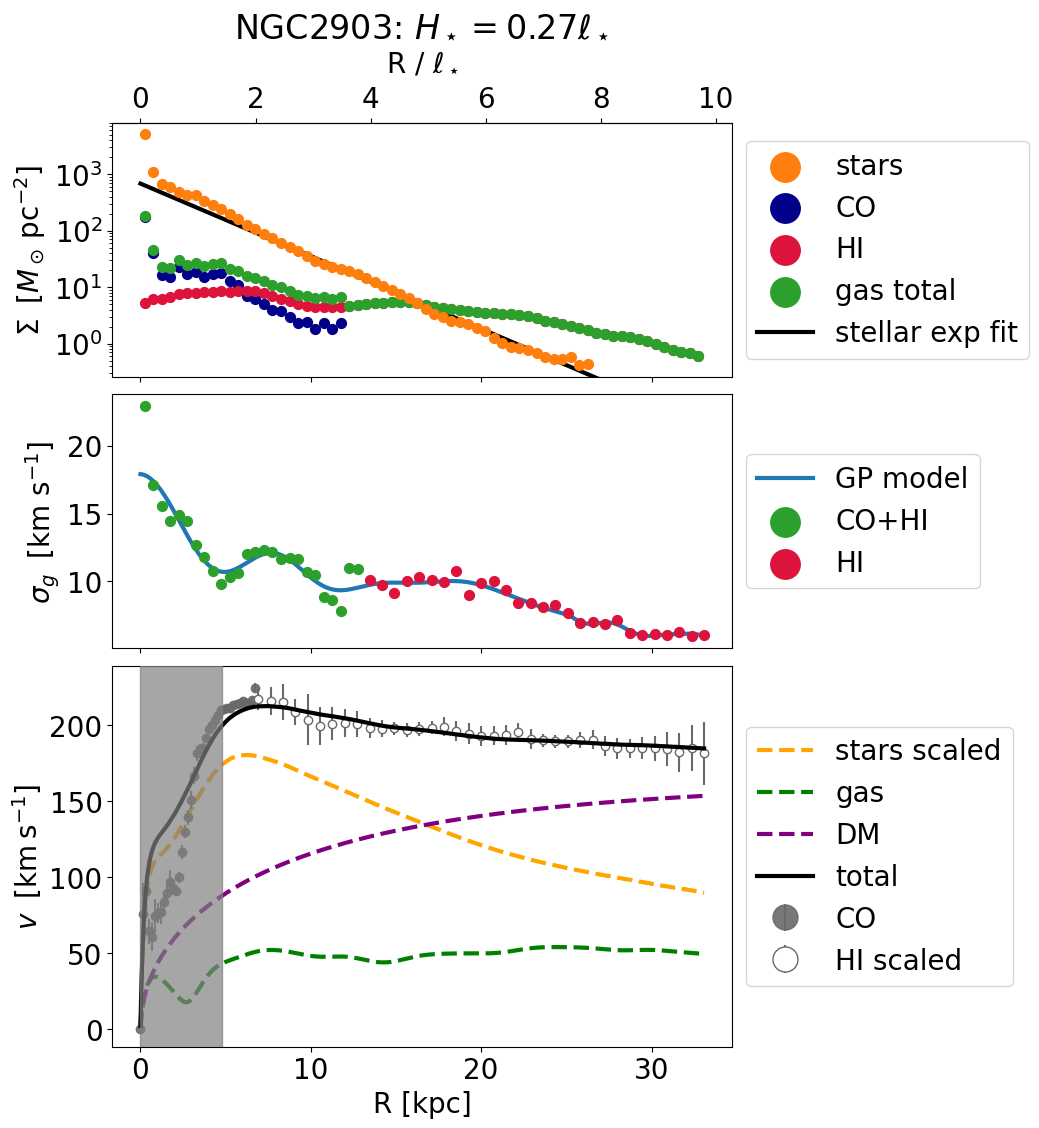}
\figsetgrpnote{Input parameters and rotation curve modeling results. From the top, the first panel plots the surface density profiles for the stars, molecular gas (CO), atomic gas (\HI), and total gas. The second panel plots the gas velocity dispersion from combined CO and \HI measurements, and the corresponding Gaussian process fit model. The third panel is identical to those shown in Figure 3.}
\figsetgrpend

\figsetgrpstart
\figsetgrpnum{12.8}
\figsetgrptitle{NGC~2997}
\figsetplot{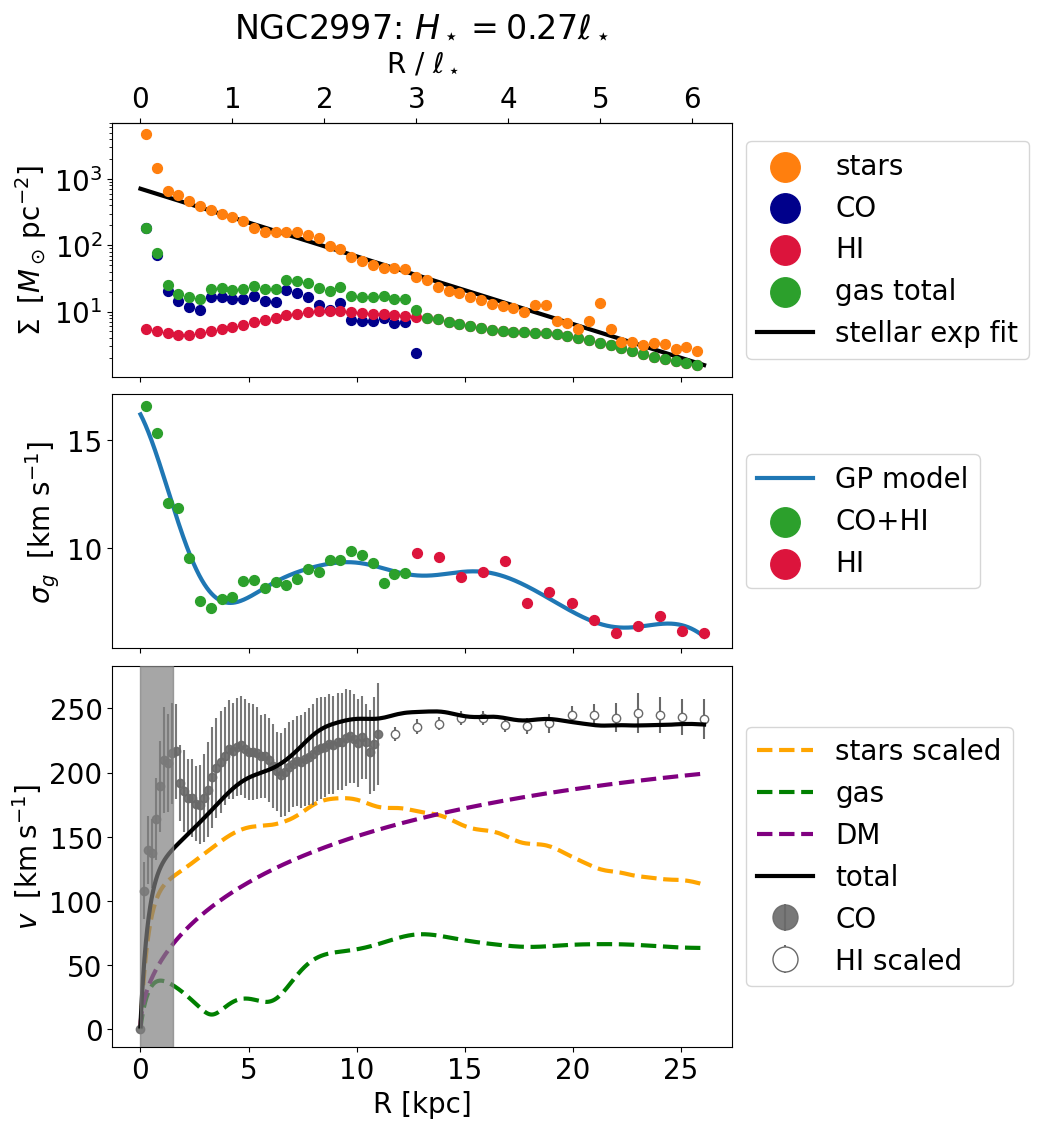}
\figsetgrpnote{Input parameters and rotation curve modeling results. From the top, the first panel plots the surface density profiles for the stars, molecular gas (CO), atomic gas (\HI), and total gas. The second panel plots the gas velocity dispersion from combined CO and \HI measurements, and the corresponding Gaussian process fit model. The third panel is identical to those shown in Figure 3.}
\figsetgrpend

\figsetgrpstart
\figsetgrpnum{12.9}
\figsetgrptitle{NGC~3137}
\figsetplot{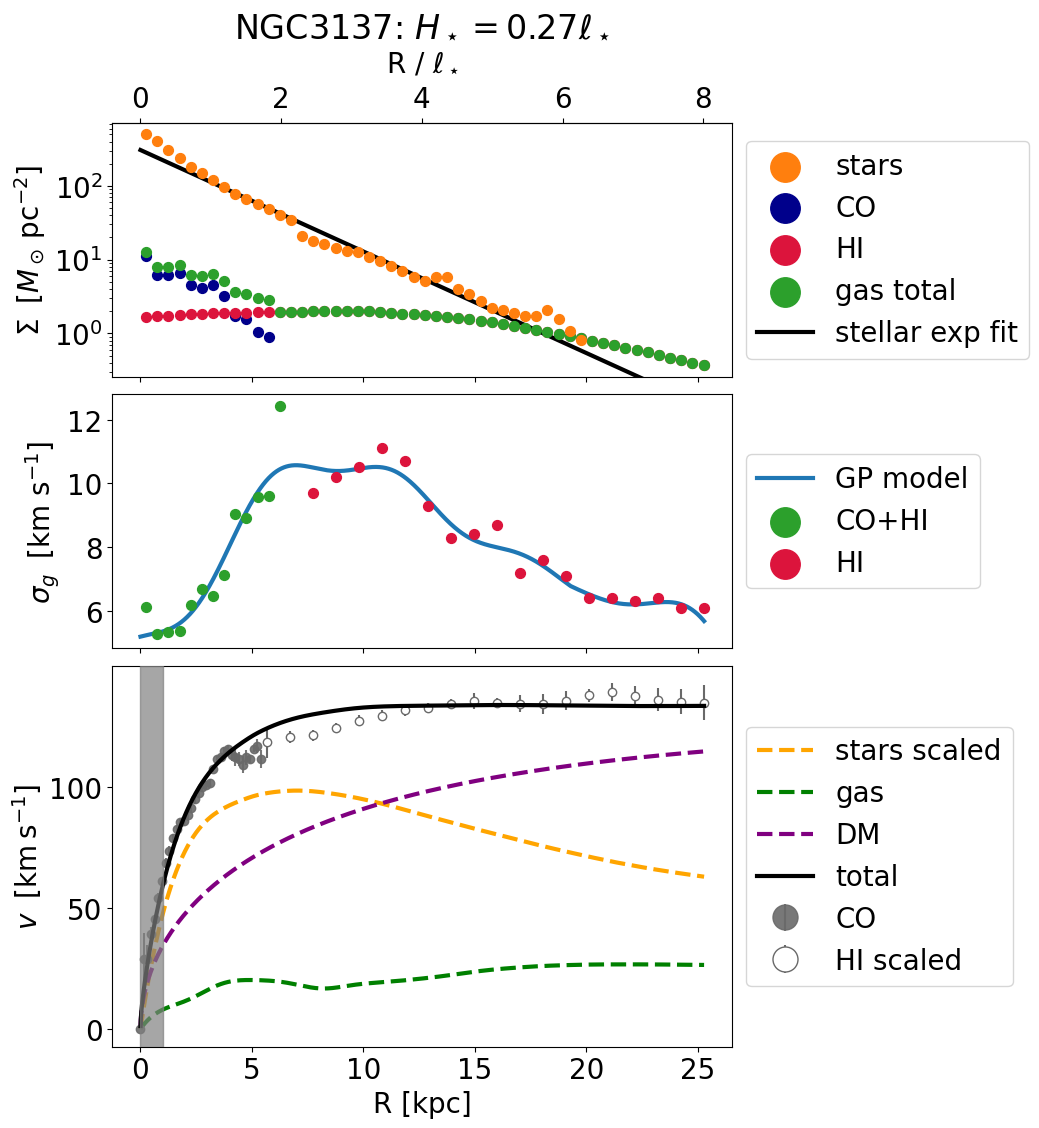}
\figsetgrpnote{Input parameters and rotation curve modeling results. From the top, the first panel plots the surface density profiles for the stars, molecular gas (CO), atomic gas (\HI), and total gas. The second panel plots the gas velocity dispersion from combined CO and \HI measurements, and the corresponding Gaussian process fit model. The third panel is identical to those shown in Figure 3.}
\figsetgrpend

\figsetgrpstart
\figsetgrpnum{12.10}
\figsetgrptitle{NGC~3507}
\figsetplot{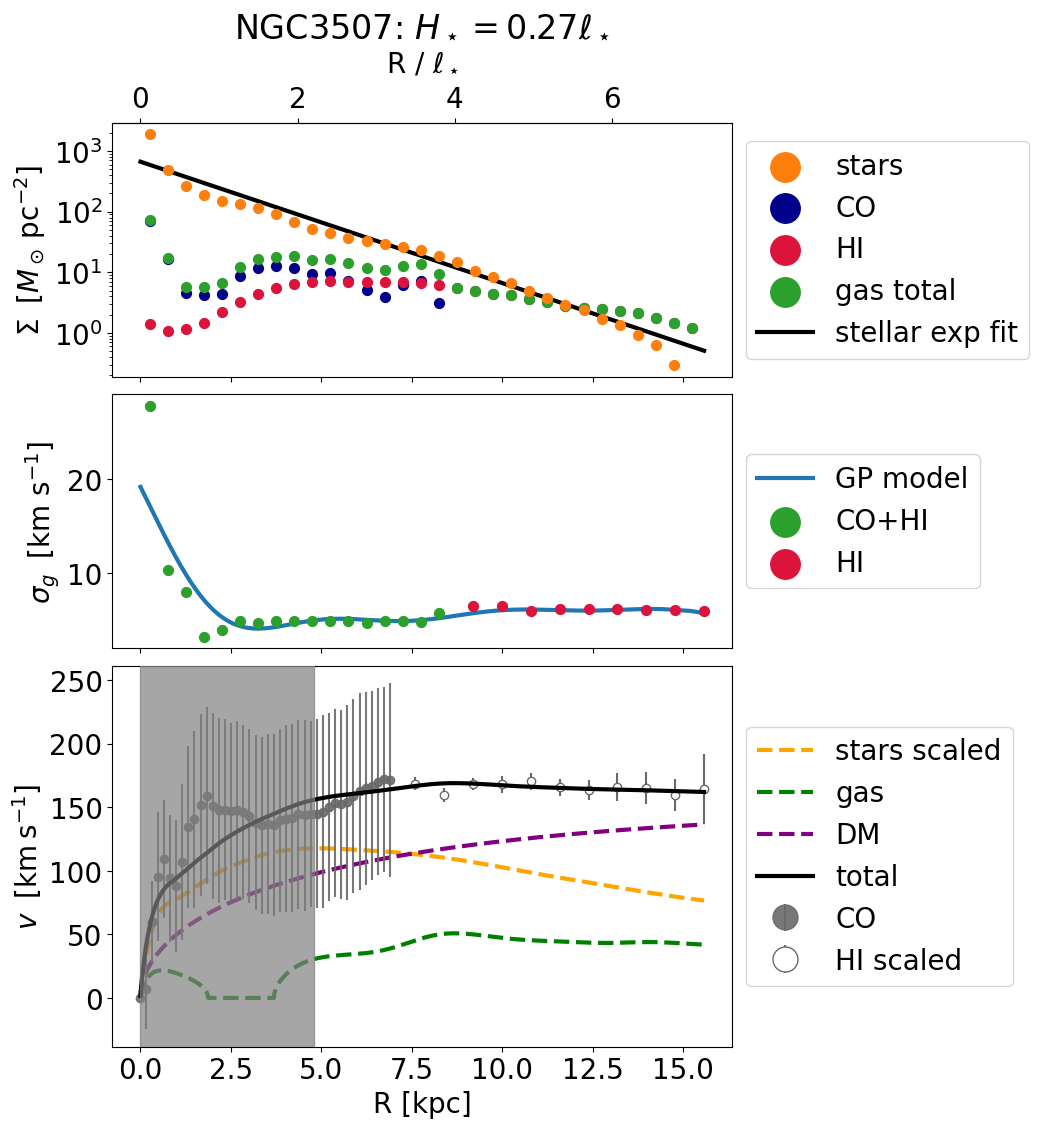}
\figsetgrpnote{Input parameters and rotation curve modeling results. From the top, the first panel plots the surface density profiles for the stars, molecular gas (CO), atomic gas (\HI), and total gas. The second panel plots the gas velocity dispersion from combined CO and \HI measurements, and the corresponding Gaussian process fit model. The third panel is identical to those shown in Figure 3.}
\figsetgrpend

\figsetgrpstart
\figsetgrpnum{12.11}
\figsetgrptitle{NGC~3511}
\figsetplot{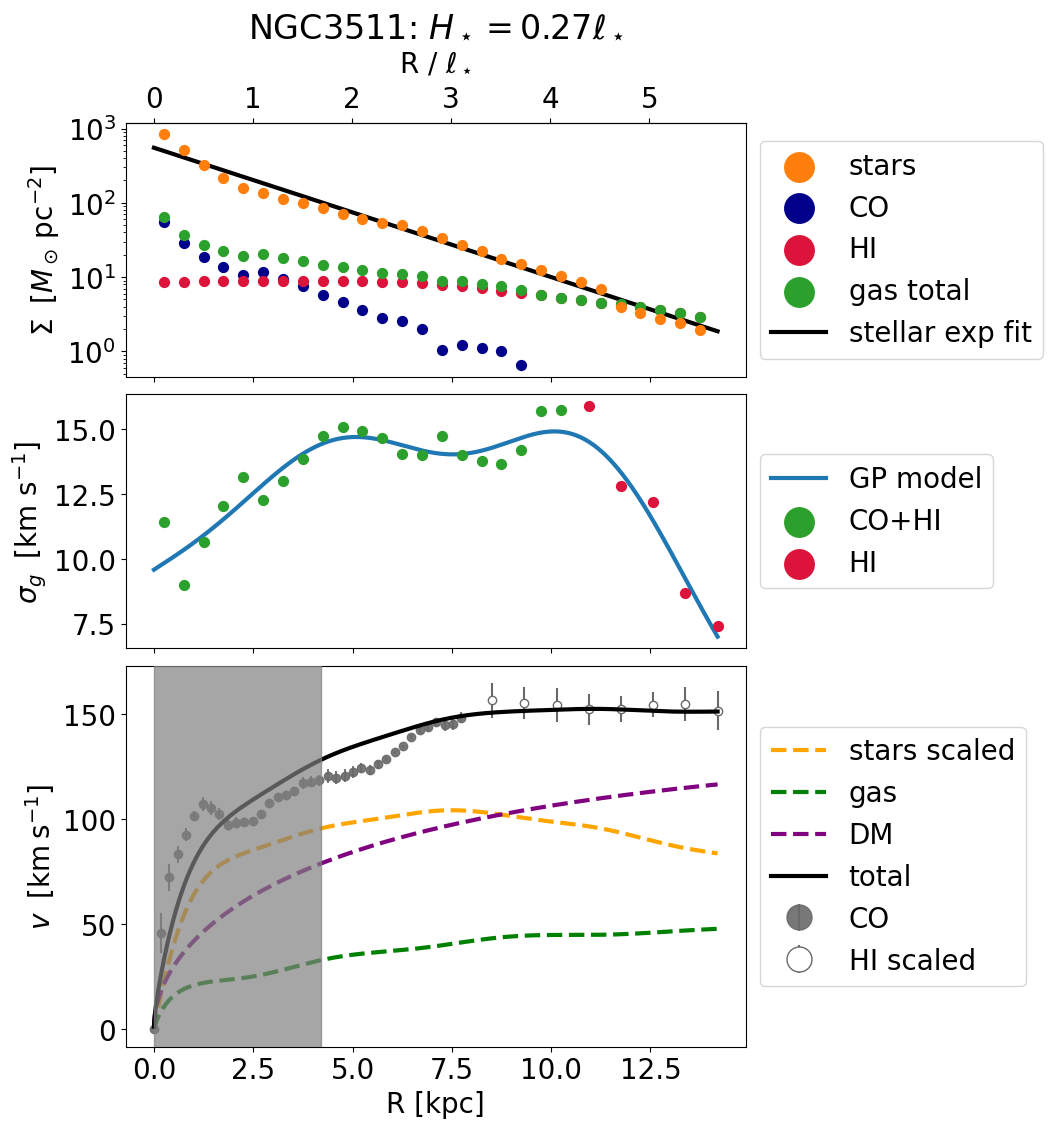}
\figsetgrpnote{Input parameters and rotation curve modeling results. From the top, the first panel plots the surface density profiles for the stars, molecular gas (CO), atomic gas (\HI), and total gas. The second panel plots the gas velocity dispersion from combined CO and \HI measurements, and the corresponding Gaussian process fit model. The third panel is identical to those shown in Figure 3.}
\figsetgrpend

\figsetgrpstart
\figsetgrpnum{12.12}
\figsetgrptitle{NGC~3521}
\figsetplot{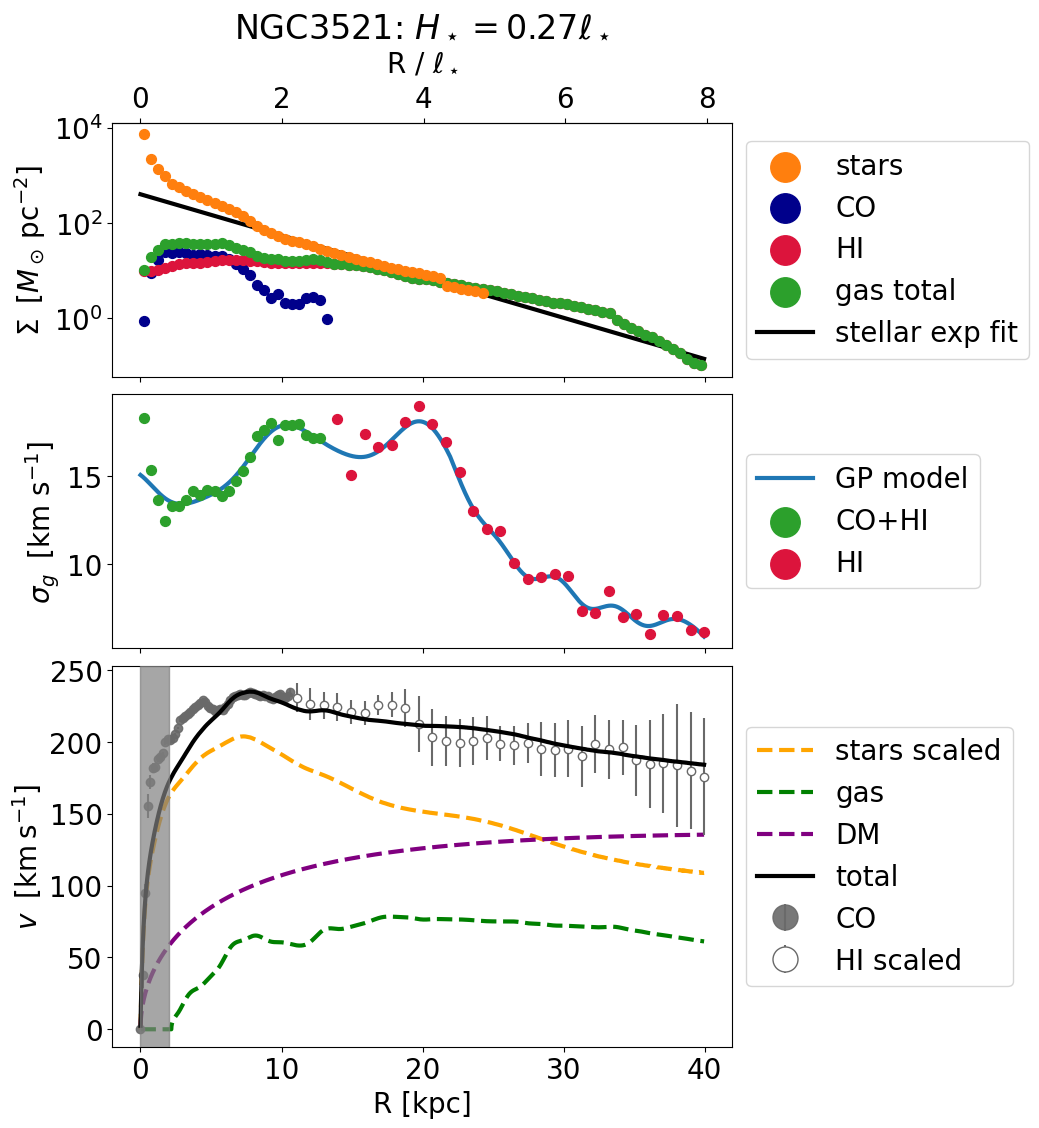}
\figsetgrpnote{Input parameters and rotation curve modeling results. From the top, the first panel plots the surface density profiles for the stars, molecular gas (CO), atomic gas (\HI), and total gas. The second panel plots the gas velocity dispersion from combined CO and \HI measurements, and the corresponding Gaussian process fit model. The third panel is identical to those shown in Figure 3.}
\figsetgrpend

\figsetgrpstart
\figsetgrpnum{12.13}
\figsetgrptitle{NGC~3621}
\figsetplot{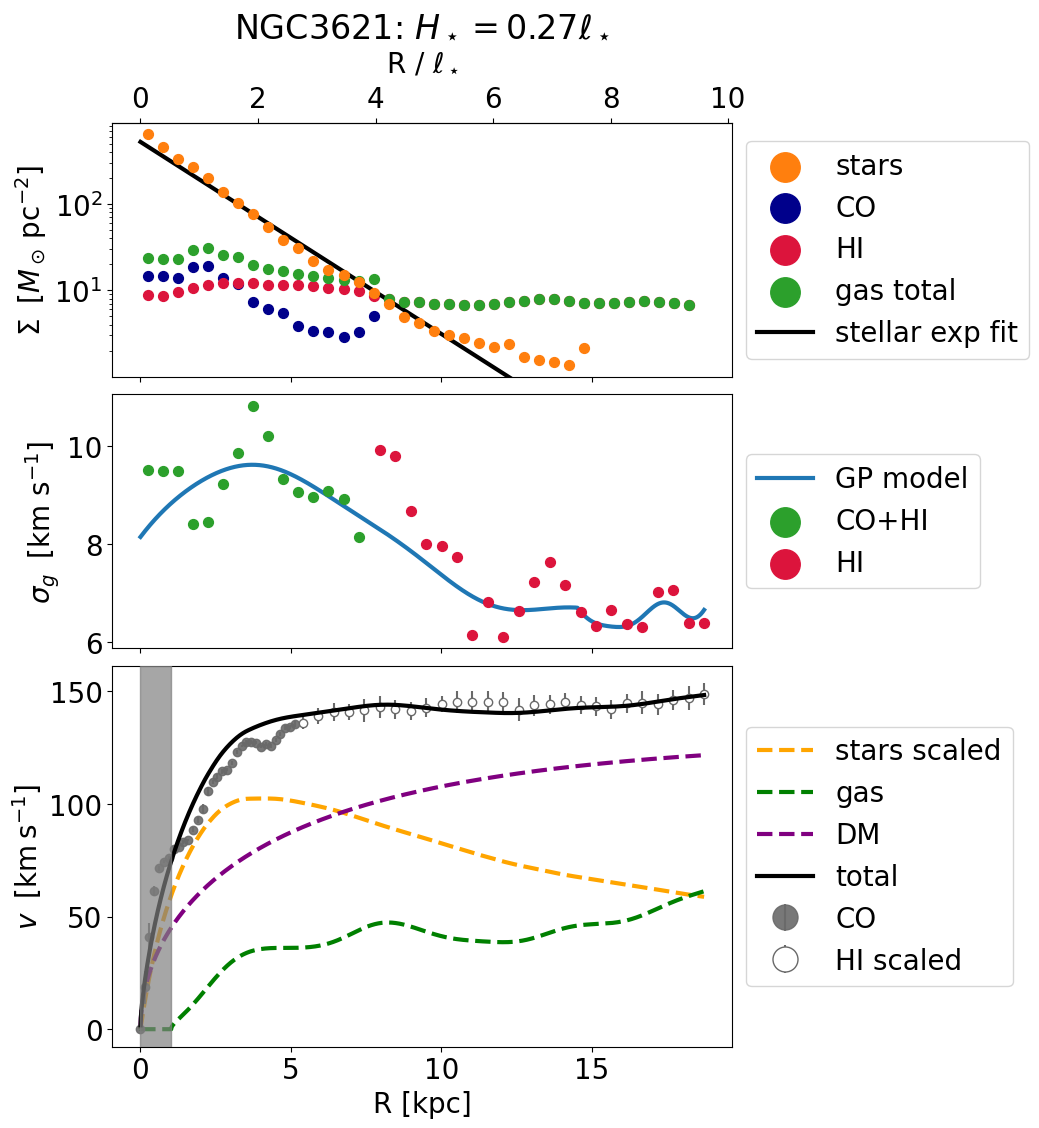}
\figsetgrpnote{Input parameters and rotation curve modeling results. From the top, the first panel plots the surface density profiles for the stars, molecular gas (CO), atomic gas (\HI), and total gas. The second panel plots the gas velocity dispersion from combined CO and \HI measurements, and the corresponding Gaussian process fit model. The third panel is identical to those shown in Figure 3.}
\figsetgrpend

\figsetgrpstart
\figsetgrpnum{12.14}
\figsetgrptitle{NGC~4303}
\figsetplot{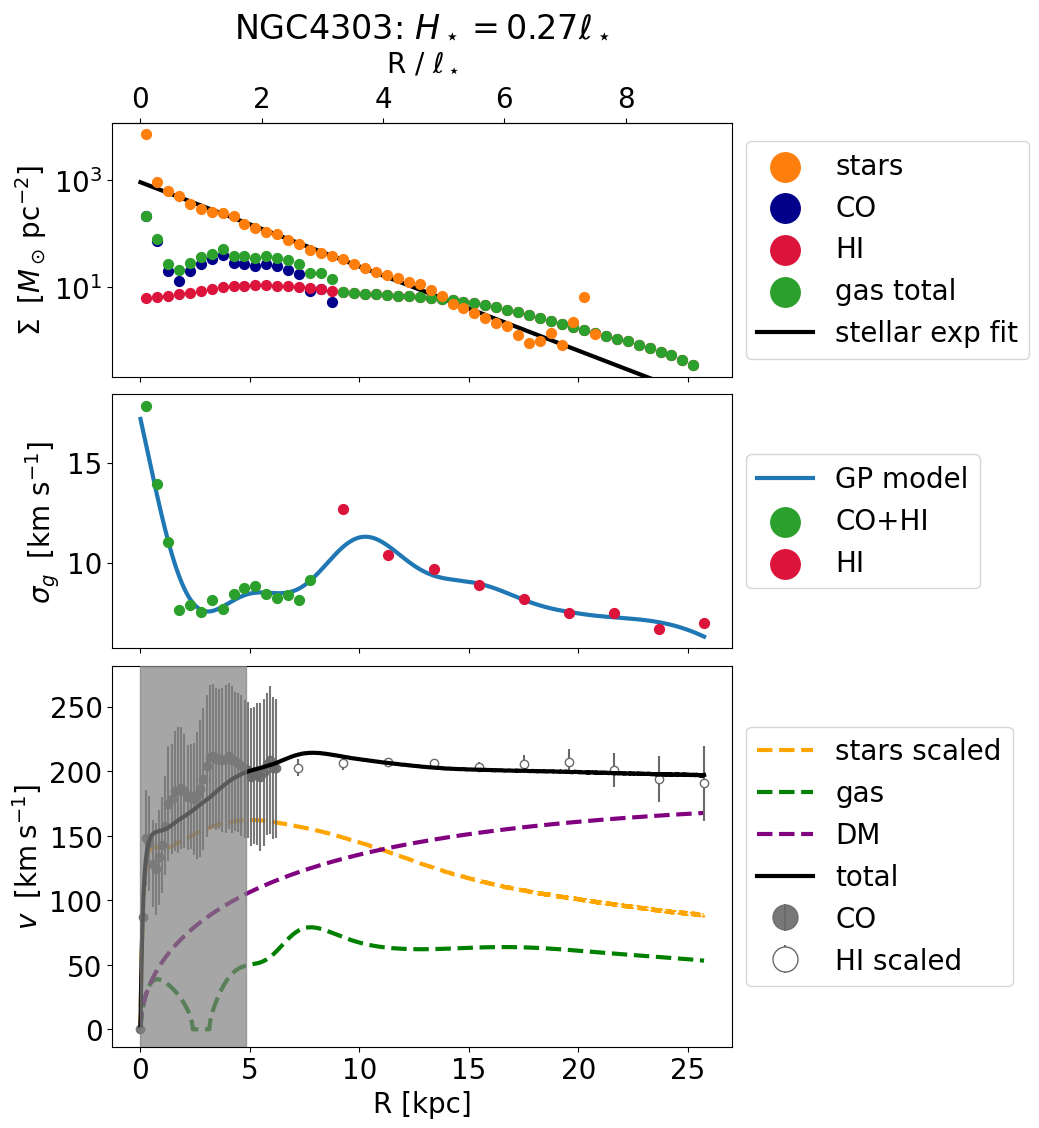}
\figsetgrpnote{Input parameters and rotation curve modeling results. From the top, the first panel plots the surface density profiles for the stars, molecular gas (CO), atomic gas (\HI), and total gas. The second panel plots the gas velocity dispersion from combined CO and \HI measurements, and the corresponding Gaussian process fit model. The third panel is identical to those shown in Figure 3.}
\figsetgrpend

\figsetgrpstart
\figsetgrpnum{12.15}
\figsetgrptitle{NGC~4571}
\figsetplot{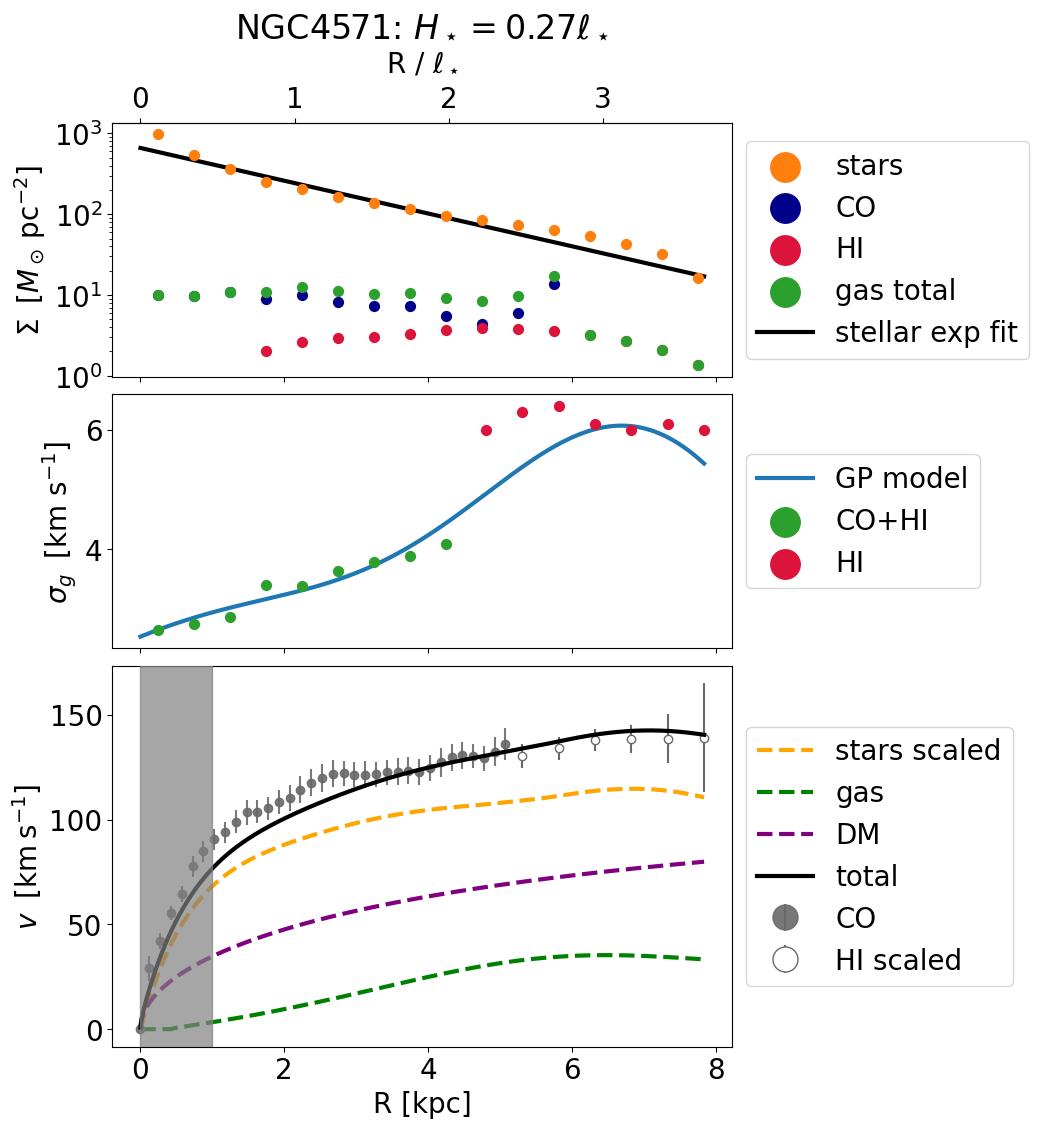}
\figsetgrpnote{Input parameters and rotation curve modeling results. From the top, the first panel plots the surface density profiles for the stars, molecular gas (CO), atomic gas (\HI), and total gas. The second panel plots the gas velocity dispersion from combined CO and \HI measurements, and the corresponding Gaussian process fit model. The third panel is identical to those shown in Figure 3.}
\figsetgrpend

\figsetgrpstart
\figsetgrpnum{12.16}
\figsetgrptitle{NGC~4781}
\figsetplot{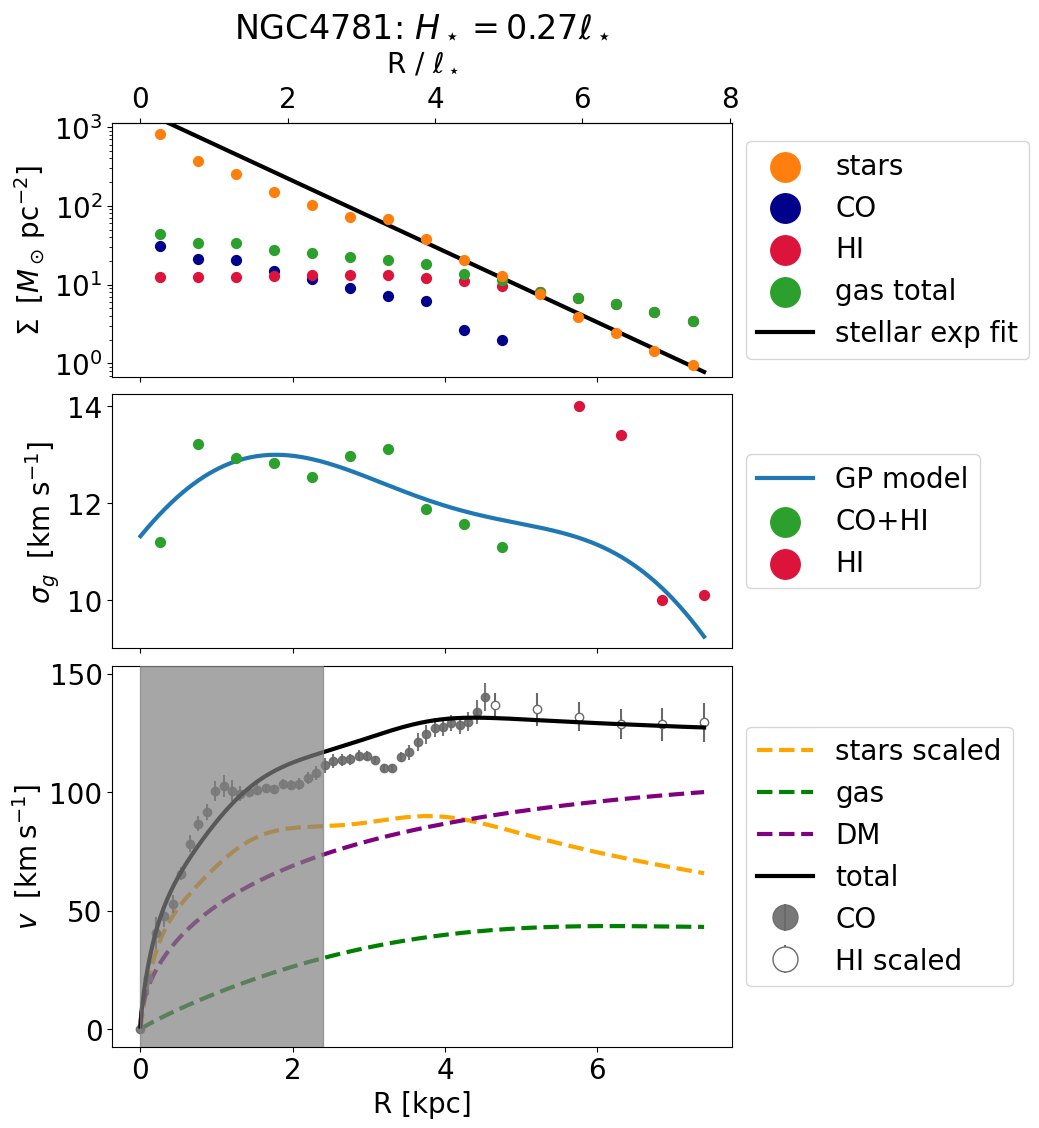}
\figsetgrpnote{Input parameters and rotation curve modeling results. From the top, the first panel plots the surface density profiles for the stars, molecular gas (CO), atomic gas (\HI), and total gas. The second panel plots the gas velocity dispersion from combined CO and \HI measurements, and the corresponding Gaussian process fit model. The third panel is identical to those shown in Figure 3.}
\figsetgrpend

\figsetgrpstart
\figsetgrpnum{A1.17}
\figsetgrptitle{NGC~5042}
\figsetplot{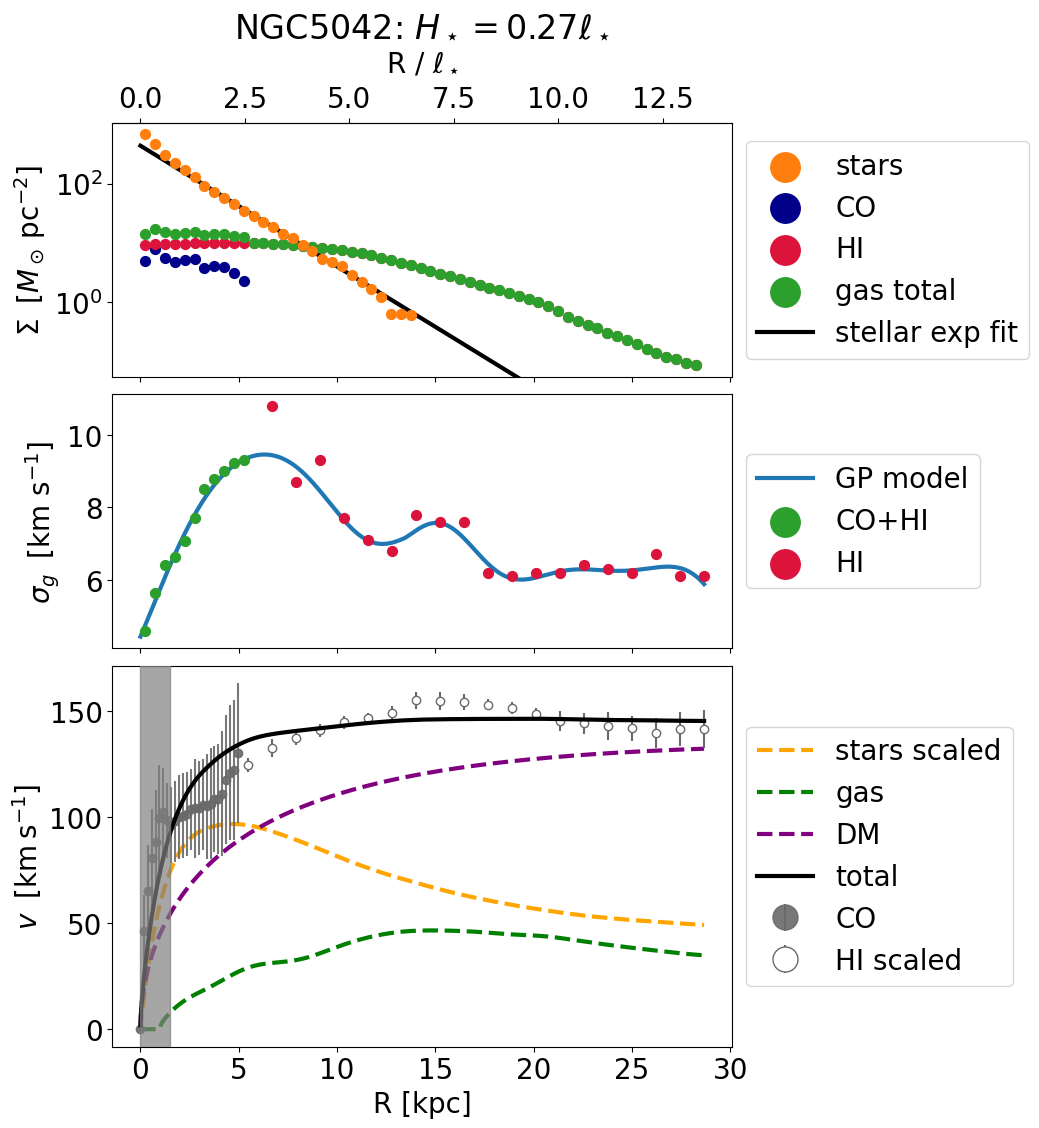}
\figsetgrpnote{Input parameters and rotation curve modeling results. From the top, the first panel plots the surface density profiles for the stars, molecular gas (CO), atomic gas (\HI), and total gas. The second panel plots the gas velocity dispersion from combined CO and \HI measurements, and the corresponding Gaussian process fit model. The third panel is identical to those shown in Figure 3.}
\figsetgrpend
\figsetend

\section{Gas Weight and Vertical Scale Height Profiles}
\label{apdx:weight}

\setcounter{figure}{0}

We report radial profiles of gas weight and vertical scale height for all 17 galaxies in \autoref{fig:spam}.

\begin{figure*}[p!]
\gridline{
\fig{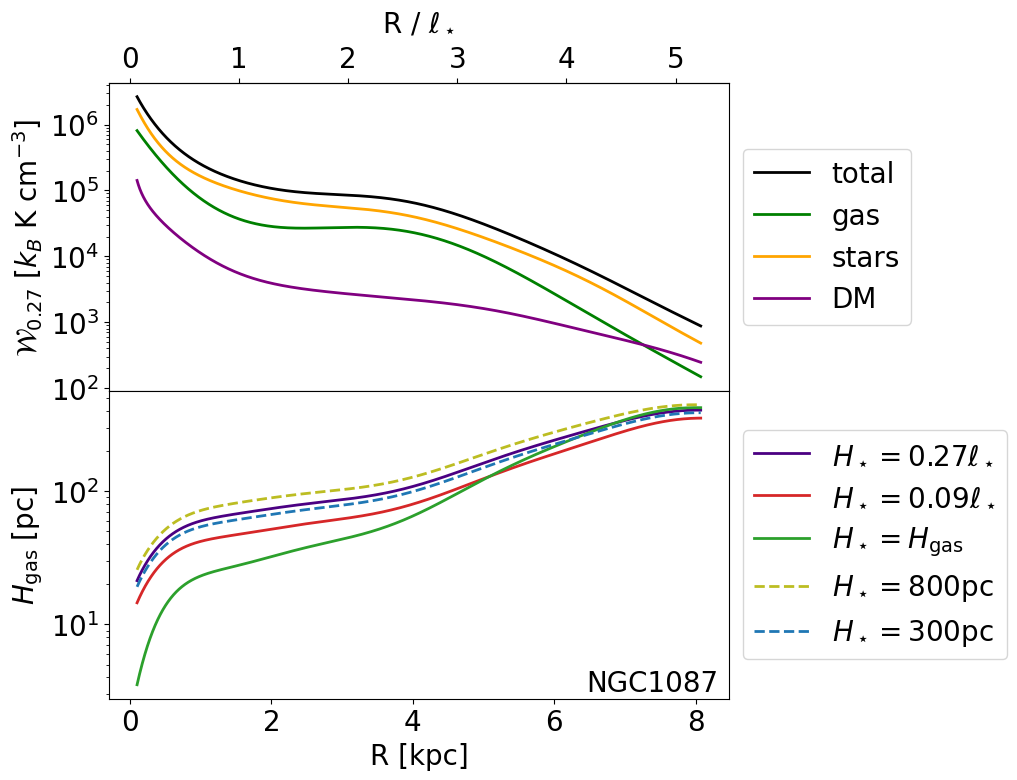}{0.49\textwidth}{NGC~1087}
\fig{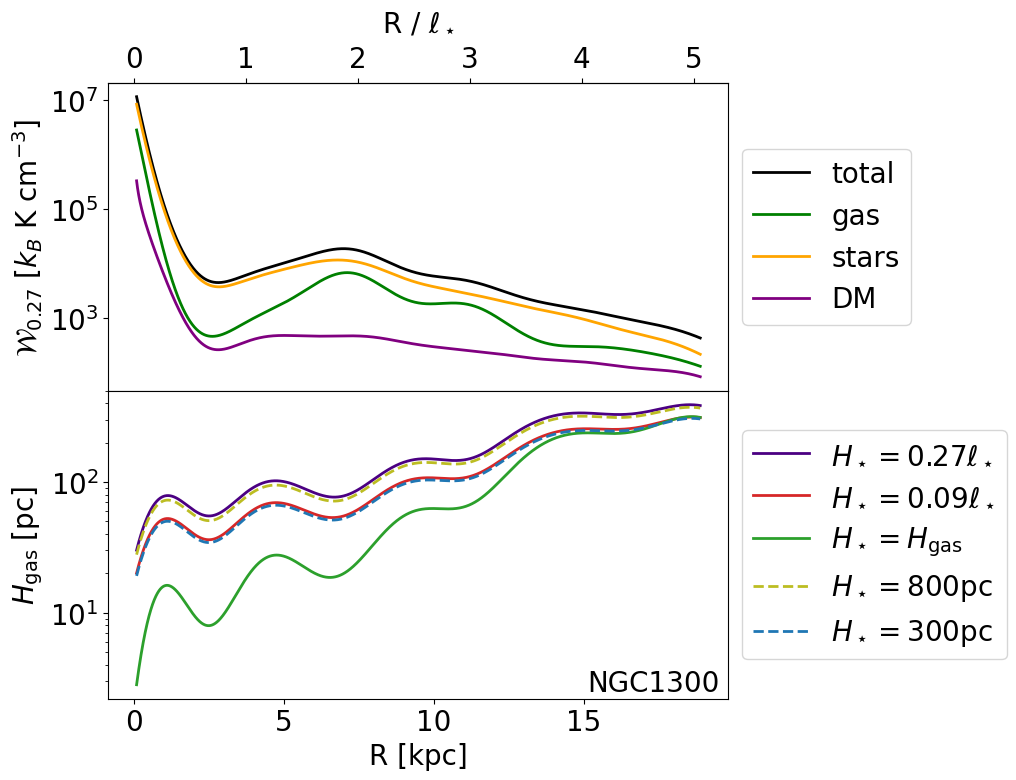}{0.49\textwidth}{NGC~1300}
}
\vspace{-1\baselineskip}
\gridline{
\fig{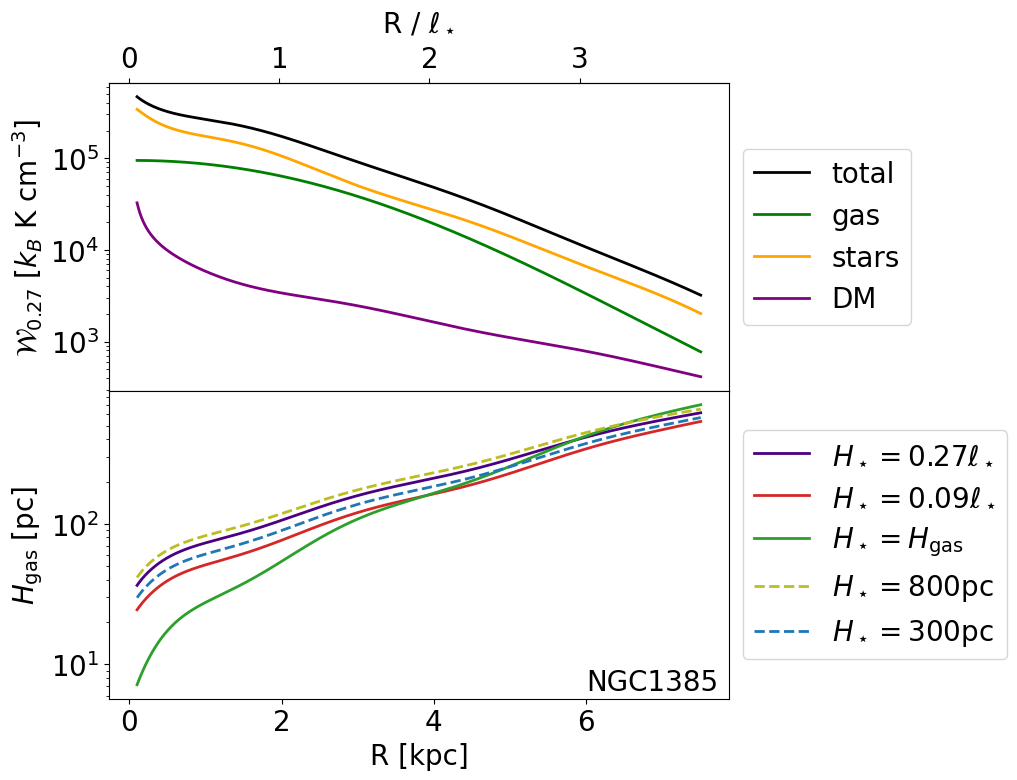}{0.49\textwidth}{NGC~1385}
\fig{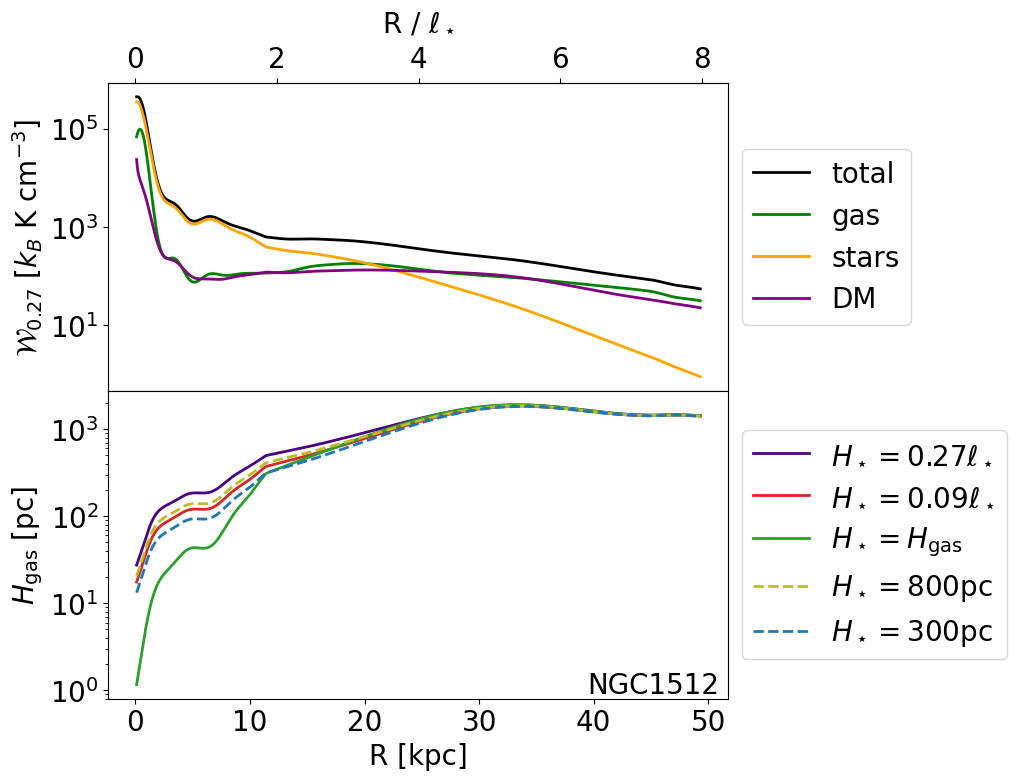}{0.49\textwidth}{NGC~1512}
}
\vspace{-1\baselineskip}
\gridline{
\fig{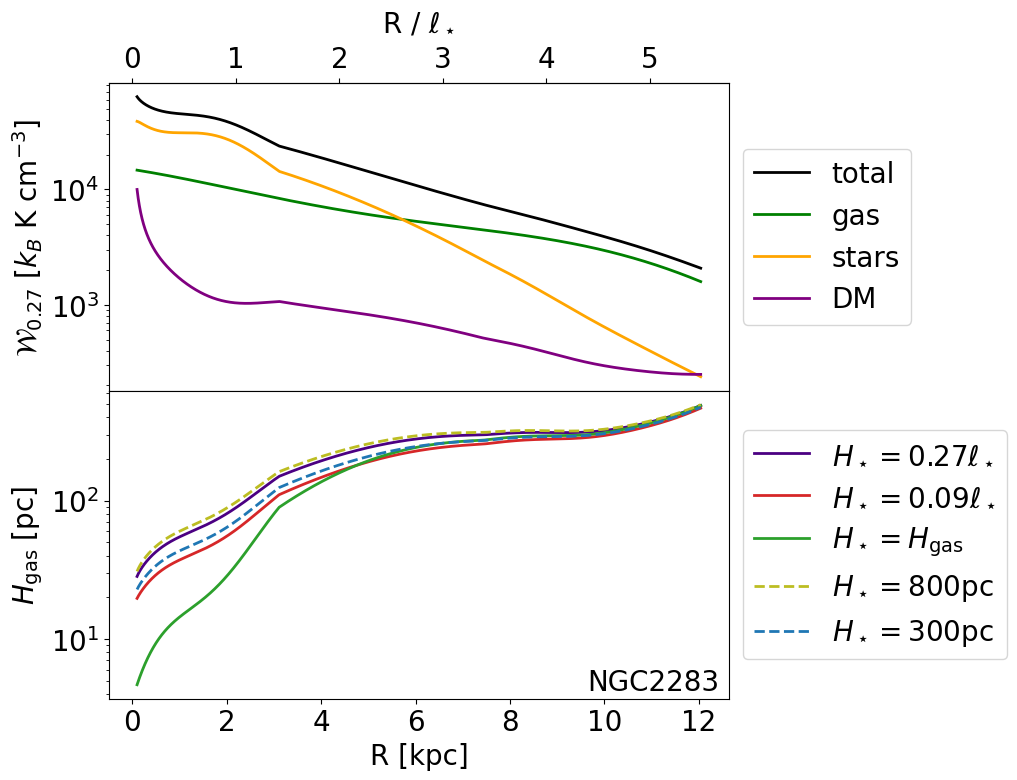}{0.49\textwidth}{NGC~2283}
\fig{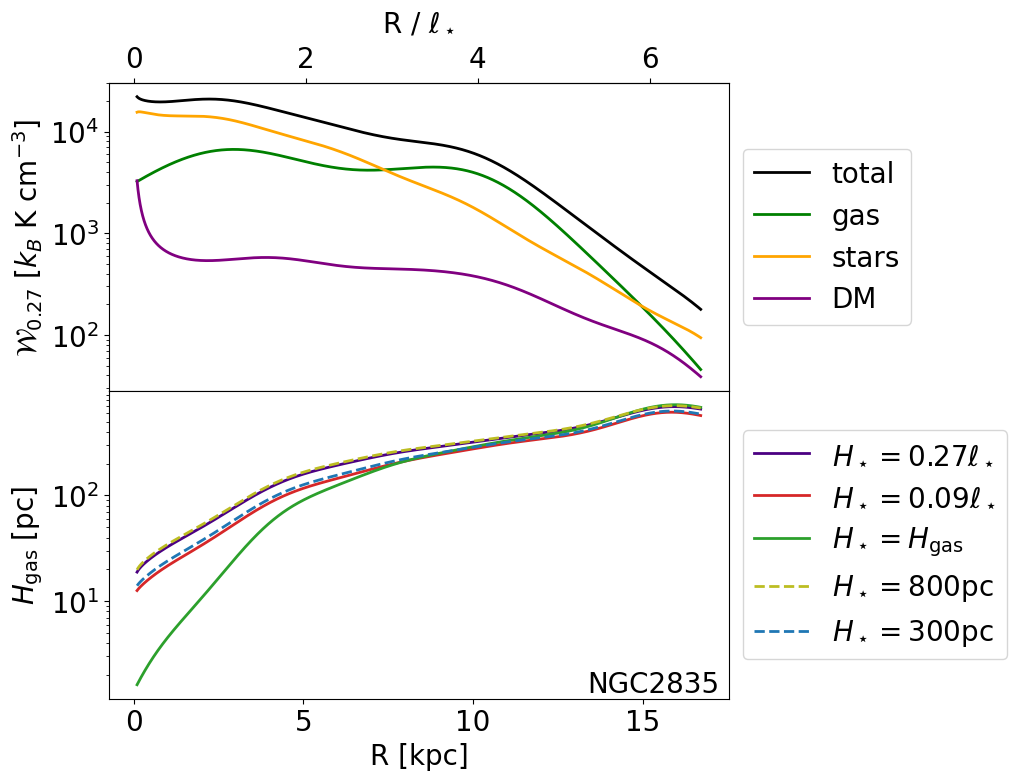}{0.49\textwidth}{NGC~2835}
}
\vspace{-0.5\baselineskip}
\caption{Gas weight and vertical scale height profiles derived in this work. The top and bottom panels correspond to \autoref{fig:weight}(a) and \autoref{fig:scaleheights}, respectively, but here showing results for all 17 galaxies.}
\label{fig:spam}
\end{figure*}

\setcounter{figure}{0}
\begin{figure*}[p!]
\gridline{
\fig{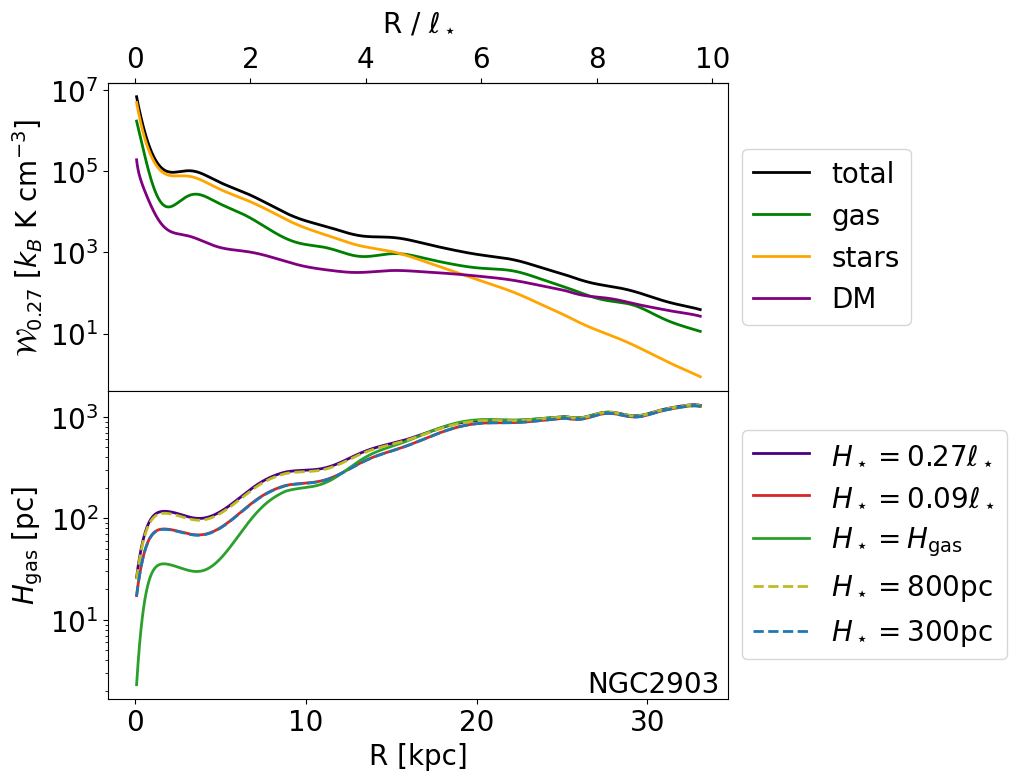}{0.49\textwidth}{NGC~2903}
\fig{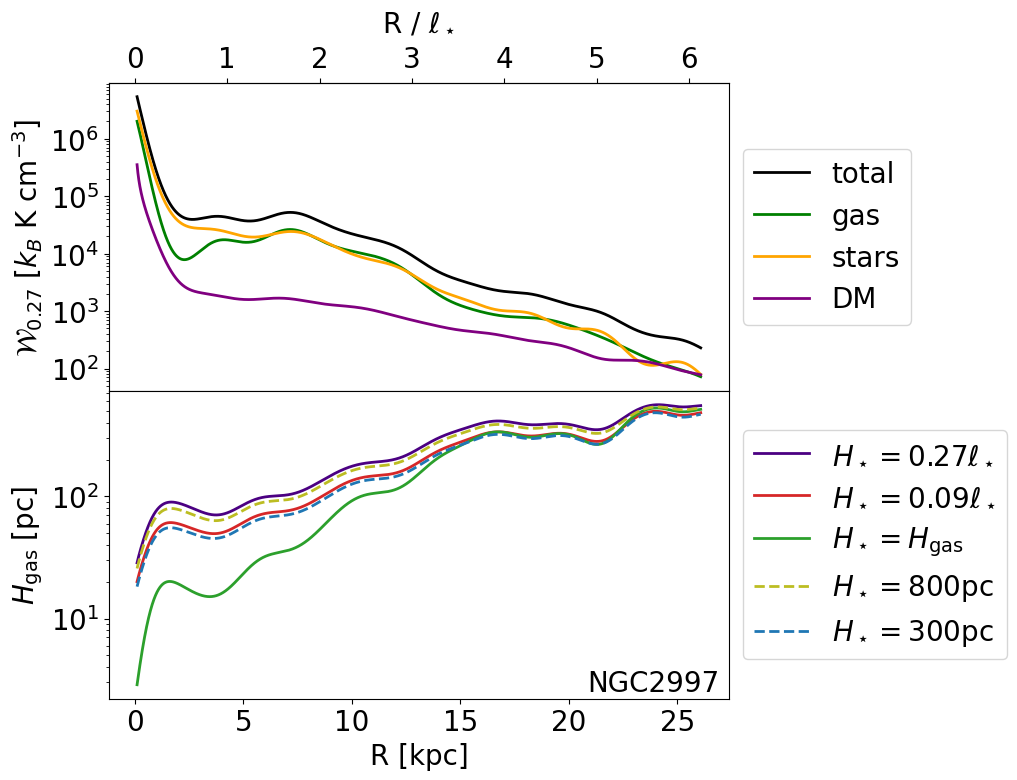}{0.49\textwidth}{NGC~2997}
}
\vspace{-1\baselineskip}
\gridline{
\fig{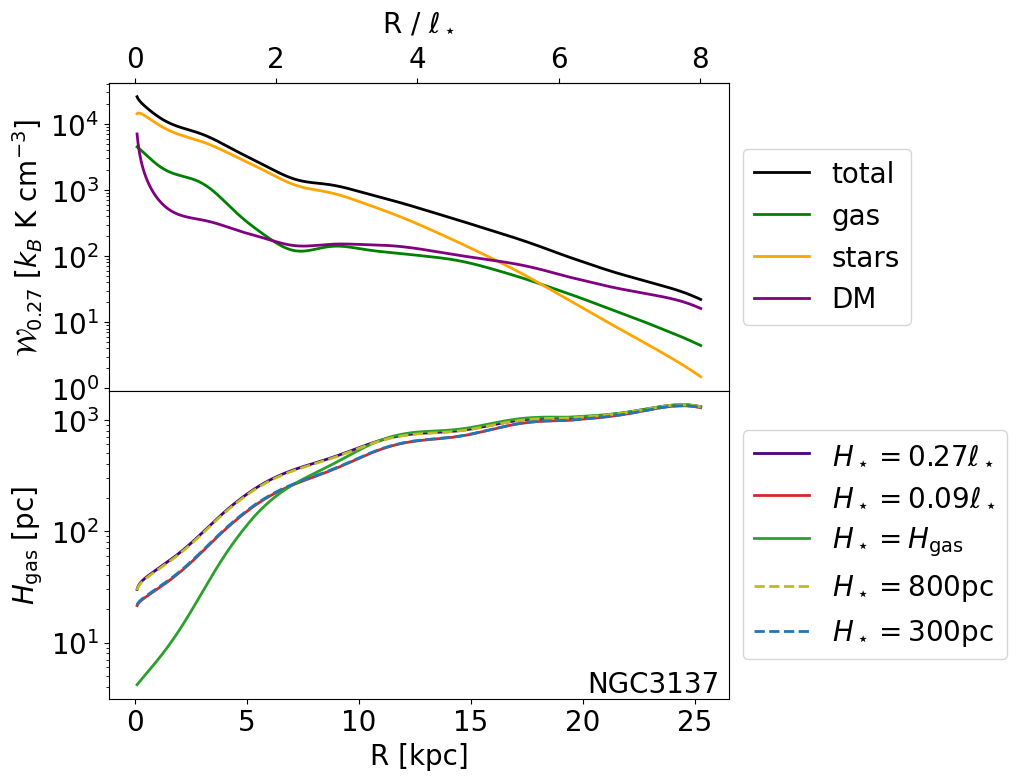}{0.49\textwidth}{NGC~3137}
\fig{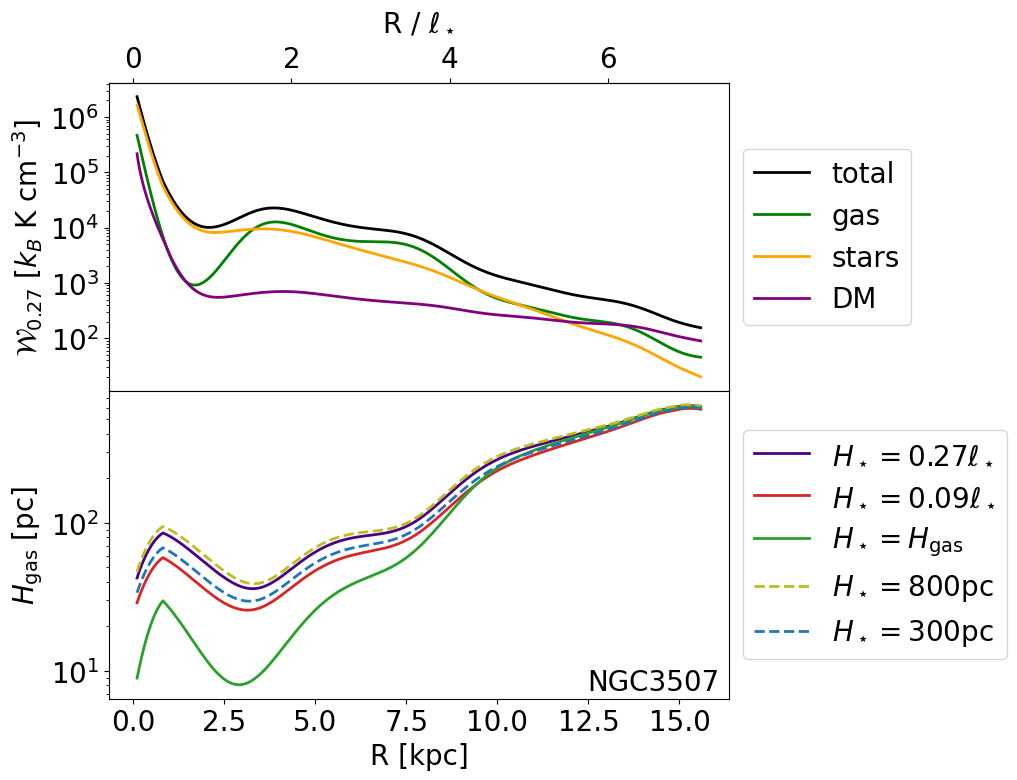}{0.49\textwidth}{NGC~3507}
}
\vspace{-1\baselineskip}
\gridline{
\fig{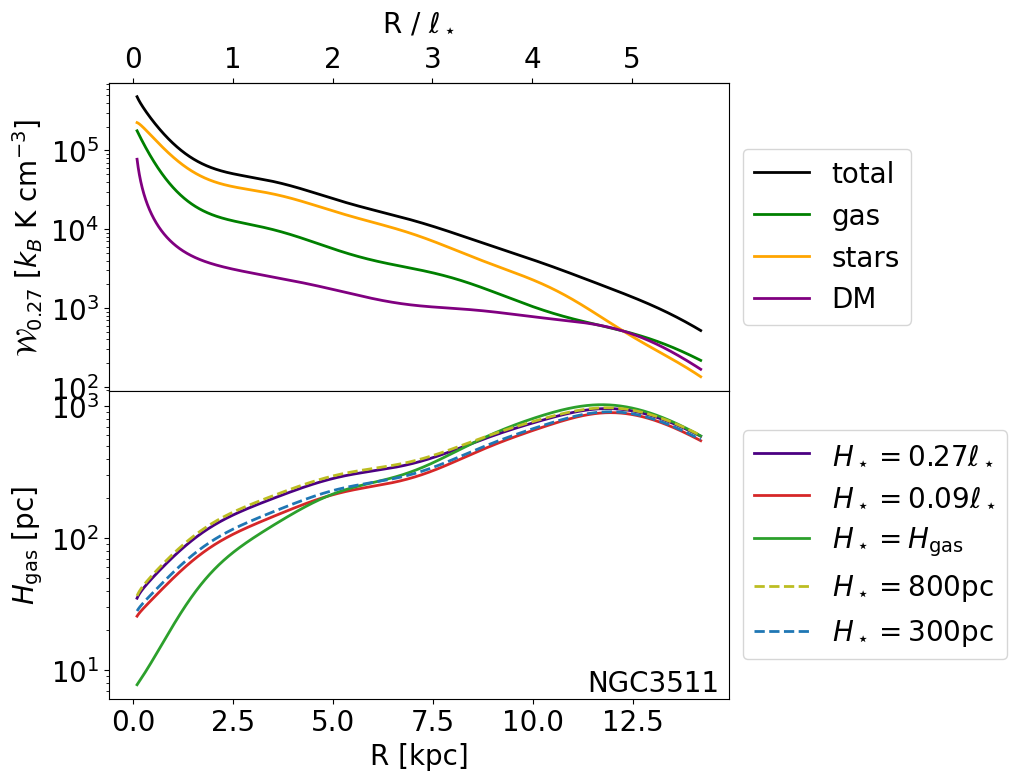}{0.49\textwidth}{NGC~3511}
\fig{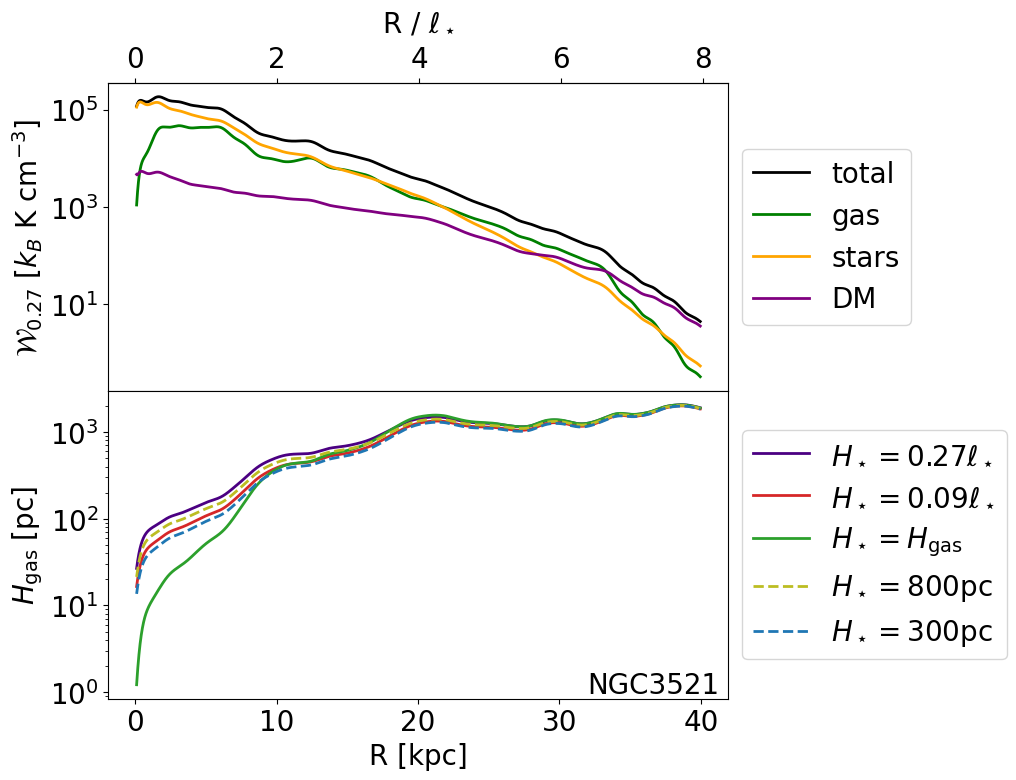}{0.49\textwidth}{NGC~3521}
}
\vspace{-0.5\baselineskip}
\caption{(continued)}
\end{figure*}

\setcounter{figure}{0}
\begin{figure*}[p!]
\gridline{
\fig{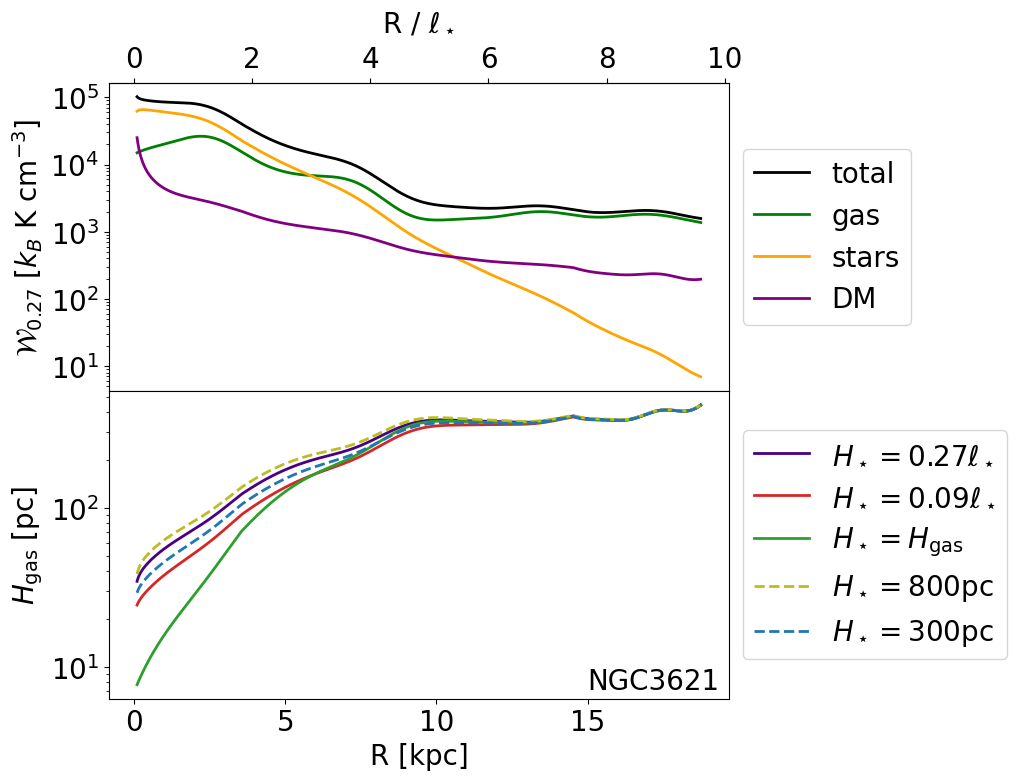}{0.49\textwidth}{NGC~3621}
\fig{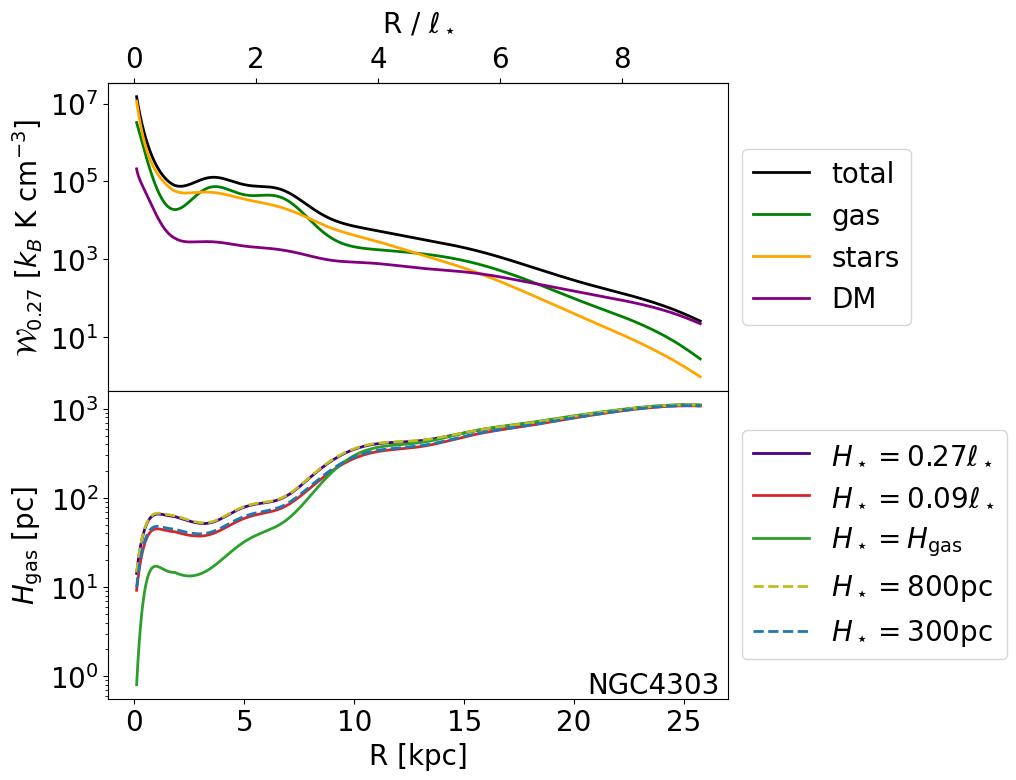}{0.49\textwidth}{NGC~4303}
}
\vspace{-1\baselineskip}
\gridline{
\fig{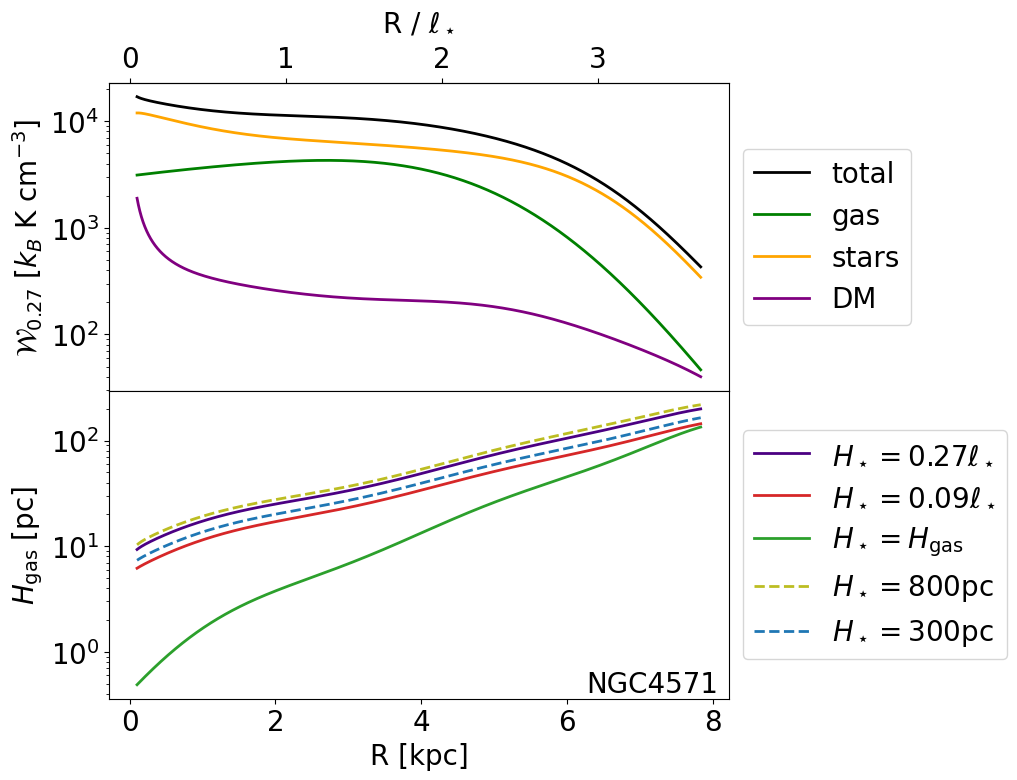}{0.49\textwidth}{NGC~4571}
\fig{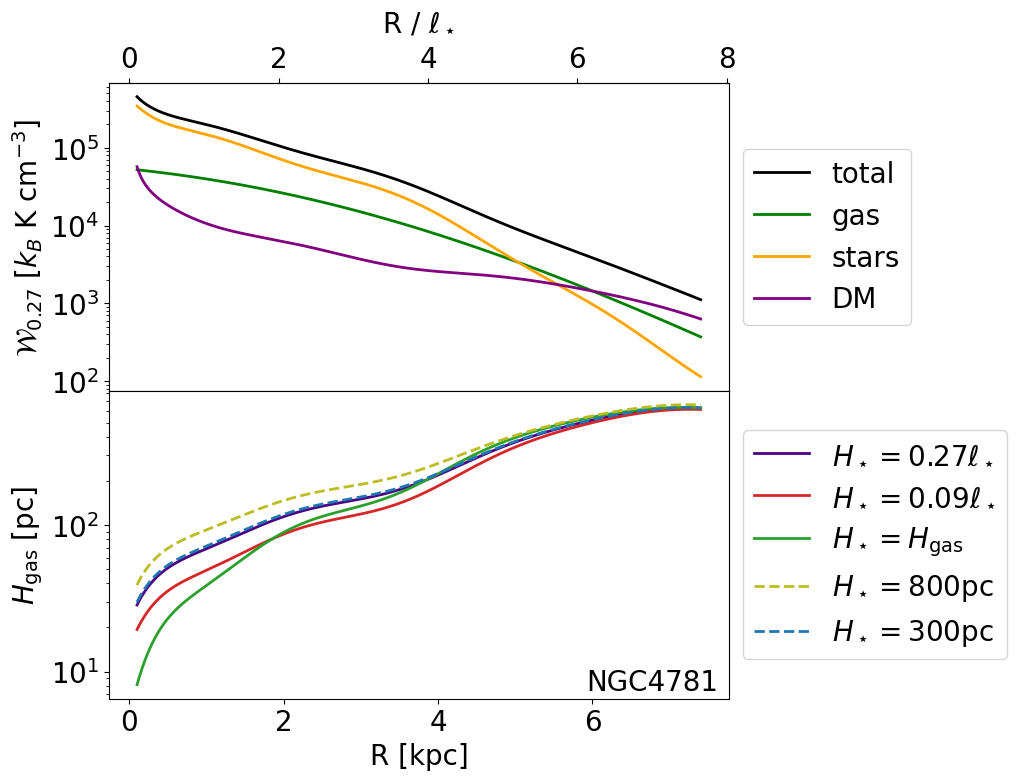}{0.49\textwidth}{NGC~4781}
}
\vspace{-1\baselineskip}
\gridline{
\fig{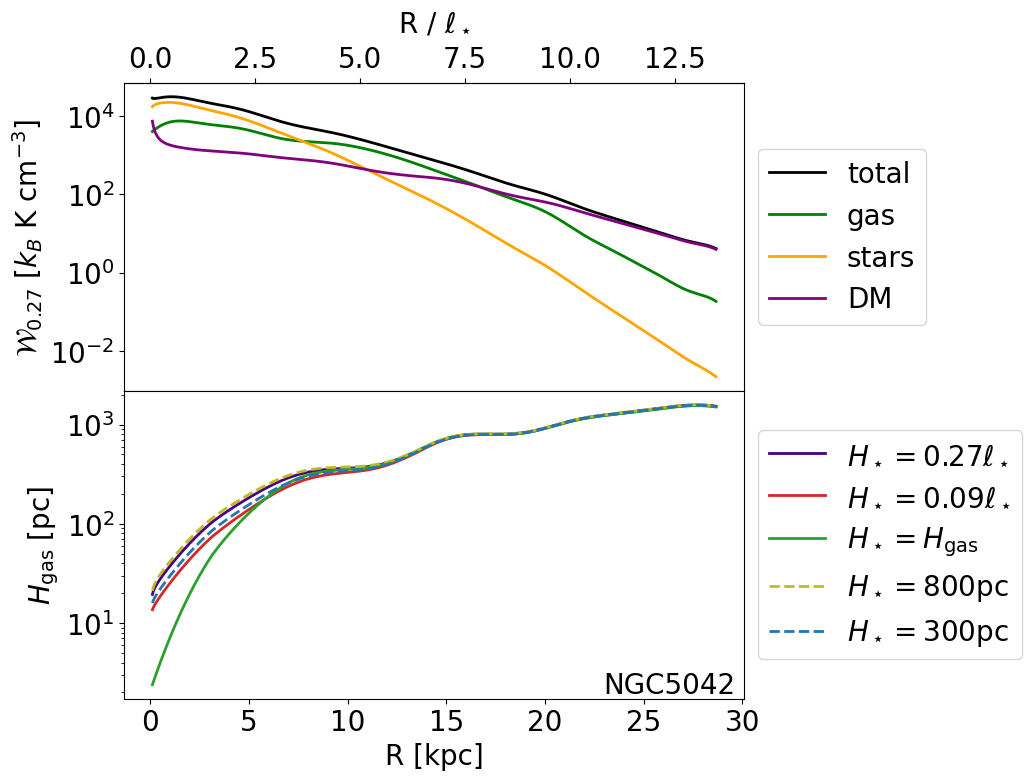}{0.49\textwidth}{NGC~5042}
}
\vspace{-0.5\baselineskip}
\caption{(continued)}
\end{figure*}

\bibliography{refs}{}
\bibliographystyle{aasjournal}

\end{document}